\documentclass[fleqn]{aa}
\usepackage[varg]{txfonts}

\usepackage{longtable}
\usepackage{rotating}
\usepackage{graphicx}
\usepackage{tikz}
\usepackage{siunitx}
\usepackage[urlcolor=blue,citecolor=blue,colorlinks=true]{hyperref}
\usepackage[colorinlistoftodos,prependcaption]{todonotes}
\usepackage[capitalise]{cleveref}
\usepackage{booktabs}

\newcommand{\mum}{\si{\micro\meter}}
\newcommand{\LAGN}{L_\mathrm{AGN}^{5100\si{\angstrom}}}
\DeclareSIUnit\angstrom{\text{Å}}
\DeclareSIUnit\erg{\text{erg}}

\begin{document} 
\title{Genuine Retrieval of the AGN Host Stellar Population (GRAHSP)}
\author{Johannes Buchner\inst{1,2}\thanks{\href{mailto:mara@mpe.mpg.de}{jbuchner@mpe.mpg.de}}
\and 
Hattie Starck\inst{1}
\and
Mara Salvato\inst{1,2}
\and 
Hagai Netzer\inst{3}
\and 
Zsofi Igo\inst{1}
\and 
Brivael Laloux\inst{4,5}
\and 
Antonis Georgakakis\inst{4}
\and
Isabelle Gauger\inst{1}
\and
Anna Olechowska\inst{1,7}
\and 
Nicolas Lopez\inst{1}
\and 
Suraj D Shankar\inst{6}
\and
Junyao Li\inst{8,9}
\and
Kirpal Nandra\inst{1}
\and
Andrea Merloni\inst{1}
}

\institute{
   Max Planck Institute for Extraterrestrial Physics, Giessenbachstrasse, 85741 Garching, Germany \and
   Excellence Cluster Universe, Boltzmannstr. 2, D-85748, Garching, Germany \and
   School of Physics and Astronomy, Tel Aviv University, Tel Aviv 69978, Israel \and
   Institute for Astronomy \& Astrophysics, National Observatory of Athens, V. Paulou \& I. Metaxa, 11532, Greece \and
   Centre for Extragalactic Astronomy, Department of Physics, Durham University, United Kingdom \and
   Institute of Astronomy and Astrophysics, University of Tübingen, Sand 1, 72076 Tübingen, Germany \and
   Ludwig-Maximilians-Universität München, Faculty of Physics, Scheinerstrasse 1, D-81679 München, Germany \and
   Department of Astronomy, University of Illinois at Urbana-Champaign, Urbana, IL 61801, USA \and
   Kavli Institute for the Physics and Mathematics of the Universe, The University of Tokyo, Kashiwa 277-8583 (Kavli IPMU, WPI), Japan
}

   \date{Received xx, Accepted xx}

  \abstract
       {
   The assembly and co-evolution of super-massive black holes (SMBH) and their host galaxy stellar population is one of the key open questions in modern galaxy evolution. Observationally constraining this question is challenging. Important parameters of galaxies, such as the stellar mass ($M_\star$) and star formation rate (SFR), are inferred by modeling the spectral energy distribution (SED), with templates constructed on the basis of various assumptions on stellar evolution. In the case of galaxies triggering SMBH activity, the active galactic nucleus (AGN) contaminates the light of the host galaxy at all wavelengths, hampering inference of host galaxy parameters. 
                  Underestimating the AGN contribution due to incomplete AGN templates results in a systematic overestimation of the stellar mass, biasing our understanding of AGN and galaxy co-evolution. This challenge has gained further impetus with the advent of sensitive wide-area surveys with millions of newly detected luminous AGN, including by eROSITA, Euclid and LSST.
      }
     {
   We aim to estimate robustly the accuracy, bias, scatter and uncertainty of AGN host galaxy parameters including stellar masses, and improve these measurements relative to previously used techniques.
      }
     {
   This work makes two important contributions. Firstly, we present a new SED fitting code, with an AGN model composed of a flexible power-law continuum with empirically determined broad and narrow lines and a FeII forest component, a flexible infrared torus that can reproduce the diverse dust temperature distributions, and appropriate attenuation on the galaxy and AGN light components. We verify that this model reproduces published X-ray to infrared SEDs of AGN to better than 20 per cent accuracy. 
   A fully Bayesian fit includes uncertainties in the model and the data, making the inference highly robust.
   The model is constrained with a fast nested sampling inference procedure supporting the many free model parameters.
   Secondly, we created a benchmark photometric dataset where optically-selected pure quasars are paired with non-AGN pure galaxies at the same redshift. Their photometry flux is summed into a hybrid (Chimera) object but with known galaxy and AGN properties. Based on this data-driven benchmark, true and retrieved stellar masses, SFR and AGN luminosities can be compared, allowing the evaluation and quantification of biases and uncertainties inherent in any given SED fitting methodology.
   }
     {
   The Chimera benchmark, which we release with this paper, shows that previous codes systematically over-estimate $M_\star$ and SFR by 0.5 dex with a wide scatter of 0.7 dex, at AGN luminosities above $10^{44}\unit{\erg\per\second}$. In 20 percent of cases, the estimated error bars lie completely outside a 1 dex wide band centered around the true value, which we consider an outlier. In contrast, GRAHSP shows no measurable biased on $M_\star$ and SFR, with an outlier fraction of only about 5 percent. 
   GRAHSP also estimates more realistic uncertainties.
   }
        {Unbiased characterization of galaxy hosting AGN enables characterization of the environmental conditions conducive to black hole growth, whether star formation is suppressed at high black hole activity, and identify mechanisms that prevent over-luminous AGN relative to the host galaxy mass. It can also shed light on the long-standing question whether AGN obscuration is primarily an orientation effect or related to phases in galaxy evolution.
   }


   \maketitle

 \section{Introduction} \label{sec:Intro}

The scaling relations of supermassive black holes (SMBH) at the center of massive, passive galaxies presented in \citet[][]{Magorrian1998}, together with the recognition that these are grown by active galactic nuclei (AGN) episodes \citep{Soltan1982} triggered a revolution in studies of galaxy evolution. It implies that galaxies and AGN are merely experiencing varying levels of SMBH accretion activity. Only combined, these phases give a full picture of galaxy evolution. Further studies extended SMBH scaling relations to all types of galaxies \cite[e.g., early- and late-type, dwarf and active galaxies][]{Gebhardt2000, Ferrarese2000, Xiao11, Gultekin11, Kormendy13, Schutte19}, finding correlations between the black hole mass ($M_\mathrm{BH}$) and host physical parameters (e.g., stellar mass and velocity dispersion). These facts, combined with the similarity of the cosmic star formation rate (SFR) and BH accretion history \citep[e.g.][]{Aird2012, MadauDickinson2014} points toward the interconnection between the SMBHs and their host and require considering the AGN phase in galaxy evolution studies. Indeed, models have been proposed in which the energy produced during SMBH growth episodes can modulate the formation of stars in the host galaxy \citep[e.g.,][]{Sanders1988}, thereby establishing AGN as an important component for our understanding of galaxy evolution \citep[e.g.,][]{DiMatteo2005,Hopkins2008}.
In turn, this complicates a task that was already difficult for inactive galaxies. Unlike redshift, estimates of galaxy physical parameters such as stellar mass, star formation rate, and bolometric luminosity cannot be measured directly. Known degeneracies like age-metallicity \cite[][]{Worthey94} or  color-redshift \cite[][]{Masters2015} are hard to break with SED fitting, in part because the number of parameters in the fit is usually larger than the number of photometric data points.

For galaxies hosting an active SMBH, the need to account for the nuclear emission at various wavelengths complicates things further. The relative host/AGN contribution is unknown, and the AGN contribution at various wavelengths changes, depending on whether or not the nuclear emission is obscured by dust. Thus, basic questions such as ``Are the physical properties different for galaxies hosting an inactive/active SMBH?'', ``Are the differences between obscured and unobscured sources dependent on orientation or evolution?'' remain unanswered.
Not correctly accounting for the nuclear component can result in inaccurate and/or bias galaxy parameters and hence interpretations, and could be the origin of conflicting results reported in the literature so far. For example, \citet{Stemo2020}, using a sample of about 2500 X-ray selected AGN, found that AGN host galaxies tend to have lower SFR than normal galaxies of equal mass. This result is in contrast with what was found in \citet{Mountrichas2022}, \citet{Pouliasis2022} \citet[][]{Suh19} for the same type of AGN. 
It has also been reported that the stellar mass of AGN host galaxies depends on their level of obscuration \citep[suggesting an evolutionary link;][]{Koutoulidis2022, Zou2019} and in other work that it does not \citep[consistent with the AGN unification model;][]{Masoura2021}. If there is an evolution in place from obscured to unobscured objects as suggested by \citet[][]{Hopkins2008}, how can it be that the hosts of obscured AGN are more massive than the host of the unobscured ones
\citep[][]{Koutoulidis2022}?
This uncertainty in the actual stellar mass of AGN host galaxies indirectly affects other parameters, such as specific accretion-rate \citep[used to estimate the rate at which the BH accrete material][]{Aird2012, Georgakakis2017a} and specific star formation.

The increased availability of wide- and all-sky areas allows the selection of bright, unobscured AGN. Contrary to typical pencil-beam surveys (e.g., Lockman Hole \citep{Fotopoulou2016}, Chandra deep-field south \citep{HsuCDFSz2013, Luo2016}), where bright, unobscured AGN are virtually absent, they are abundant in all-sky surveys. To mention already existing surveys, these include optical spectroscopic by the Sloan Digital Sky Survey SDSS \citep{Eisenstein2011SDSS}, photometry imaging by the Dark Energy Spectroscopic Instrument Legacy Surveys \citep{Dey2019}, mid-infrared by WISE \citep{Wright2010, Mainzer2011, Meisner2023}, astrometry by Gaia \citep[e.g.,][]{Gaia2023a}, X-rays by 4XMM \citep{Webb2020}, XMMSlew \citep{Saxton2008XMMSL1}, and eROSITA \citep{Predehl2021}. 
 Therefore, a major effort is required to develop algorithms that provide more reliable physical parameters for unobscured AGN, finally allowing for inclusive galaxy evolutionary studies. In recent years an increasing number of algorithms have been developed (and often made public) for this purpose, such as AGNFitter \citep[][]{CalistroRivera2016}, MAGPHYS \citep[][]{daCunha2008}, PROSPECTOR \citep[][]{Johnson2021}, CIGALE \citep[][]{Boquien2019}, X-CIGALE \citep[extending CIGALE to X-ray emission;][]{Yang2020}, FAST \citep{Aird2017}, among others \citep[see review by][]{Pacifici2023}. Figure 1 of \cite{Thorne2022} nicely summarises the basic ingredients of the most used codes\footnote{More illustrations are available at \url{https://www.astrojess.com/graphics/interactive-sed-diagram}.}.

This paper demonstrates that the existing approaches are insufficient in the new parameter space, and present a new code, GRAHSP (Genuine Retrieval of the AGN Host Stellar Population). GRAHSP differs substantially through a more complete model description of the AGN component and more realistic uncertainty estimation.
 GRAHSP has been developed by the eROSITA team specifically for measuring the physical parameters of the millions of AGN that eROSITA started to detect in 2020 \citep[see][that presents the first all-sky survey data release]{Merloni2024}. 
The reliability of measurements obtained with GRAHSP at different host-to-AGN ratios has been tested thoroughly by measuring the physical parameters of AGN constructed by combining galaxies with known stellar mass from COSMOS  with a QSO. We have defined this testing sample as the CHIMERA sample, an ideal benchmark for reference. For non-AGN galaxies we verify that GRAHSP results are consistent with published results. This ensures that the use can be generalised to any sample of extragalactic sources.
In addition to stellar mass, GRAHSP estimates key galaxy parameters such as the star-formation rate, AGN parameters such as the bolometric luminosity, power-law continuum slope, torus covering factor, and the system's dust attenuation.

In \cref{sec:Method} we present GRAHSP. The samples and data sets used in this work are described in \cref{sec:Data}, which includes AGN-free galaxy samples  and infrared, X-ray, and optically selected AGN samples. \Cref{sec:data:Chimera} describe a novel benchmark data set, where the true decomposition between galaxy and AGN light is known by construction. In \cref{sec:Test}, the model is vetted with initial tests. Based on the benchmark data set, section~\ref{sec:Results} demonstrates the key result: GRAHSP's stellar mass estimates are unbiased and outperform literature methods, even in the difficult case of unobscured, luminous AGN.
Throughout the paper, we assume AB magnitudes unless stated otherwise and consistently use the cosmology of \cite{Planck18}.\section{Method \label{sec:Method}}

The modeling of multi-wavelength photometry of AGN candidates can be driven by several goals: (1) the determination of redshifts (photo-z); (2) the identification of whether an AGN is present; (3) the measurement of AGN parameters, such as luminosities, and (4) the measurement of galaxy parameters.

If the goal is to estimate redshifts, the colour-redshift relation can be mapped with a small set of galaxy, AGN, and hybrid templates \citep[e.g.][]{Salvato2009, Salvato2019}. If the goal is to identify AGN, light at any wavelength in excess of that expected by galaxy emission needs to be detected, e.g., by the statistical preference for an additional blue power-law (associated with the accretion disk) or the presence of mid-infrared bump components (related nuclear dust heated by the AGN). Two commonly adopted models for these two components are those proposed by \cite{Richards2006} and \cite{dale2014ApJ...784...83D}, respectively; see \cite{Thorne2022} for a recent work. The goal can also be achieved with diagnostic colour plots in the UV \citep[e.g.][]{Richards2009} or infrared \citep[e.g.][]{Lacy2004}.

If the goal is the inference of AGN parameters, such as those of a clumpy torus \citep{Nenkova2008a,Honig2010,Stalevski2016} or the estimation of the black hole spin assuming a certain accretion flow model \citep{Netzer2014}, one may assume an AGN model and interpret the inferred parameters within this framework for the scope of the study.
The most challenging goal is the robust inference of galaxy parameters under the presence of AGN light. This is the goal tackled in this work.

Suppose, for example, the case of photometry from an extremely luminous quasar being modeled by an AGN template library\footnote{Current codes all try to account for AGN contribution. Modelling without an AGN, can be considered the extreme case of the described scenario.}. If any residuals remain at this stage due to an imperfect AGN modelling, and a galaxy template library is added and optimized to fit the residuals, the stellar population will be extremely massive (to reach quasar luminosities) and either red (to fix residuals in the near-infrared) or blue (to fix residuals in the UV-optical). The inferred parameters will be driven by the AGN mis-modelling and have nothing to do with the underlying galaxy.
In \cref{sec:Results}, we will show that this hypothetical scenario occurs in real applications leading to biased measurements of key physical parameters.
While the inference would be wrong, this may not be detectable by residual diagnostics such as reduced $\chi^2$, since there may not be any residuals remaining. Also, the true stellar mass cannot be easily verified by other, independent methods.
The over-estimated galaxy template normalisation cannot be resolved by refining the AGN template normalisation further with data from longer and shorter wavelengths, because the AGN template is already in the right place. This includes incorporating X-ray information \citep[e.g.,][]{Yang2020}.
This failure mode has the worrying potential to induce spurious correlations between host galaxy mass and AGN accretion rate.

In addition to the challenges above, AGN are ubiquitously variable \citep{Bershady1998,Klesman2007,Sesar2007SDSSvariability,Trevese2008}. While galaxies are, for the most part, static light emitters on human time scales; and their SEDs can be modelled with a single static template, AGN break this assumption. Therefore, residuals caused by non-simultaneously collected photometry should also be considered in the modelling.

\begin{figure*}[ht]
    \centering
    \includegraphics[width=\textwidth]{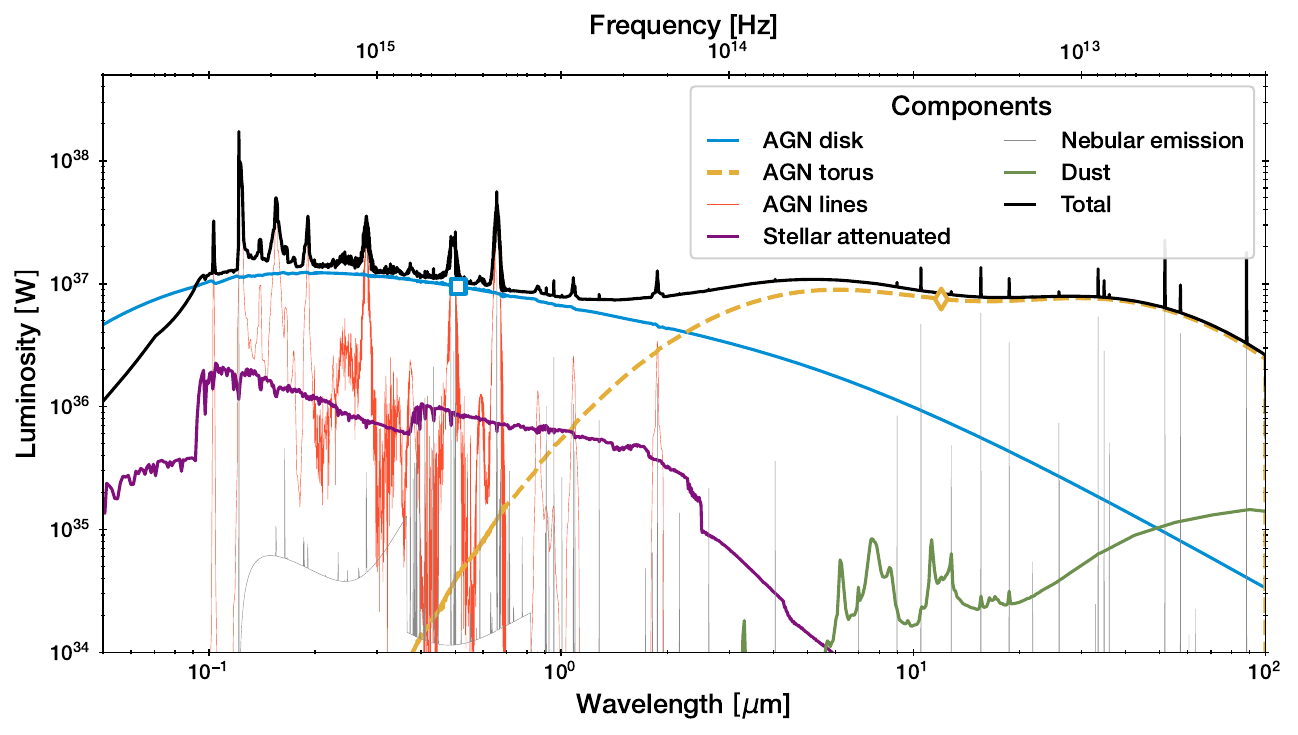}
    \caption{Overview of how the individual model components contribute to the summed emission (black).
    The AGN power-law continuum (blue; accretion disk, or BBB) is enhanced by emission lines including an iron forest (red). The disk model is normalised at the monochromatic luminosity at 5100$\,\si{\angstrom}$ (blue square), here $\LAGN=\qty{e44}{\erg\per\second}=\qty{e37}{\watt}$. The torus (yellow dashed curve) typically dominates the disk continuum above approximately $1\si{\micro\meter}$. The torus is normalised by the luminosity ratio 5100$\,\si{\angstrom}$ to 12$\si{\micro\meter}$ (yellow diamond).
    The galaxy components include a stellar population (purple), nebular emission lines (gray at the bottom), and galaxy dust emission (green at the bottom right).
    }
    \label{fig:modelcomp}
\end{figure*}
\begin{figure*}[ht]
    \centering
    \includegraphics[width=\textwidth]{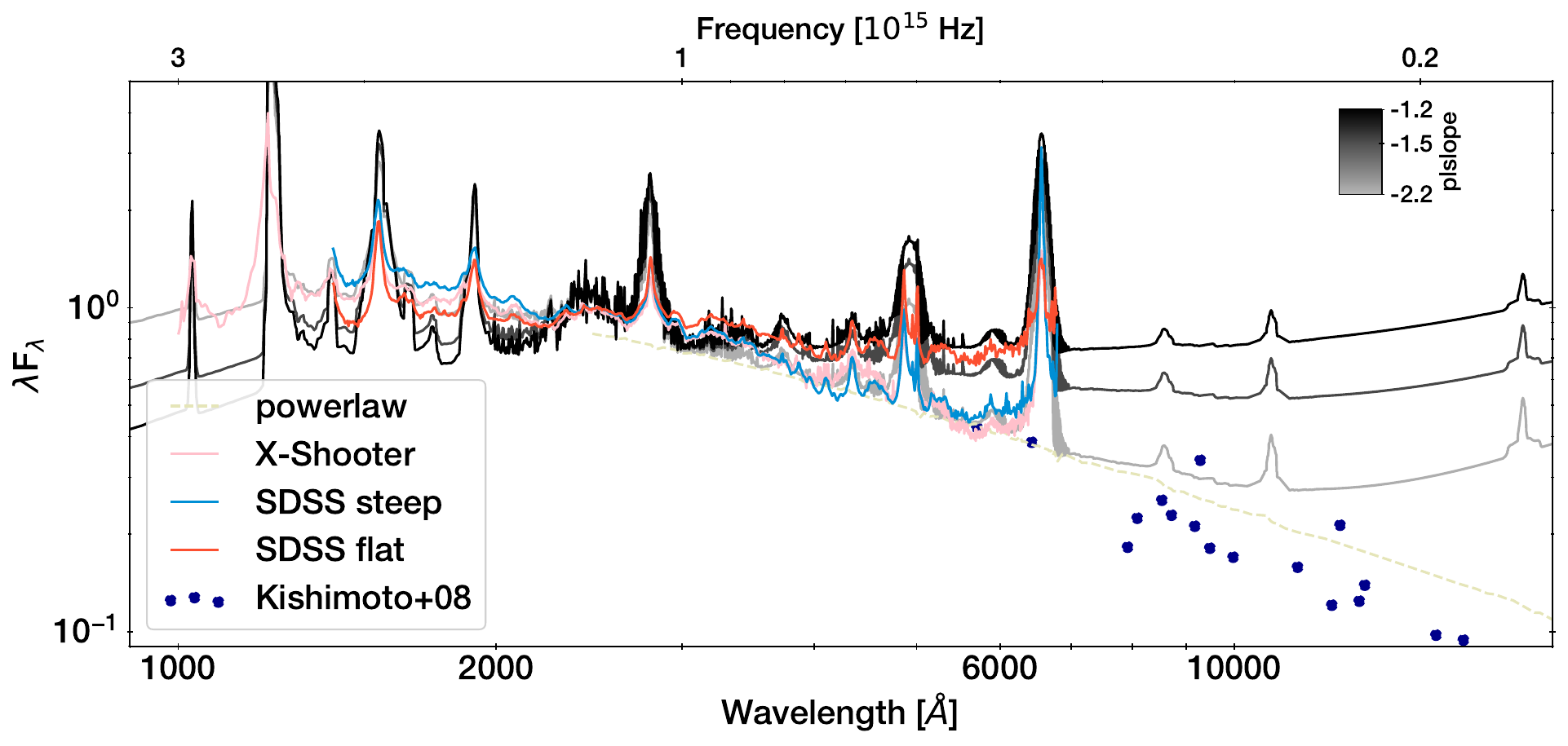}
    \caption{Detailed view of the optical continuum and emission lines. The model continuum bending power-law (yellow dashed line) reproduces the polarization measurements of \cite{Kishimoto2008} (dark blue data points). The full model including a torus and emission lines is shown in solid lines.
    Variations of the power-law slope (light grey to black) reproduce SDSS steep and flat unabsorbed spectra from \cite{Richards2006} and \cite{Selsing2016XShooter} (red, blue and pink lines). Towards the infrared, the torus component dominates the continuum.
    }
    \label{fig:optagn}
\end{figure*}

In the absence of a physical model that accurately reproduces the AGN-plus galaxy photometry without significant residuals, two broad approaches have been followed in the literature:
(1) The photometry can be degraded (by increasing the errors) until the model provides a sufficient description. This approach is implemented in CIGALE v2022.1 \citep{newcigale2022ApJ...927..192Y}, for example. (2) A suitable AGN template library can be derived empirically, for instance, from spectroscopy. Some examples of this approach include the infrared modelling of \cite{Mullaney2011,Mor2012,Kirkpatrick2012}, and the recent parametric Seyfert 1 model of \cite{Temple2021}.
These templates still have to be constructed to be flexible enough to capture the full diversity of the AGN phenomenon. While such an empirical AGN model may not be directly associated with physical properties, it is effective at ``deleting'' the AGN light contribution so that the host galaxy light can be analysed, with realistic uncertainty estimates.
The necessary flexibility of the AGN model leads to many free fitting parameters, in addition to the galaxy parameters. It is then easily possible that the number of parameters exceeds the number of data points. Such over-parametrisation leads to degeneracies which need to be explored computationally. Bayesian inference methods can address this \citep[see e.g.][]{CalistroRivera2016}.

We are primarily following the second approach. In \cref{sec:Model}, we describe our flexible parametric AGN model and detail its components. \Cref{subsec:calibration} validates the model's capability to predict high-quality AGN-galaxy hybrid spectra. The parameter ranges are also calibrated there. Model parameters are typeset in \texttt{monospace}, as are code module names in the section headers.
For analysing photometry, GRAHSP builds on some of the open source infrastructure from CIGALE and accepts similar input and configuration files.
However, the computational methods for achieving rapid model evaluation are substantially different (\cref{sec:computationengine}).
For fitting, GRAHSP is paired with an advanced inference procedure for uncertainty quantification, which considers model limitations and AGN variability (\cref{sec:Inference}).

\subsection{Model\label{sec:Model}}

The spectral model implemented in GRAHSP consists of several components. The AGN components include a continuum (\cref{sec:activatepl}) with emission lines (\cref{sec:activatelines}) and an infrared torus (\cref{sec:activatetorus}). Bolometric AGN luminosities are defined in \cref{bolometric}. The galaxy components are presented in \cref{sec:galmodel}. Both the AGN and galaxy components can undergo dust attenuation (\cref{sec:attenuation}) and finally redshifting (\cref{sec:redshifting}).
\Cref{fig:modelcomp} gives a visual overview of the model components, which are now introduced in order.

\subsubsection{The big blue bump continuum (\texttt{activatepl})\label{sec:activatepl}}

We adopt a flexible model motivated by theoretical and observational considerations. Thin accretion disk models \citep[][]{Shakura1973} predict a power-law emission spectrum in the UV to the optical range with a smooth UV bend-over that depends on the spin, black hole mass, and Eddington ratio \citep[e.g.][]{Netzer2014}. In observed SEDs, the bend towards the far-UV is wider than predicted \citep[e.g.][]{Blaes2001}, i.e. it is UV-brighter.
This can be partially addressed by assuming strongly inhomogeneous thin disks that locally fluctuate in temperature \citep[][]{Dexter2011}. 
Because of these complications, we avoid assuming a specific physical accretion flow model and instead model the big blue bump (BBB) phenomenologically. 
From Ly$\alpha$ to $1\mum$, observed stacked spectra \citep[e.g.][]{Zheng1997HST,Selsing2016XShooter} show a power-law spectral density $F_\lambda\propto\lambda^\alpha_\lambda$ with typical indices $\alpha_\lambda$ between -1.3 and -2.2. There is substantial object-to-object diversity \citep[][]{Richards2006} not attributed to dust attenuation. Additionally, the power-law continuum can change its slope on time scales of hundreds of days, as demonstrated by optical difference spectra in \cite{Ruan2014}.
Above $1\unit{\micro\meter}$, emission by the torus (discussed further in \cref{sec:activatetorus}) adds to the SED of the big blue bump. Only thanks to polarization-spectroscopy by \cite{Kishimoto2008} could this transition be directly disentangled free of any additional contributions by reprocessing (emission lines, dusty torus). They showed that the big blue bump emission continues a power-law decline at least until $2\mum$ with $\alpha_\lambda\approx-1.7$ for the observed source.

To encompass this diversity in continua, we adopt the smooth bending power-law (SBPL) parameterisation of \cite{Ryde1999}. This formulation transitions from a power-law with index $\alpha_1$ (\texttt{uvslope} parameter) to one with index $\alpha_2$ (\texttt{plslope}) at a break wavelength $\lambda_\mathrm{break}$ (\texttt{plbendloc}). The width of the bend can be controlled with the parameter $\Lambda$ (\texttt{plbendwidth}).
The key \texttt{L\_AGN} parameter $\lambda \LAGN$ sets the power-law normalisation at $\lambda_0=5100\si{\angstrom}$. The luminosity spectral density is:
\begin{align}
L(\lambda)=\LAGN \times \left(\frac{\lambda}{\lambda_0}\right)^{(\alpha_1 + \alpha_2 + 2)/2} \times 
 \left(\frac{e^{q} + e^{-q}}{e^{q_\mathrm{piv}} + e^{-q_\mathrm{piv}}}\right)^{\Lambda\times(\alpha_2 - \alpha_1)/2} \times \frac{\lambda_0}{\lambda}\label{eq:sbpl}
\end{align}
with $q=\ln\left(\lambda/\lambda_\mathrm{break}\right)/\Lambda$ and $q_\mathrm{piv}=\ln\left(\lambda_0/\lambda_\mathrm{break}\right)/\Lambda$.
We verified that recent state-of-the-art relativistic accretion disk simulations \cite[e.g.,][]{Hagen2023reldisk} can be approximated with \cref{eq:sbpl}. The smooth bend-over towards the UV is illustrated by the blue curve in Figure~\ref{fig:modelcomp}.

\subsubsection{AGN emission lines \texttt{activatelines}\label{sec:activatelines}}
AGN emission lines are superimposed on the continuum. Their contribution to the observed narrow or broad-band photometry can be substantial \citep[see e.g.][]{Temple2021}\footnote{Already the same consideration was done in the past for the contribution of emission lines in the SED fitting of normal galaxies \citep[e.g.,][]{Ilbert2006, Cardamone2010,HsuCDFSz2013,Ananna2017}.}. 
The line luminosity of H$_{\mathrm \beta}$ is set by default to 2 per cent of $\LAGN$ for the broad line component and to 0.2 per cent for the narrow line component. This can be modified by the \texttt{linestrength\_boost\_factor} parameter.
The luminosity ratio of the other lines relative to H$_{\mathrm \beta}$ is set based on the broad and narrow line list of \citet[][]{Netzer1990lines}, listed in \cref{tab:linelist}. H$\gamma$ is added based on \cite{Rakshit2020}.
With the luminosity ratio defined, the full-width-half-maximum (FWHM; \texttt{linewidth} parameter) of the lines can be chosen somewhat arbitrarily between hundreds and tens of thousands $\si{\km\per\s}$, since this cannot be distinguished with photometry.

\begin{table}
\centering
    \caption{List of broad and narrow emission lines, with their centre wavelength, and normalisations relative to H$\beta$. Based on \cite{Netzer1990lines}. These lines are considered in GRAHSP, in addition to FeII.}
\begin{tabular}{l r c c}
Line & $\lambda$ [$\si{\angstrom}$]  & broad & narrow \\ 
\hline 
\hline
Ly$\beta$ & 1026 & 1 & 0.1  \\ 
OVI & 1035 & 3 & 0.3  \\ 
Ly$\alpha$ & 1215 & 10 & 35  \\ 
NV & 1240 & 3 & 3  \\ 
OI & 1304 & 0.5 & 0  \\ 
CII & 1336 & 0.3 & 0  \\ 
SiIV,OIV] & 1400 & 1.5 & 1  \\ 
NIV] & 1486 & 0.7 & 1  \\ 
CIV & 1549 & 5 & 8  \\ 
HeII & 1640 & 0.6 & 0.6  \\ 
OIII] & 1663 & 0.5 & 0.5  \\ 
NIII] & 1750 & 0.4 & 0.4  \\ 
CIII]+SiIII & 1909 & 3 & 3  \\ 
MgII & 2798 & 3 & 1  \\ 
{[NeV]} & 3426 & 0 & 1  \\ 
{[OII]} & 3727 & 0 & 3  \\
{[NeIII]} & 3869 & 0 & 0.5  \\ 
H$\gamma$ & 4861 & 0.25 & 0.25  \\ 
HeII & 4686 & 0.1 & 0.1  \\ 
H$\beta$ & 4861 & 1 & 1  \\ 
{[OIII]} & 4959 & 0 & 3  \\ 
{[OIII]} & 5007 & 0 & 9  \\ 
HeI & 5876 & 0.15 & 0.15  \\ 
{[FeVII]} & 6087 & 0 & 0.1  \\ 
{[OI]} & 6300 & 0 & 0.6  \\ 
{[FeX]} & 6374 & 0 & 0.04  \\ 
H$\alpha$ & 6563 & 4 & 3  \\ 
{[NII]} & 6583 & 0 & 1  \\ 
{[SII]} & 6716 & 0 & 1  \\ 
{[SII]} & 6731 & 0 & 1  \\ 
CaII & 8498 & 0.1 & 0  \\ 
CaII & 8622 & 0.1 & 0  \\ 
{[SIII]} & 9069 & 0 & 0.2  \\ 
{[SIII]} & 9532 & 0 & 0.2  \\ 
HeI & 10830 & 0.3 & 0.3  \\
P$\alpha$ & 18750 & 0.4 & 0.3  \\ 
\hline
    \end{tabular}
    \label{tab:linelist}
\end{table}

In addition to the individual lines, a FeII forest template is added. Following \cite{Merloni2010}, we select from \cite{Bruhweiler2008} the template with density $n_\mathrm{H}=10^{11}\,\si{\per\cm^3}$, microturbulence $\xi=20\,\si{\km\per\s}$ and ionising flux $\phi_\mathrm{H}=10^{20.5}\,\unit{\cm^{-2}\second^{-1}}$, and de-redshift it from $z=0.004$. The template luminosity is normalised at $\lambda=4593.4\,\si{\angstrom}$ relative to the H$_{\mathrm \beta}$ line luminosity (see above) with ratio $A_\mathrm{FeII}$ (\texttt{AFeII}). We also tested inclusion of a Balmer continuum component, however, the improvements to the presented results were negligible.

\Cref{fig:modelcomp} illustrates the importance of these features. The total SED (black) is elevated in the optical wavelengths above the power-law component (blue) by the Fe forest and lines (red). \Cref{fig:optagn} compares the AGN model with different slope (gray shades) to the empirical steep and flat quasar templates by \cite{Richards2006} (blue and red curves) and the higher redshift stacked quasar spectrum of \cite{Selsing2016XShooter} (pink curve). The key emission features, shown here with $A_\mathrm{lines}=1$, $A_\mathrm{FeII}=6$ and $W_\mathrm{lines}=10000\unit{\kilo\meter\per\second}$, are present in the model and reproduce the observations. At the shortest wavelengths below $2000\unit{\angstrom}$, the current FeII template is potentially missing emission.
At wavelengths above $1\unit{\micro\meter}$, the power law continues downwards (dashed line) as required by the spectral-polarimetric observations (blue circles), which were anchored to the dashed line in this figure at the lowest measured wavelengths. Above $1\unit{\micro\meter}$, the emission begins to be dominated by the emission of the torus, which is discussed in the next section.

\subsubsection{AGN torus \texttt{activategtorus}\label{sec:activatetorus}}

\begin{figure*}[ht]
    \centering
    \includegraphics[width=\textwidth]{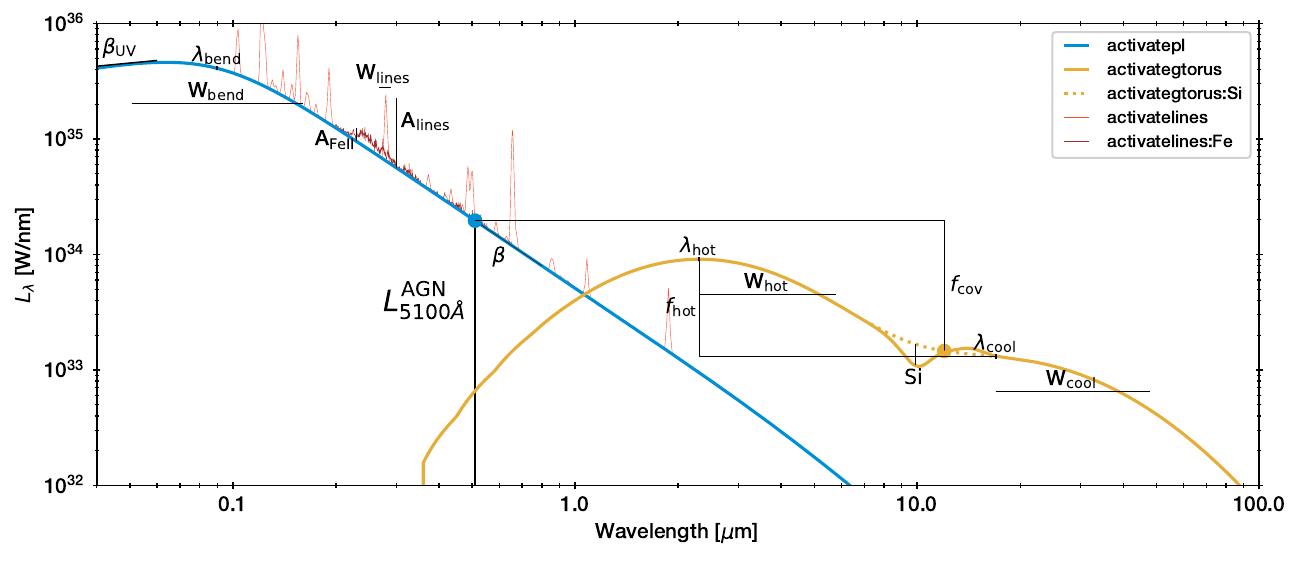}
    \caption{Overview of the AGN model parameters and how they configure the spectrum, shown here in $L_\lambda$ with arbitrary units.
    The power law (blue) is normalised at $5100\unit{\angstrom}$, where it has a power law slope of $\beta$. The power law bends over at $\lambda_\mathrm{bend}$ towards the UV, where it has slope $\beta_{UV}$. The width of this transition is set by $W_\mathrm{bend}$.
    Emission lines with FWHM $W_\mathrm{lines}$ and an FeII template are added and can be further scaled by $A_\mathrm{lines}$ and $A_\mathrm{FeII}$, respectively.
    The torus component (dark yellow) is normalised by the ratio of $12\mu{}m$ and $5100\unit{\angstrom}$ $\lambda L_\lambda$ luminosities, $f_\mathrm{cov}$ (see \cref{eq:fcov}). It consists of the sum of two log-quadratic curves, with width $W_\mathrm{cool}$ and $W_\mathrm{hot}$ and location $\lambda_\mathrm{cool}$ and $\lambda_\mathrm{hot}$. The peak-to-peak $\lambda L_\lambda$ ratio is set by $f_\mathrm{hot}$ (see \cref{eq:fhot}).
    The depth of the Si feature, in emission if positive or in absorption if negative (here: -1), is set by Si.
    The flexibility of these 15 parameters is restricted in \Cref{subsec:calibration}.
    }
    \label{fig:AGNcomponents}
\end{figure*}

The AGN infrared emission is associated with the reprocessing of UV emission by dust \citep[see e.g. the review by][]{Netzer2015}. The infrared emission is composed of at least two components, one dominating in the mid-infrared (cold dust) and one dominating in the near-infrared (hot dust). 
At long wavelengths, the infrared emission has recently been imaged in nearby galaxies \citep[e.g.][]{Garcia-Burillo2021} and is associated with parsec-scale, cold molecular dust (the torus). 
In addition to the cold dust, a near-universal component is hot dust \citep[e.g.][]{Mor2012}. Through interferometry observations, the hot dust was associated with polar regions relative to the torus \citep[][]{Tristram2014,Asmus2016}. 
However, \cite{Lyu2017} found that there is a substantial population (30-40 per cent) of AGN deficient in warm or hot dust. This suggests that the covering factors, geometries, and/or dust properties are diverse.

The properties of the dust, chemical composition and grain size distribution are also not fully understood \citep[e.g.][]{Hoenig2013}. In empirical studies, the SED is often approximated by a grey body. Towards short wavelengths, such parameterisations are flexible, while at the longest wavelength, typically a $\lambda^{-4}$ power-law decline is assumed (Rayleigh-Jeans tail).
The average shape at the longest wavelength is debated, for example, in \cite{Xue2011} and \cite{Symeonidis2022}. Direct observations with Herschel by \cite{Bernhard2021} show a wide, log-quadratic bend peaking around $30-50\unit{\micro\meter}$ that is completely dominated by the host galaxy already at $\geq70\unit{\micro\meter}$.

We use a parameterised empirical template that can emulate this diversity.
The hot dust components can be approximated by black bodies with a temperature distribution. The peak and low-wavelength distribution is then well described by a log-quadratic curve. This functional form is sub-optimal at the longest wavelength, where the Rayleigh-Jeans law motivates a linear rather than quadratic decline in a log-log plot.
However, the tail of the torus hot dust component is typically dominated by cold galaxy-scale dust emission. As a result, deviations of the log-quadratic at the long-wavelength regime are negligible for the modelling (we verify this below in \cref{subsec:calibration}). The same argument applies to the tail of the hot dust, which is dominated by the cold dust component.
We therefore use the sum of two log-quadratic curves for the hot and cold dust, which are:
\begin{equation}
\begin{split}
L_\mathrm{cool} & \propto \exp[-(\lambda - \lambda_\mathrm{COOL})^2 / (2W_\mathrm{COOL}^2)],\\
L_\mathrm{hot} &\propto \exp[-(\lambda - \lambda_\mathrm{HOT} \, \, \,)^2 /(2W_\mathrm{HOT}^2)].
\label{eq:dust}
\end{split}
\end{equation}
The parameters include the peak $\lambda_\mathrm{COOL}$ (\texttt{COOLlam}) and width $W_\mathrm{COOL}$ (in dex; \texttt{COOLwidth}) of the Gaussian-like form for the spectrum (and analogously for the hot dust component). These parameters are related to the temperature distribution of the dust.
The normalisation of the hot dust peak relative to the cold dust peak in $\lambda L_\lambda$ is a parameter that we term the hot factor:
\begin{equation}
f_\mathrm{hot}=\frac{\lambda_\mathrm{HOT} L_\mathrm{HOT}}{\lambda_\mathrm{COOL}}.
\label{eq:fhot}
\end{equation}
With the two log-quadratic curves fixed, the normalisation of the torus to the AGN powerlaw has to be defined. 
The ratio of the near to mid-infrared continuum torus template amplitude to that of the UV to optical power-law template, $f_\mathrm{cov}$ (\texttt{fcov}), is commonly referred to as the torus covering factor:
\begin{equation}
\frac{(\lambda L_\lambda)(12 \mum)}{(\lambda L_\lambda)(5100\si{\angstrom})}=2.5 \times f_\mathrm{cov}.
\label{eq:fcov}
\end{equation}
However, the translation to a geometric covering factor of a physical structure is complicated by anisotropy in the emission profile and distribution of dust clumps  \citep{Stalevski2016}.
\Cref{subsec:calibration} below demonstrates that the model can reproduce torus spectral templates derived by other works, including those of \cite{Lyu2017}, \cite{Mor2012}, \cite{Mullaney2011} and \cite{Kirkpatrick2012}, as well as individual AGN in the local Universe.

An important feature in the mid-infrared is the absorption and emission by Si dust at $12\mum$. Because the Si feature is not as pronounced as smooth models suggest, this has been interpreted as evidence for a clumpy torus \citep{Nenkova2008}. However, it was later shown that similar effects can be produced with smooth geometries \citep{Feltre2012}. \cite{Goulding2012} showed that deep Si absorption features are associated with edge-on galaxies, suggesting that non-nuclear dust imprints this feature.

We allow the model to place Si in absorption or emission. An Si template is created by taking the difference between the average templates of faint (on average in absorption) and luminous (on average in emission) AGN in the 8 to 18$\mum$ range from \cite{Mullaney2011}. The template is normalised at $12\mum$. The \texttt{Si} parameter then controls the amplitude of this contribution, with zero corresponding to it not being present.
The effect of this parameter is very localised, as shown by comparing the dotted and solid dark yellow curves in \cref{fig:AGNcomponents}. Indeed, this parameter is inconsequential if no photometry filter covers the 8 to $18\mum$ rest-frame wavelengths.

\subsubsection{Bolometric luminosities}\label{bolometric}

The luminosity integrated over the entire wavelength range is a crucial measure of the radiative energy budget in the circum-nuclear environments of supermassive black holes. However, since some radiation is absorbed and then re-emitted (as emission lines or in the torus), some care is needed to define what to compute to avoid double-counting.
Furthermore, because our conversion between fluxes and luminosities assumes the luminosity distance, we compute isotropic luminosities, i.e., under the assumption of an isotropic radiation profile. Converting to more realistic, anisotropic emission luminosities requires a physical model \citep[see, e.g.,][for a detailed discussion]{Stalevski2016}.

GRAHSP computes two bolometric AGN luminosities. Both are intrinsic, i.e., before applying attenuation, rest-frame, isotropic luminosities.
The first, \texttt{lumBolBBB}, integrates all AGN SED components except for the torus upwards of 91.2nm. The substantial contribution of the ionising far-UV and X-rays are not included in \texttt{lumBolBBB}. This is because the wavelength range is rarely directly measured. The user should apply a model-dependent correction factor to obtain the true bolometric luminosity.
The second is the bolometric luminosity of the torus, \texttt{lumBolTOR}, integrated over the entire wavelength range. While \texttt{lumBolBBB} is primarily informed by data in UV to optical rest-frame wavelengths, \texttt{lumBolTOR} is informed by infrared data.
Both bolometric luminosities are reported, as well as the ratio between them, \texttt{ratioTORBBB}=\texttt{lumBolTOR}/\texttt{lumBolBBB}. How to interpret such a ratio of instantaneous, isotropic luminosities as a covering factor of a light-year-sized torus that reprocesses the radiation from an anisotropically emitting accretion disk is discussed extensively in \cite{Stalevski2016}.

\subsubsection{Galaxy model\label{sec:galmodel}}

For the host stellar population, we adopt standard CIGALE modules \cite{Boquien2019,newcigale2022ApJ...927..192Y}. These implement stellar population synthesis (SPS) by \cite{Bruzual2003} (\texttt{bc03}) and \cite{maraston2005MNRAS.362..799M} (\texttt{m2005}) combined with a parametric star formation history (SFH) model.
The stellar population is scaled to a total stellar mass of $M_\star$, a key parameter setting the normalisation of the galaxy emission spectrum.
In the SPS, a metallicity needs to be assumed.
The effects of metallicity on the resulting SED are well known to be degenerate with stellar age. Instead of exploring this degeneracy, we follow previous work and choose to measure the effective stellar age, assuming a fixed metallicity, and interpret the results in this context.

\begin{figure}[ht]
    \centering
    \includegraphics[width=1\columnwidth]{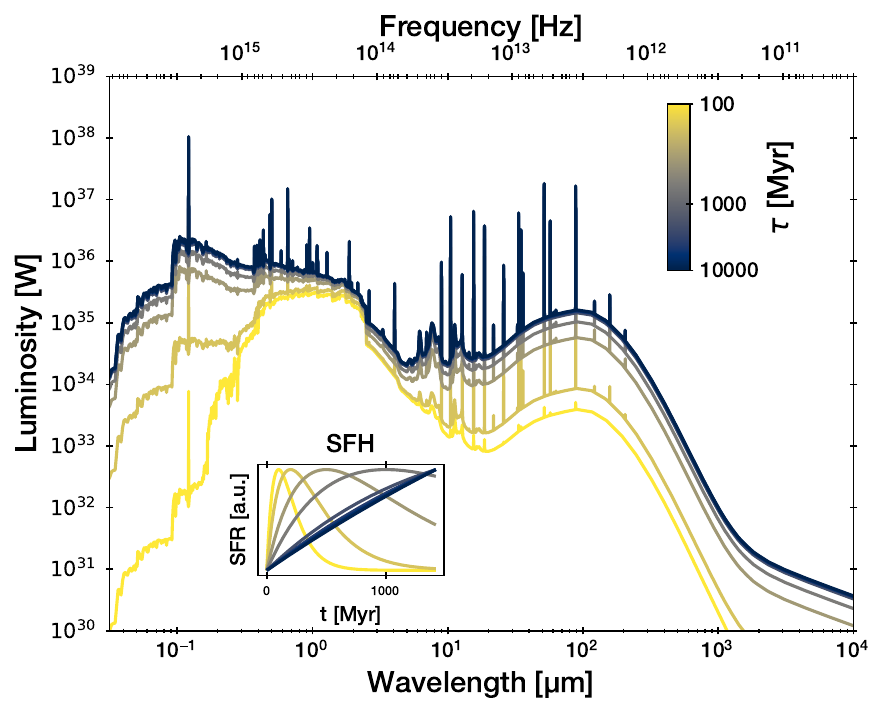}
    \caption{
    Example galaxy SEDs of stellar populations with different star formation histories. The star formation rate (see inset) rises linearly towards the present (right), with an exponential cutoff timescale $\tau$. At low $\tau$, the yellow SFR curve truncates quickly, i.e., is dominated by old stars. The corresponding yellow SED in the main panel peaks between $0.3-3\unit{\micro\meter}$. The blue curve in the inset corresponds to continuously rising star formation. The corresponding blue SED in the main panel is dominated by luminous young stars, nebular emission lines and infrared dust emission.
    Minimal attenuation, E(B-V)=0.01, is applied.
    }
    \label{fig:sfh}
\end{figure}

The galaxy evolution field has recently developed non-parametric SFHs \citep[e.g.][]{Iyer2019}, which can approximate complex SFHs seen in cosmological simulations. \cite{Leja2019} demonstrated that non-parametric SFHs can more faithfully reconstruct the stellar mass build-up even in individual galaxies, given precision photometry.
However, both parametric and non-parametric approaches agree within 0.1\,dex on present-time properties, such as the current stellar mass and the average star formation rate (SFR) in the last 100 Myr \citep{Leja2019}. In this work \cite[see also][]{Ciesla2015}, we demonstrate that contamination by AGN light creates much larger uncertainties when inferring SFR.
To compare results on stellar mass from the literature,
it is beneficial to stick with a simple parametric SFH. In GRAHSP, any SFH models implemented in CIGALE \citep[see][]{Boquien2019} can be enabled. For this work, we adopt a tau-delayed SFH (\texttt{sfhdelayed}), where the SFH rises linearly with time $t$ and is then truncated with an exponential cut-off time scale $\tau$:
\begin{equation}
    \mathrm{SFR}(t)\propto \frac{t}{\tau^2} \exp(-t/\tau)
\label{eq:sfh}
\end{equation}
\Cref{fig:sfh} illustrates galaxy spectra of different SFHs (inset) for completeness. A maximum age for the oldest stars, $t_0$, can be set (\texttt{main\_age}), which defines the start of the SFH. 

We point out that this parameterisation induces a characteristic SFR prior. The current SFR (averaged over the last 10 Myr for example) can be read off as the right-most point of each curve in the inset of \cref{fig:sfh}. The maximum (highest $\tau$) values are near SFR(10Myr)=$M_\star$/10Myr (and similar for SFR defined over other time windows), where most stellar mass was built relatively recently. A uniform or log-uniform grid on $\tau$ then induces an SFR prior that declines exponentially to arbitrarily low SFR (lowest $\tau$). The endpoints of the inset of \cref{fig:sfh} indeed show a bimodality, with few points in the middle. This means that even if a random $\tau$ and $M_\star$ were picked, a main sequence can emerge. A red cloud can emerge by imposing a SFR floor.

As for the AGN component, emission lines may contribute a substantial flux in galaxy spectra \citep[][]{Ilbert2006, HsuCDFSz2013}. Nebular emission lines are added with the \texttt{nebular} CIGALE module \citep{Boquien2013,Boquien2019}. Their effect is clearly seen in recently star-forming templates in \cref{fig:sfh}.

For gauging the relative importance of AGN and host galaxy emission, GRAHSP also computes an AGN fraction. Following \cite{dale2014ApJ...784...83D}, \texttt{fracAGNDale} is the AGN luminosity fraction from $5-20\unit{\micro\meter}$. Additionally, the bolometric AGN fraction \texttt{fracAGNTOR} is the ratio of \texttt{lumBolTOR} to the bolometric galaxy luminosity. These are also computed before applying attenuation (next section, \cref{sec:attenuation}), and for consistency, the bolometric galaxy luminosity does not include galaxy dust re-emission.
From our key model parameters, $\LAGN$ and $M_\star$, we can also consider a rough contrast ratio of AGN to host galaxy, after converting the stellar mass with a mass-to-light ratio $\Upsilon$:
\begin{equation}
    \lambda = \LAGN / M_\star \times \Upsilon 
\label{eq:lambda}
\end{equation}
For simplicity, for $\Upsilon$ we adopt the solar bolometric mass-to-light ratio $\Upsilon_0=M_\odot/L_\odot=M_\odot/\qty{3.83e33}{\erg\per\second}$, which is close to the Milky Way V-band mass-to-light ratio $\Upsilon_\mathrm{V,MW}=1.5$ \citep[e.g.,][]{Flynn2006}.
If we assume the same bolometric corrections for X-ray and $5100\unit{\angstrom}$ (justified in \cref{subsec:xray}), from black hole mass scaling relations $\lambda=2.7$ is the approximate Eddington limit \citep{Aird2017}.

\subsubsection{Attenuation model (\texttt{biattenuation}) \label{sec:attenuation}}
\begin{figure}[t]
    \centering
    \includegraphics[width=\columnwidth]{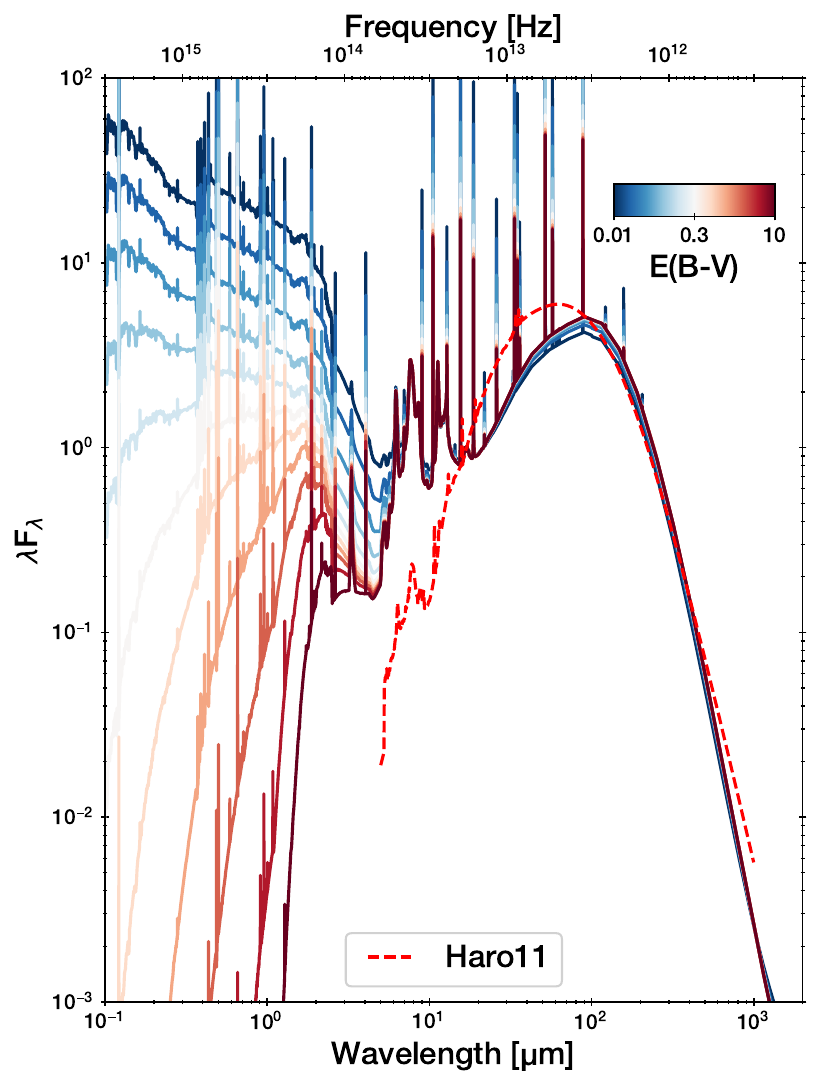}
    \caption{
    Effect of attenuation on the galaxy model. Models are shown from intrinsic (dark blue) to strongly attenuated (dark red). For illustration, the extremely attenuated local low-metallicity star-bursting galaxy Haro~11 from \cite{Lyu2016} is overplotted as a dashed red curve.
    }
    \label{fig:galebv}
\end{figure}

\begin{figure}[t]
    \centering
    \includegraphics[width=\columnwidth]{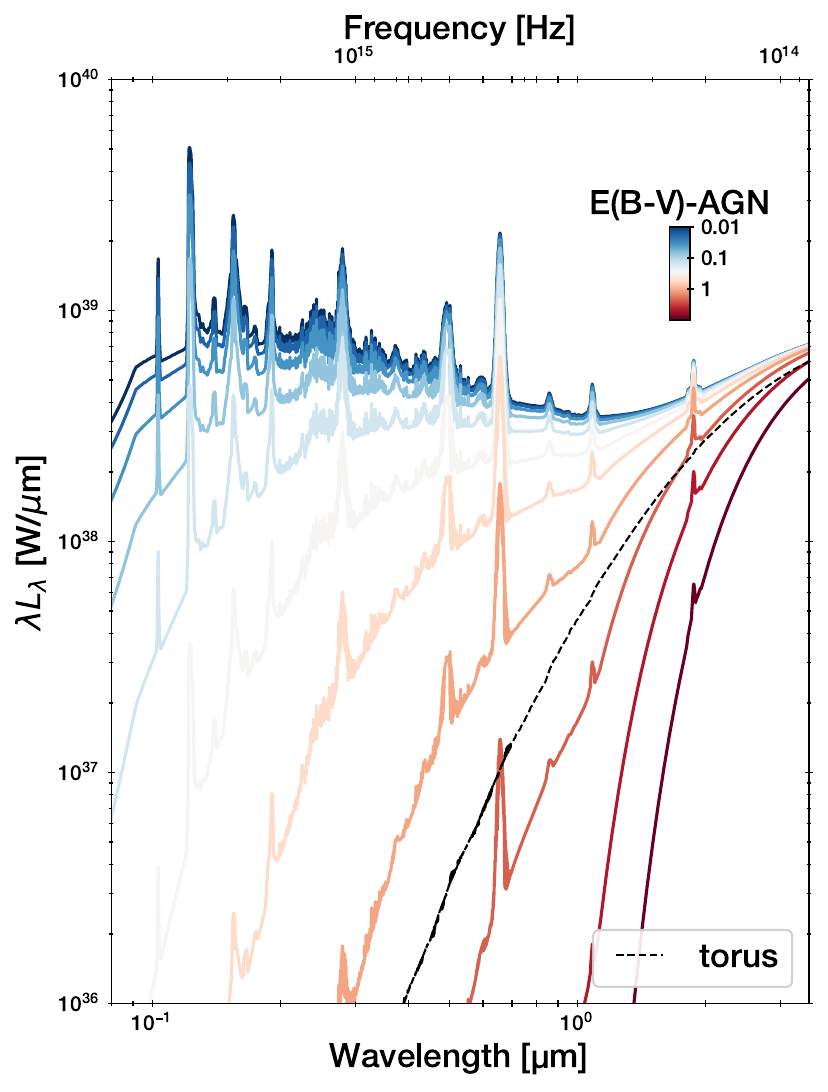}
    \caption{
    Effect of attenuation on the AGN model. Models are shown from intrinsic (blue top curves) to strongly attenuated (red bottom curves). The intrinsic torus model component (see \cref{sec:activatetorus}) is shown in dashed black. The dark red curves illustrate that the most heavily attenuation can also suppress the torus.
    }
    \label{fig:agnebv}
\end{figure}

The spectra of AGN and galaxies are frequently attenuated by dust along the line of sight. Dust attenuation can be modeled by a variety of empirical laws.
\cite{Salvato2009} and \cite{Hopkins2004}, analysing large photometric and spectroscopic AGN samples, respectively, find a preference for the dust attenuation law of \cite{Prevot1984} derived from the Small Magellanic Cloud (SMC), $A_\text{SMC}(\lambda)$. There is still debate whether steeper and even entirely feature-less (power-law like) attenuation may be preferable \citep{Fynbo2013,Zafar2015}. Because the reddening of the continuum is dependent on the assumed intrinsic AGN continuum model, for which we chose a flexible parameterisation, we remain with a SMC-like dust attenuation model. The level of attenuation is parameterised with E(B-V):
\[L_\lambda'(\lambda)=L_\lambda(\lambda)\times10^{E(B-V)\times A_\text{SMC}(\lambda)/-2.5}.\]
We approximate the SMC attenuation curve as a broken power law:
\[A_\text{SMC}(\lambda) = N \times \left( \frac{\lambda}{\lambda_\mathrm{break}} \right)^\gamma \]
where $\gamma=\gamma_\mathrm{OPT}$ (default: -1.2) below $\lambda_\mathrm{break}$ and $\gamma=\gamma_\mathrm{NIR}$ (default: -3) above. The normalisation is $N=1.2$ at $\lambda_\mathrm{break}=1100\unit{\nano\meter}$.

In our implementation, the AGN and galaxy components are attenuated differently.
The galaxy attenuation level is parameterized by the \texttt{E(B-V)} color excess parameter, and the total galaxy luminosity absorbed is recorded. Enforcing energy balance, the luminosity is then re-emitted in the infrared following the \cite{dale2014ApJ...784...83D} model of galactic dust emission (\texttt{galdale2014}). \Cref{fig:galebv} illustrates this, and how the model approximates a dusty star-burst galaxy.
For the AGN light, the situation is different \citep[see also][]{CalistroRivera2016}. The dust emission is already empirically modeled with the torus component (\cref{sec:activatetorus} above). The energy may not be balanced due to variability and anisotropy, as the line-of-sight absorption differs from geometrically averaged absorption. Since the AGN is embedded in the host galaxy dust but also nuclear dust, it may undergo additional attenuation.
Therefore, we attenuate the AGN components not only with \texttt{E(B-V)}, but with \texttt{E(B-V)} + \texttt{E(B-V)-AGN}, where \texttt{E(B-V)-AGN} is a parameter giving the nuclear attenuation color excess.

The \texttt{E(B-V)-AGN} parameter also allows the AGN model to transition from a Sy1 to a Sy2. \Cref{fig:agnebv} illustrates how the blue continuum, its lines and ultimately also the torus are attenuated by increasing values of \texttt{E(B-V)-AGN}. In our model, narrow and broad emission lines are both attenuated to the same extent, while in reality, narrow emission lines should remain visible in Sy2. However, this approximation is sufficient because in the relevant wavelength range, the photometry of Sy2 galaxies is dominated by host galaxy light \citep{Hickox2017}.

\subsection{Model validation and parameter range calibration \label{subsec:calibration}}

\begin{figure*}
    \centering
    \includegraphics[width=0.899\textwidth]{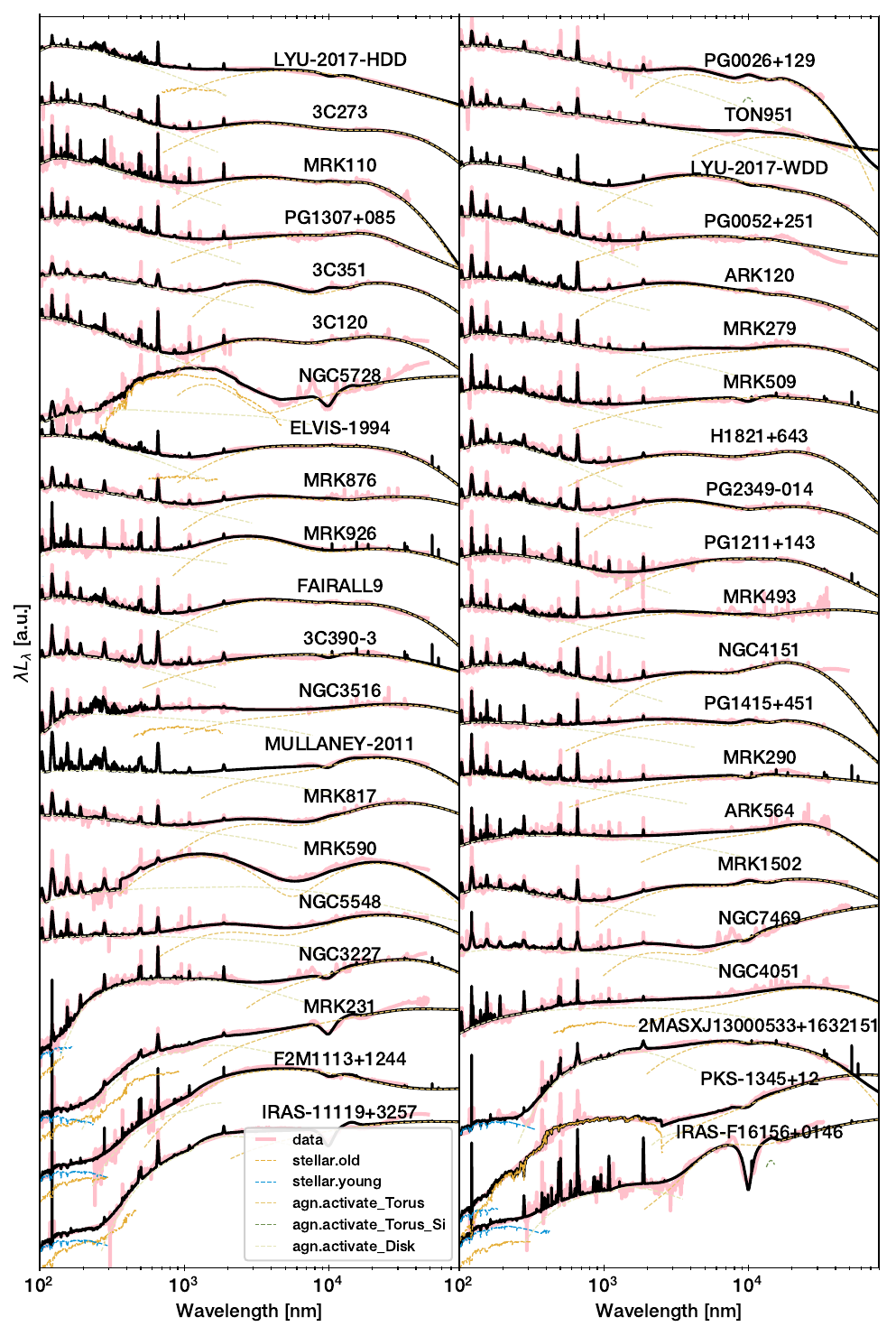}
    \caption{
    Atlas of AGN spectra (red) from \cite{Brown2019}. The black curve shows our best-fit GRAHSP model. Dashed curves show model components.
    }
    \label{fig:browncomparison}
\end{figure*}

\begin{sidewaystable*}[hp]
    \caption{Best fit model parameters for the AGN SED Atlas of \cite{Brown2019}. The first row indicates the parameter; the second row shows whether the logarithm of the parameter is listed, the third row indicates the units. The last few rows give the minimum, maximum and 10\% and 90\% quantiles of each column.}
    \centering
    {\small
    \setlength{\tabcolsep}{4pt}

\begin{tabular}{l|cccc|cccccccc|cc|ccccccc}
 & $M_\star$ & $\tau$ & age & $\alpha$ & L$_\mathrm{AGN}$ & A$_\mathrm{Fe II}$& A$_\mathrm{lines}$ & W$_\mathrm{line}$ & $\beta$ & $\lambda_\mathrm{bend}$ & W$_\mathrm{bend}$ & $\beta_\mathrm{UV}$ & E$_\mathrm{B-V}$ & E$_\mathrm{B-V}^\mathrm{AGN}$ & Si & $f_\mathrm{cov}$ & $\lambda_\mathrm{cool}$ & W$_\mathrm{cool}$ & $\lambda_\mathrm{hot}$ & W$_\mathrm{hot}$ & $f_\mathrm{hot}$ \\
& log & log & log & & log & log & log & log & & log & log & & log & log & & &   & log &   & log & \\
Galaxy & $M_\odot$ & Myr & Myr &  &  erg/s & & & km/s & & nm & dex & & mag & mag & & & $\mu$m  & dex &  $\mu$m  & dex & \\
\hline
\hline
IRAS-11119+3257 & 10.1 & 2.9 & 3.4 & 1.5 & 45.8 & 0.7 & -0.3 & 3.7 & -1.6 & 2.1 & -1.7 & -0.5 & -0.9 & 0.1 & -2.2 & 0.2 & 30.0 & 0.6 & 3.8 & 1.0 & 9.9\\
IRAS-F16156+0146 & 8.1 & 1.0 & 1.4 & 1.5 & 44.5 & 0.2 & 0.5 & 3.4 & -2.2 & 1.6 & -2.0 & 1.3 & -0.7 & -0.1 & -2.9 & 0.8 & 26.4 & 0.9 & 6.4 & 0.2 & 0.3\\
F2M1113+1244 & 9.9 & 3.1 & 2.7 & 2.6 & 46.0 & 1.3 & -0.5 & 2.8 & -1.6 & 2.3 & -2.0 & -0.1 & -0.9 & -0.1 & -0.8 & 0.8 & 30.0 & 0.8 & 1.7 & 0.6 & 4.0\\
PKS-1345+12 & 11.0 & 2.0 & 3.0 & 2.2 & 44.6 & 1.8 & -0.3 & 3.9 & -1.4 & 1.9 & -0.3 & -0.7 & -0.8 & 0.9 & -0.5 & 0.9 & 28.1 & 0.6 & 2.6 & 0.6 & 0.0\\
MRK231 & 10.6 & 2.5 & 3.3 & 2.7 & 45.1 & 0.6 & -0.4 & 3.9 & -1.7 & 2.2 & -0.7 & 1.7 & -1.1 & -0.1 & -1.6 & 0.7 & 30.0 & 0.2 & 5.1 & 1.0 & 2.5\\
2MASXJ13000533+1632151 & 7.5 & 1.7 & 1.4 & 1.8 & 45.4 & 1.1 & -0.5 & 4.1 & -2.5 & 2.3 & -0.2 & -1.0 & -0.8 & 0.1 & -0.4 & 0.1 & 18.8 & 0.3 & 4.4 & 0.5 & 2.5\\
NGC3227 & 7.0 & 3.5 & 3.0 & 1.0 & 42.8 & 0.9 & -0.3 & 3.6 & -1.9 & 2.3 & -1.2 & 1.9 & -0.8 & -0.6 & -0.8 & 0.2 & 19.1 & 0.4 & 4.9 & 0.4 & 0.2\\
NGC4051 & 8.3 & 1.2 & 3.7 & 1.3 & 41.7 & -0.5 & -0.5 & 3.4 & -1.1 & 2.3 & -2.0 & -1.1 & -1.6 & -1.2 & 0.0 & 0.7 & 12.9 & 0.5 & 2.7 & 0.4 & 0.2\\
NGC5548 & 7.8 & 1.3 & 3.3 & 2.2 & 43.1 & 0.8 & -0.4 & 3.8 & -1.2 & 1.7 & -1.8 & 0.8 & -2.0 & -1.6 & -0.1 & 0.9 & 12.4 & 0.5 & 1.3 & 0.4 & 0.4\\
NGC7469 & 8.5 & 3.7 & 3.1 & 1.9 & 43.7 & 1.2 & -0.4 & 4.1 & -1.4 & 1.7 & -1.2 & 1.1 & -2.1 & -1.4 & -0.5 & 1.0 & 24.3 & 0.9 & 1.7 & 0.3 & 0.1\\
MRK590 & 9.1 & 1.5 & 4.1 & 1.8 & 42.9 & 0.7 & -0.2 & 4.1 & -1.0 & 1.7 & -1.2 & 0.9 & -1.3 & -2.1 & 0.1 & 1.0 & 15.3 & 0.4 & 0.9 & 0.4 & 1.6\\
MRK1502 & 8.2 & 2.3 & 4.1 & 1.7 & 44.5 & 1.3 & -0.3 & 3.9 & -1.7 & 2.1 & -0.9 & -0.2 & -2.0 & -1.9 & 0.7 & 1.0 & 13.0 & 0.5 & 2.3 & 0.4 & 0.6\\
MRK817 & 7.7 & 2.1 & 3.9 & 1.9 & 44.0 & 0.9 & -0.4 & 3.8 & -1.5 & 1.9 & -1.7 & -0.0 & -2.0 & -2.0 & 0.3 & 0.6 & 17.9 & 0.5 & 2.0 & 0.4 & 0.3\\
ARK564 & 9.0 & 1.6 & 3.2 & 1.3 & 43.6 & 1.4 & -0.5 & 3.5 & -1.4 & 1.9 & -1.7 & 0.3 & -1.9 & -1.2 & 0.1 & 0.4 & 21.5 & 0.3 & 3.9 & 0.6 & 0.6\\
MRK290 & 8.2 & 2.6 & 2.7 & 2.2 & 43.4 & 1.3 & -0.3 & 3.7 & -1.6 & 2.0 & -0.6 & 2.0 & -1.9 & -2.0 & -0.5 & 0.6 & 28.4 & 0.9 & 0.9 & 1.0 & 5.3\\
NGC3516 & 9.3 & 1.6 & 3.4 & 1.9 & 42.9 & 1.9 & -0.5 & 3.8 & -1.3 & 2.2 & -1.9 & 1.3 & -1.9 & -1.7 & 0.1 & 0.6 & 14.0 & 0.6 & 1.3 & 0.6 & 0.4\\
PG1415+451 & 8.9 & 2.5 & 4.1 & 1.9 & 44.4 & 0.5 & -0.5 & 3.5 & -1.6 & 2.0 & -1.2 & 0.1 & -2.1 & -1.5 & 0.7 & 0.5 & 14.8 & 0.4 & 1.1 & 0.6 & 0.7\\
3C390-3 & 7.4 & 3.2 & 1.3 & 1.0 & 44.1 & -0.9 & 0.1 & 3.9 & -2.3 & 2.0 & -0.8 & -0.6 & -1.1 & -1.9 & -0.6 & 0.7 & 23.2 & 0.2 & 1.0 & 0.9 & 3.6\\
NGC4151 & 7.8 & 1.6 & 2.9 & 3.0 & 42.9 & 1.3 & -0.2 & 3.7 & -2.2 & 1.8 & -0.9 & -0.4 & -1.1 & -2.1 & 0.1 & 0.5 & 15.0 & 0.3 & 2.6 & 0.4 & 0.5\\
FAIRALL9 & 9.2 & 2.1 & 3.7 & 2.3 & 44.3 & 1.4 & -0.2 & 3.8 & -2.0 & 2.0 & -0.5 & 0.8 & -2.1 & -1.6 & 0.1 & 0.7 & 14.3 & 0.4 & 2.0 & 0.5 & 1.3\\
MRK493 & 8.3 & 2.1 & 4.0 & 2.1 & 43.5 & 1.6 & -0.5 & 3.6 & -1.8 & 2.0 & -1.5 & 0.5 & -2.1 & -1.8 & 0.3 & 0.5 & 18.4 & 0.6 & 1.1 & 0.6 & 0.8\\
MRK926 & 9.7 & 3.0 & 3.4 & 2.0 & 44.1 & -0.1 & -0.2 & 3.6 & -1.6 & 2.1 & -0.9 & 0.0 & -2.1 & -1.3 & 0.0 & 0.4 & 10.0 & 0.6 & 2.0 & 0.4 & 2.3\\
PG1211+143 & 8.3 & 1.3 & 1.9 & 1.2 & 44.8 & -0.4 & -0.3 & 3.5 & -2.2 & 2.0 & -1.3 & 1.5 & -1.9 & -1.1 & -0.1 & 0.4 & 19.3 & 0.2 & 3.4 & 0.6 & 5.5\\
MRK876 & 8.7 & 2.7 & 4.1 & 1.9 & 45.0 & -0.6 & -0.3 & 3.8 & -1.9 & 2.1 & -1.8 & -0.6 & -2.0 & -1.5 & 0.2 & 0.3 & 14.1 & 0.6 & 2.3 & 0.4 & 1.0\\
PG2349-014 & 8.0 & 2.8 & 4.1 & 1.9 & 44.9 & 1.4 & -0.3 & 3.9 & -1.8 & 2.1 & -0.7 & -0.3 & -1.9 & -2.0 & -0.2 & 0.4 & 14.5 & 0.4 & 1.9 & 0.5 & 1.3\\
H1821+643 & 8.5 & 2.7 & 3.2 & 1.8 & 46.0 & 1.2 & -0.1 & 3.8 & -2.7 & 2.2 & -0.2 & 0.2 & -2.0 & -1.9 & 0.1 & 0.5 & 16.6 & 0.3 & 1.4 & 0.6 & 0.7\\
NGC5728 & 10.4 & 2.0 & 3.7 & 2.6 & 42.6 & 0.0 & -0.5 & 4.1 & -1.3 & 1.7 & -2.0 & -0.1 & -2.1 & -1.2 & -1.6 & 1.0 & 12.2 & 0.8 & 1.2 & 0.2 & 1.5\\
MRK509 & 7.5 & 1.7 & 1.5 & 2.6 & 44.3 & 1.3 & -0.2 & 3.7 & -1.9 & 2.0 & -0.5 & 1.4 & -2.1 & -1.8 & -0.7 & 0.5 & 21.3 & 0.4 & 1.1 & 0.9 & 3.3\\
3C120 & 7.9 & 1.8 & 3.3 & 1.0 & 44.1 & 1.2 & -0.1 & 3.6 & -2.5 & 2.0 & -0.3 & -0.6 & -2.1 & -1.9 & 0.1 & 0.6 & 12.1 & 0.4 & 2.0 & 0.4 & 0.8\\
MRK279 & 8.7 & 1.9 & 3.6 & 2.2 & 44.0 & 0.6 & -0.3 & 3.8 & -1.6 & 2.1 & -0.9 & 0.1 & -2.1 & -1.8 & -0.1 & 0.3 & 13.8 & 0.4 & 3.0 & 0.4 & 0.6\\
3C351 & 8.1 & 2.8 & 3.8 & 2.2 & 45.9 & 1.0 & -0.5 & 4.1 & -1.8 & 2.0 & -1.9 & 0.3 & -1.9 & -1.5 & 0.6 & 0.3 & 14.1 & 0.4 & 2.4 & 0.3 & 1.0\\
ARK120 & 8.8 & 2.3 & 4.0 & 2.4 & 44.3 & 1.3 & -0.1 & 3.8 & -2.0 & 1.9 & -0.7 & 0.6 & -1.3 & -2.1 & 0.2 & 0.3 & 11.1 & 0.4 & 1.6 & 0.5 & 2.1\\
PG1307+085 & 10.3 & 1.7 & 3.3 & 2.3 & 44.8 & 1.5 & -0.3 & 3.6 & -2.2 & 2.0 & -0.1 & 0.3 & -1.8 & -2.0 & -0.1 & 0.4 & 16.1 & 0.2 & 1.6 & 0.7 & 1.3\\
PG0052+251 & 9.7 & 1.1 & 2.5 & 1.7 & 44.9 & 1.0 & -0.1 & 3.8 & -2.7 & 2.3 & 0.1 & -1.0 & -1.9 & -2.1 & -0.3 & 0.3 & 28.6 & 0.9 & 1.2 & 0.8 & 5.1\\
MRK110 & 10.1 & 1.6 & 4.1 & 1.4 & 43.8 & 1.3 & -0.1 & 3.3 & -2.4 & 2.1 & -0.8 & -0.1 & -1.9 & -1.4 & 0.2 & 0.3 & 14.0 & 0.3 & 1.9 & 0.5 & 1.3\\
3C273 & 7.0 & 2.2 & 2.7 & 2.0 & 46.0 & 1.1 & -0.4 & 3.5 & -2.6 & 2.1 & -0.2 & -0.1 & -2.1 & -1.6 & 0.0 & 0.2 & 17.4 & 0.4 & 1.7 & 0.5 & 1.5\\
TON951 & 9.7 & 1.4 & 4.1 & 2.0 & 44.6 & -0.2 & -0.5 & 3.9 & -1.6 & 2.2 & -0.9 & -1.2 & -2.1 & -1.7 & 0.1 & 0.1 & 12.5 & 1.0 & 7.2 & 0.5 & 6.0\\
PG0026+129 & 7.8 & 2.3 & 3.5 & 2.8 & 45.2 & 0.5 & -0.5 & 3.8 & -2.0 & 2.0 & -0.4 & -2.0 & -1.5 & -1.7 & 1.1 & 0.1 & 13.4 & 0.3 & 3.0 & 0.3 & 1.3\\
\hline
\hline
min & 7.0 & 1.0 & 1.3 & 1.0 & 41.7 & -0.9 & -0.5 & 2.8 & -2.7 & 1.6 & -2.0 & -2.0 & -2.1 & -2.1 & -2.9 & 0.1 & 10.0 & 0.2 & 0.9 & 0.2 & 0.0\\
10\% & 7.5 & 1.3 & 2.0 & 1.2 & 42.9 & -0.2 & -0.5 & 3.5 & -2.6 & 1.7 & -1.9 & -0.9 & -2.1 & -2.0 & -0.8 & 0.2 & 12.4 & 0.3 & 1.1 & 0.3 & 0.3\\
90\% & 10.2 & 3.0 & 4.1 & 2.7 & 45.8 & 1.6 & -0.1 & 4.1 & -1.3 & 2.3 & -0.2 & 1.5 & -0.9 & -0.1 & 0.3 & 0.9 & 28.4 & 0.9 & 4.3 & 0.9 & 5.3\\
max & 11.0 & 3.7 & 4.1 & 3.0 & 46.0 & 1.9 & 0.5 & 4.1 & -1.0 & 2.3 & 0.2 & 2.0 & -0.7 & 0.9 & 1.1 & 1.0 & 30.0 & 1.0 & 7.2 & 1.0 & 9.9\\
\hline
\hline
\end{tabular}
    }
    \label{fig:browntable}
\end{sidewaystable*}

\begin{table*}[ht]
    \caption{Best-fit model parameters for the AGN SED Atlas of \cite{Brown2019}. Only AGN shape parameters are shown. The first row represents an average AGN, while the next two rows represent warm- and hot-dust-deficient sub-populations. These differ in the torus properties listed in the last six columns, which give the covering factors and log-quadratic shape parameters of the cool and hot components. The last row is an AGN torus template starting at 6$\mu{}m$, therefore the first columns are blanked out.}
    \centering
    {
    \setlength{\tabcolsep}{4pt}

\begin{tabular}{l|cccccccc|cc|ccccccc}
 & A$_\mathrm{Fe II}$& A$_\mathrm{lines}$ & W$_\mathrm{line}$ & $\beta$ & $\lambda_\mathrm{bend}$ & W$_\mathrm{bend}$ & $\beta_\mathrm{UV}$ & E$_\mathrm{B-V}$ & E$_\mathrm{B-V}^\mathrm{AGN}$ & Si & $f_\mathrm{cov}$ & $\lambda_\mathrm{cool}$ & W$_\mathrm{cool}$ & $\lambda_\mathrm{hot}$ & W$_\mathrm{hot}$ & $f_\mathrm{hot}$ \\
& log & log & log & & log & log & & log & log & & &   & log &   & log & \\
Template & & & km/s & & nm & dex & & mag & mag & & & $\mu$m  & dex &  $\mu$m  & dex & \\
\hline
\hline
Elvis-1994 & 1.8 & -0.5 & 3.6 & -2.7 & 1.7 & 0.2 & 1.9 & -2.0 & -2.0 & 0.1 & 0.6 & 14.0 & 0.4 & 2.5 & 0.5 & 1.1\\
Lyu-2017-WDD & 0.6 & -0.5 & 3.5 & -2.3 & 2.0 & -1.4 & 0.3 & -2.0 & -1.2 & -0.4 & 0.2 & 18.8 & 0.3 & 2.4 & 0.5 & 3.1\\
Lyu-2017-HDD & 1.9 & -0.5 & 3.6 & -2.6 & 2.1 & -1.2 & -0.5 & -1.2 & -1.7 & -0.5 & 0.2 & 16.0 & 0.5 & 1.8 & 0.6 & 6.3\\
Mullaney-2011 & -- & -- & -- & -- & -- & -- & -- & -- & -1.8 & -0.7 & 0.6 & 14.7 & 0.4 & 2.5 & 0.5 & 0.3\\
\hline
\hline
\end{tabular}
    }
    \label{fig:templatetable}
\end{table*}

\begin{figure}[ht]
    \centering
    \includegraphics[width=\columnwidth]{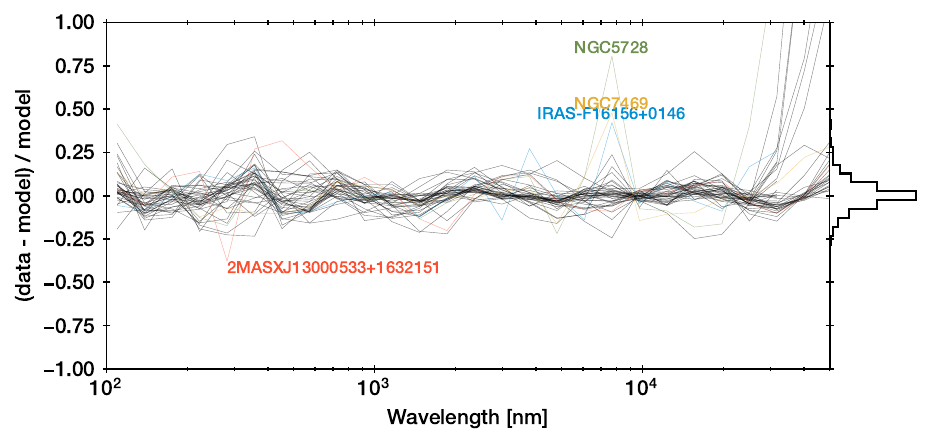}
    \caption{
    Relative residuals of the fits of \cref{fig:browncomparison} are shown as curves. Four individual cases with the largest deviations are highlighted in colours and labelled.
    The distribution of residuals is presented on the right-hand side as a histogram.
    The vast majority of residuals (96.6\%) are concentrated to smaller than 20 per cent deviations.
    }
    \label{fig:brownresid}
\end{figure}

This section tests and validates the performance of the GRAHSP model.
Additionally, for each model parameter we identify the range required to reproduce the observations, to be used as prior knowledge when fitting lower-quality and noisier photometric observations typical of distant AGN samples.
For this purpose, consistent, high-quality UV to far-IR spectra of galaxies hosting AGN are needed.
\cite{Brown2019} compiled spectroscopic and photometric data from 0.09 to 30$\mu m$ on a diverse set of 41 local AGN.
For each AGN, they carefully cross-calibrated and aperture-corrected the observations into one continuous broad-band SED.
\Cref{fig:browncomparison} shows these data-driven model spectra in red.
This AGN SED atlas was created to demonstrate the diversity of galaxies with AGN, including different host-AGN contrasts and infrared torus shapes. We use the AGN SED Atlas to identify reasonable ranges for each fitting parameter to restrict our otherwise very flexible model.
We also include the empirical AGN templates of \cite{Elvis1994}, \cite{Mullaney2011} and the hot-dust deficient and warm-dust deficient templates of \cite{Lyu2017} derived from 87 Palomar-Green quasars.

We optimize our model parameters to best approximate each spectrum.
The best-fit model is overlaid in black in  \Cref{fig:browncomparison}. Individual model components are presented as dashed curves. Overall, the fit is very good, and the model captures the diversity of AGN-galaxy SED shapes.
The model quality can be quantified by looking at the fit residuals in \Cref{fig:brownresid}.  These were computed by averaging the model and data in windows of $\Delta \lambda = 0.2 \lambda$, as relevant to broad-band photometry.
The residuals remain below 20 per cent across the entire wavelength range considered, with few (3.4\%) exceptions. In 2MASXJ13000533+1632151, the reddened galaxy continuum is not perfectly approximated near 300nm. IRAS-F16156+0146 shows an extremely deep Si absorption feature, which is not modelled well in the wings. NGC~7469 and NGC~5728 show complex galaxy polycyclic aromatic hydrocarbon emission features which are not fully captured by the \cite{dale2014ApJ...784...83D} templates adopted here.

The best-fit model parameters for each AGN are listed in \Cref{fig:browntable}. The diversity of the AGN Atlas can only be captured by allowing all listed parameters to vary. However, we can use the typical values to restrict the allowed parameter range. The last rows in \Cref{fig:browntable} list the 10 and 90 per cent quantile for each parameter. For example, for the attenuation of the AGN, E(B-V) values ranging from 0.01 to 1 mag are found, and for the galaxy, E(B-V) values ranging from 0.01 to 0.1 mag are found.
The best-fit model parameters for four empirical templates are listed in \Cref{fig:templatetable}. Here we see that the normal AGN template of \cite{Elvis1994} has a covering factor of 60 per cent with cool and hot torus components centred at 14 and 2.5$\mu{}m$, of widths 0.4 and 0.5 dex, respectively. This is very similar to the \cite{Mullaney2011} template in the last row. In contrast, the warm-dust deficient template of \cite{Lyu2017} is decomposed with a narrower cold dust energy distribution, while the hot-dust deficient template contains a hot dust component centred at shorter wavelengths. The amplitude factor of the hot dust component (last column) also varies from 0.3 to 6. We cannot rule out the possibility that parameter degeneracies influence the values listed in \Cref{fig:templatetable}, however, the found ranges from \Cref{fig:browntable} and \Cref{fig:templatetable} motivate the parameter ranges when fitting the model to data. To achieve efficient fits with the remaining 21 free model parameters requires an advanced inference engine.

\subsection{Inference engine\label{sec:computationengine}}

\subsubsection{Photometric flux prediction\label{sec:redshifting}}
The composed model is redshifted and converted into observable flux $F_\lambda$ using the luminosity distance $D_L$: 
$F_\lambda(\lambda)=L_\lambda(\lambda)/(4\pi D_L^2)/(1+z)$. 
For these steps, the implementation is taken from CIGALE \citep[][\texttt{redshifting} module]{Boquien2013}, which also applies absorption by the inter-galactic medium \citep{Madau1992}.
Finally, the flux of a photometric filter band $i$ of interest is predicted with the filter transmission curve $T(\lambda)$ as
$F_i=\lambda_\mathrm{pivot}^2 \int {F(\lambda) T(\lambda)\,d\lambda}$
with the pivot wavelength 
$\lambda_\mathrm{pivot}=\left(\int{T(\lambda)\times\lambda d\lambda}\right)/\left(\int{T(\lambda)/\lambda \,d\lambda}\right)$.
See appendix A2 of \cite{Bessel2012photfilter} for a discussion why the pivot wavelength, rather than the mean filter wavelength, is relevant here.
The model filter flux $F_i$ is a function of all model parameters, $F_i=f(L_\mathrm{AGN}, M_\star, \dotsc)$.

\subsubsection{Computational optimisations\label{sec:computational}}

\begin{figure}
    \centering
    \includegraphics[width=\columnwidth]{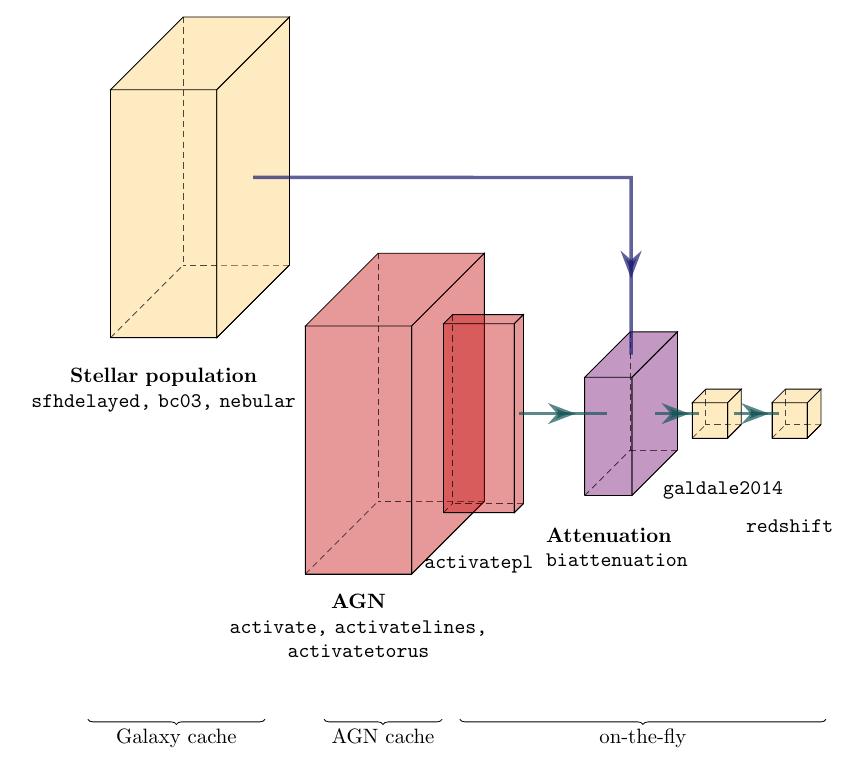}
    \caption{GRAHSP computation pipeline. One sub-pipeline (yellow) produces the galaxy SED (large yellow box), one sub-pipeline (red large box) produces the AGN SED. Each sub-pipeline has a separate cache. Subsequent modules are applied on-the-fly, including the AGN power-law, attenuating the AGN and the galaxy components, summing and transforming into observed-frame fluxes.}
    \label{fig:pipeline}
\end{figure}

GRAHSP includes several optimisations that allow for the fitting of a single galaxy in 2-3 minutes on a typical single computing core. This includes the generation of visualisations of the SED fit and the inferred parameter uncertainties. This speed enables processing large samples, or experimentation and variation of models and data processing of individual objects.
For rapid evaluation of a model with a substantial number of components, and fitting that fully explores the degeneracies between the many model parameters, GRAHSP departs substantially from existing approaches. We review some of these first.

In CIGALE, the model is implemented as a pipeline of modules that successively build up the SED. Each module has parameters. Each parameter can take a set of possible values, which is specified by the user as a list of numbers. CIGALE then creates an n-dimensional grid of all possible parameter combinations and instantiates each model spectrum. Because of the curse of dimensionality and limited computing resources (in memory and computation budget), this requires the user to sparsely choose the grid points for each parameter. This problem is exacerbated by the fact that an AGN+galaxy model is a linear combination of two models. However, CIGALE is unaware of this and thus builds the full model grid inefficiently, with the \texttt{frac\_agn} specifying the relative normalisation of the AGN and galaxy components. The AGNFitter \citep{CalistroRivera2016} and FortesFit \citep{Rosario2019fortesfit} instead first build two grids and linearly combine them during fitting, with stellar mass and AGN luminosity as normalisations. 
The parameterisation can be important because a log-uniform prior on stellar mass and AGN luminosity, each, prefers a much wider specific accretion rate distribution than setting a uniform prior on \texttt{frac\_agn}. The latter can induce a bias to a narrow range of specific accretion rates, which may affect galaxy-AGN co-evolution studies.
To test a new model variant (e.g., for a new filter set and/or redshift) with the two-grid precomputation approach, the user needs to write custom code to create the grids, and to balance the possible science goals with the grid size and computational cost. Another approach is to generate models on-the-fly, including instantiating the stellar population, attenuation, redshifting, and filter flux computation. Crucially, this enables full exploration of degenerate parameters, such as the E(B-V) and power-law slopes, with arbitrarily fine resolution.

GRAHSP takes a hybrid approach. It uses the pipeline-of-modules approach of CIGALE, but computes models on-the-fly. Models are cached using a hash map, which retrieves the output of the last module where all parameters matched. Therefore, while sampling, GRAHSP progressively builds up an implied grid. However, the grid is not necessarily fully constructed. The cache keeps only a user-defined maximum number of models (the \texttt{CACHE\_MAX} environment variable), dismissing from memory models that have become uninteresting as the fit proceeds. The caching speeds up computation when the available machine memory is large.
More specifically, the main run script of GRAHSP,  \texttt{dualsampler.py}, entertains two pipelines, as illustrated in \Cref{fig:pipeline}. One pipeline creates the AGN components with mock galaxy components and a second one the galaxy components with mock AGN components. The mocked components only permit a single value for each module parameter. This means that the implied grid for each of the two pipelines is, as in AGNfitter or FortesFit, only concerned with either the galaxy or AGN parameter space. In the final model evaluation, the AGN components from the AGN pipelines and the galaxy component from the galaxy pipeline are scaled and added together. These pipelines include the module list that can be freely chosen by the user at run time.
After the combination of the galaxy and AGN components, several modules can be applied relatively cheaply on-the-fly afterwards (see \cref{fig:pipeline}). This includes flux computation, \texttt{redshifting}, \texttt{biattenuation} and also \texttt{activepl}. The expensive computation of attenuation for each filter band is disabled during fitting, and re-enabled during the analysis of the results.
Since these modules are evaluated on-the-fly, their parameter grids can be arbitrarily fine. This enables a detailed exploration of the degeneracies between continuum shape (\texttt{activatepl}) and its attenuation (\texttt{biattenuation}). It also enables the incorporation of redshift uncertainties, either as a free fitting parameter or with an informative prior from photometric redshift estimation.

GRAHSP is also parallelised. SED analysis can be trivially split across computing cores without communication when analysing more objects than computing cores (known as an embarrassingly parallel problem). The user can set the number of cores to use (\texttt{--cores} command-line option) and choose among several parallelisation backends provided by the \texttt{joblib} and \texttt{multiprocessing} python libraries.
Here, a further modification to CIGALE is necessary. CIGALE stores its database of model templates in a \texttt{sqlite} file on disk. However, mutual locking of multiple parallel runs accessing the \texttt{sqlite} file can cause substantial delays on shared file systems, as typically happens in computing centres. To avoid this, if the environment variable \texttt{DB\_IN\_MEMORY} is set to 1, the database ($2\si{\giga\byte}$) is copied to memory on start-up, which brings substantial speed-ups on large machines.

\subsubsection{Module Pipeline \label{pipeline}}
The order of execution of the modules is: \texttt{sfhdelayed}, \texttt{bc03} (or \texttt{m2005}), \texttt{nebular, activate, activatelines, activatetorus, activatepl, biattenuation, galdale2014, redshifting}. The module \texttt{galdale2014} has to be placed after \texttt{biattenuation} to receive the attenuated luminosity.

\subsubsection{Parameters and priors \label{sec:params}}

\Cref{tab:parameters} summarizes the modules (in square brackets) and their parameters introduced in the previous sections.
\Cref{fig:modelcomp} gives an impression of the shapes of the model components with arbitrarily chosen galaxy and AGN parameters. For stellar mass, AGN luminosity and redshift, a continuous parameter space is explored, and the user can specify priors through the data file and command line options. The redshift can be constrained with a fixed (spectroscopic) value, or with uncertainties\footnote{Supporting arbitrary photometric redshift probability distributions is planned for future releases.} (photometric), or left unconstrained. The AGN luminosity can be constrained by providing a log-normal flux prior, for example from an X-ray detection. 
To test fits without a host galaxy, the upper limit on the stellar mass can be lowered through a command-line option.
The other parameters are specified through a configuration file (\texttt{pcigale.ini}). The meaning of the AGN parameters is illustrated in \cref{fig:AGNcomponents}. The effect of the two E(B-V) parameters is shown in \cref{fig:galebv,fig:agnebv}.
As in CIGALE, the specified grid parameter values also define a prior. \texttt{numpy} functions can be used to generate e.g., log-uniform grids.

The values chosen for the tests in this work are listed in the right-most column of \cref{tab:parameters}. For the galaxy, these follow previous CIGALE-based fits by \cite{Ciesla2015} and \cite{Yang2020}. For the AGN component, we generally follow the ranges found in \cref{subsec:calibration}.

\begin{table*}[]
\centering
    \caption{Modules and their parameters. GRAHSP allows the user to specify the parameter ranges or values as they see fit. The last column lists the choices for this work. In cases of discrete parameters, grid values are specified. For continuous parameters, probability density functions are adopted.}
\begin{tabular}{l c p{7cm}}
     Parameter & Unit & Prior \\
     \hline
     \hline
     Galaxy components: & & \\
     \hline
     \texttt{stellar\_mass} & $M_\odot$ & log-uniform between $10^5$ and $10^{\mathtt{mass\_max}} M_\odot$ \\
     \texttt{mass\_max} & $\log M_\odot$  & 15 (command-line parameter) \\
     \addlinespace[4pt]
     \textbf{\texttt{[sfhdelayed]}} & & \\
     \texttt{tau\_main} & Myr & 100, 200, 500, 1000, 3000, 5000, 7000, 10000 \\
     \texttt{age} & Myr & 158 to 10000 in 18 log-uniform steps \\
     \texttt{sfr\_A} &  & 1 \\
     \addlinespace[4pt]
     \textbf{\texttt{[m2005]}} & & \\
     \texttt{imf} & & 0 (Salpeter) \\
     \texttt{metallicity} & & 0.02 (solar) \\
     \texttt{separation\_age} & Myr & 10 \\
     \addlinespace[4pt]
     \textbf{\texttt{[nebular]}} & & \\
     \texttt{logU} &  & -2.0 \\
     \texttt{f\_esc} & & 0.0 \\
     \texttt{f\_dust} & & 0.0 \\
     \texttt{lines\_width} & km/s & 300 \\
     \addlinespace[4pt]
     \textbf{\texttt{[galdale]}} & & \\
     \texttt{alpha} & & uniform between 0.75 and 2.75 in steps of 0.0625 \\
     \hline
     \hline
     AGN components & & \\
     \hline
     \textbf{\texttt{[activate]}} & & \\
     \texttt{L\_AGN} & erg/s & log-uniform between $10^{38}$ and $10^{50}$, \textit{or} \\
     & & log-normal with parameters taken from the data file columns `\texttt{FAGN}' (mean) and `\texttt{FAGN\_err}' (standard deviation), shifted with the luminosity distance at \texttt{redshift}. \\
      & & \\
     \textbf{\texttt{[activatelines]}} & & \\
     \texttt{AFeII} & & 0.6, 1, ..., 32 (10 logarithmic steps) \\
     \texttt{Alines} & & 0.3, 0.5, 0.7, 1, 1.5, 2, 4, 10, 20 \\
     \texttt{linewidth} & km/s & 10000 \\
     \addlinespace[4pt]
     \textbf{\texttt{[activategtorus]}} & & \\
     \texttt{fcov} & & uniform between 0.05 and 0.95 in steps of 0.05 \\
     \texttt{Si} & & uniform between -4 and +4 in steps of 0.2 \\
     \texttt{COOLlam} & & uniform between 10 and 30 in steps of 0.01 \\
     \texttt{COOLwidth} & & uniform between 0.2 and 0.65 in steps of 0.05 \\
     \texttt{HOTfcov} & & 0.04, 0.1, 0.2, 0.4, 0.6, 0.8, 1.0, 1.2, 1.4, 1.6, 2.0, 2.5, 3, 5, 10 \\
     \texttt{HOTlam} & & uniform between 1 and 5.5 in steps of 0.01 \\
     \texttt{HOTwidth} & & uniform between 0.2 and 0.65 in steps of 0.05 \\
     \addlinespace[4pt]
     \textbf{\texttt{[activatepl]}} & & \\
     \texttt{uvslope} & & 0 \\
     \texttt{plslope} & & uniform between -2.7 and -1 in steps of 0.01 \\
     \texttt{plbendloc} & nm & 50.  80.  90. 100. 120. 150. \\
     \texttt{plbendwidth} & dex & log-uniform between 0.1 and 10 with 10 steps \\
     \hline
     \hline
     attenuation \& redshifting & & \\
     \hline
     \textbf{\texttt{[biattenuation]}} & & \\
     \texttt{Law} & & Prevot \\
     \texttt{E(B-V)} & & log-uniform between 0.01 and 10 with 80 steps \\
     \texttt{E(B-V)-AGN} & & log-uniform between 0.01 and 0.1 with 80 steps \\
     \textbf{\texttt{[redshifting]}} & & \\
     \texttt{redshift} & & \textit{spec-z mode}: fixed value, from `redshift' column of input file (used in this work) \\
      & & \textit{photo-z mode}: if `redshift\_err' is also given: Weibull distribution matched in mean and standard deviation to `redshift' and `redshift\_err', except \\
      & & \textit{no-z mode}: if `redshift\_err' is negative: uniform distribution between 0.001 and 6. \\
     \hline
     \hline
\end{tabular}
    \label{tab:parameters}
\end{table*}

\subsection{Fitting method \label{sec:Inference}}

This section describes the procedure for constraining model parameters from observational data.
In many fields of astronomy, low signal-to-noise ratio data can be interpreted by fitting parametric physical models using Bayesian fitting algorithms. In contrast, in SED fitting and today's precision photometry surveys, it is common that the data quality exceeds that of the model. This situation is known as inference under model mis-specification. There is little benefit from advanced algorithms when the results are primarily limited by the quality of the model. Instead, it is more important to achieve fast fits, test robustness by varying assumptions, and understand the systematics in the results. In this work, we include improvements in the model (previous section), the likelihood (section~\ref{sec:likelihood}), model uncertainties (section~\ref{sec:modelerr}) and the Bayesian sampling algorithm exploring the parameter space (section~\ref{sec:nestedsampling}). Finally, section~\ref{sec:posterioranalysis} illustrates the outputs of the fit.

\subsubsection{Likelihood \label{sec:likelihood}}

First, we review the $\chi^2$ likelihood implemented in CIGALE and LePhare \citep{Arnouts1999}.
Assuming photometric fluxes $f_i$ were measured with associated uncertainties $\sigma_i$ for a set of bands $i\in\mathcal{B}$, the usual Gaussian likelihood $\mathcal{L}$ is adopted:
\begin{align}
{\mathcal{L}}=\prod_i\frac{1}{\sqrt{2\pi\sigma^2_i}}\times\exp\left[{-\frac{1}{2}\left(\frac{F_i-f_i}{\sigma_i}\right)^2}\right].\label{eq:likelihood}
\end{align}
If the $\sigma_i$ are fixed during the analysis, one can rewrite this as 
\begin{align}
\chi^2=-2\times \log {\mathcal{L}}=\sum_i{\left(\frac{F_i-f_i}{\sigma_i}\right)^2}+\mathrm{const}.\label{eq:chi2}
\end{align}
To maximise the likelihood (and minimise the $\chi^2$), the optimal model normalisation factor can be found analytically.
The profile likelihood variant $\chi_\mathrm{prof}^2$ is then:
\[
    \chi_\mathrm{prof}^2=\sum_i{\left(\frac{s\times F_i-f_i}{\sigma_i}\right)}
\]
with
\[
    s=\sum_i\frac{F_i\times f_i}{\sigma_i^2}/\sum_i\frac{F_i^2}{\sigma_i^2}.
\]
This is how CIGALE and LePhare optimise the normalisation analytically without exploring this additional parameter. In particular, for each model, CIGALE computes the optimal normalisation, and scales proportional parameters, including the stellar mass and AGN luminosity. Thus, the posterior distribution of stellar masses is determined primarily by the model template shapes, but does not fully consider the flux uncertainties. This can be demonstrated by fitting a single photometry data point with 20\% flux error in CIGALE with a single template model (all parameters fixed). The uncertainties on the stellar mass should then trivially be 20\%. However, CIGALE finds an uncertainty of 0\%, which is then  heuristically raised to 5\%\footnote{\href{https://gitlab.lam.fr/cigale/cigale/-/blob/805612292/pcigale/managers/results.py\#L141}{see line 141 in pcigale/managers/results.py, commit 805612292}}. 
This shows that the CIGALE uncertainties on normalisation-related parameters can be severely under-estimated. 
In GRAHSP, we instead use a Bayesian approach and resolve this issue with the full likelihood (eq.\ref{eq:likelihood}) instead of a profile likelihood. The normalisation(s) are free model parameter(s), as e.g., in FortesFit or AGNFitter.

GRAHSP supports flux upper limits. Upper limits are a property of the detection process \citep[see][]{Kashyap2010}, and describe the probability distribution that a source of a flux $f$ escaped detection by chance. We assume that the cumulative probability distribution up to flux $f$ can be described by the cumulative probability density of a Gaussian with mean $f_j$ and width $\sigma_j$, for each band $j\in\mathcal{B}$ with a non-detection. These are consistently incorporated in the likelihood as additional multiplication terms:
\begin{align}
{\mathcal{L}}'={\mathcal{L}}\times\prod_j\int_0^{f_j}{\frac{1}{\sqrt{2\pi\sigma_j^2}}\times\exp{\left[-\frac{1}{2}\left(\frac{F_j-f}{\sigma_j}\right)^2\right]}\, df}.\label{eq:uplimlikelihood}
\end{align}
This is the same treatment as implemented in CIGALE. Users provide upper limits by setting in the data file the flux uncertainty to $-\sigma_j$ (negative values) and the flux to $f_j$. Forced photometry at a position known from another band, may indicate a non-significant flux detection for that band. Following \cite{Kashyap2010}, such upper bounds are treated as normal flux measurements (see the beginning of this section). In addition to measurement uncertainties, additional systematic uncertainties have to be considered.

\subsubsection{Variability and model uncertainties \label{sec:modelerr}}

This section discusses the three types of uncertainties that GRAHSP considers: measurement uncertainties, model uncertainties, and stochasticity introduced by AGN variability.
In the implementation, the total $\sigma$ used in the Gaussian likelihood (eq.~\ref{eq:likelihood} and \ref{eq:uplimlikelihood}) is a combination with Bienaymé's identity of three contributions: 
\begin{align}
\sigma_\mathrm{tot}=\sqrt{\sigma_\mathrm{obs}^2+\sigma_\mathrm{sys}^2+\sigma_\mathrm{var}^2}.\label{eq:errorbudget}
\end{align}
The flux measurement uncertainty $\sigma_{obs}$ is provided by the input photometric catalogue. The model uncertainty $\sigma_{sys}$ is computed for each filter as:
\begin{equation}
\sigma_{i,\mathrm{sys}} = f_\mathrm{systematic\_deviation}\times F_\mathrm{obs} + f_\mathrm{sys}\times F_{i,\mathrm{AGN}}.
\label{eq:systematicterm}
\end{equation}
The first term is already used in CIGALE to account for systematic uncertainty in the galaxy model, with $f_\mathrm{systematic\_deviation}$ set to 10 per cent by default and also here. In addition, in GRAHSP the systematic fractional AGN model uncertainty, $f_\mathrm{sys}$, is a free-fitting parameter (labeled \texttt{systematics} in the output).
It scales the uncertainty with the photometric band's model flux, $F_{i,\mathrm{AGN}}$, considering only AGN components.
Large values of $f_\mathrm{sys}$ provide small $\chi^2$ values, however, due to the normalisation terms in eq.\ref{eq:likelihood}, these are disfavored by their low probability density.
We assign $f_\mathrm{sys}$ an exponential prior distribution centered at zero with a user-definable scale, set to \texttt{systematics\_width}=0.2\% in this work. The heavy tails allow poorly fitted data to increase the parameter to values much larger than \texttt{systematics\_width} \citep[see e.g. discussion in][]{Gelman2006halfcauchy}.

The inclusion of this systematic uncertainty has important implications. In standard practice, when the model produces a poor fit (large $\chi^2$ values), the model parameter uncertainties are typically too small. The fit then either has to be discarded (like in LePhare) or the resulting parameter distributions cannot be trusted. Here, instead, when the model fits the data poorly, $f_\mathrm{sys}$ becomes large, which in turn produces larger uncertainties on the fit parameters. Therefore, even in situations where the model poorly describes the data, the estimates and uncertainties by GRAHSP can be meaningful. This enables systematic analyses of samples and the direct use of the fitting results. Additionally, $f_\mathrm{sys}$ can indicate when the model is a poor fit.

The third term in \cref{eq:errorbudget} is the variance introduced by AGN variability. It is active when \texttt{variability\_uncertainty} is set to \texttt{True} in the configuration.
This term is also proportional to the AGN model fluxes, $\sigma_{var}=F_{i,\mathrm{AGN}}\times F_\mathrm{var}(L_\mathrm{AGN})$. From studying the year-to-year multi-band photometry variability of a sample of X-ray selected AGN in Pan-STARRS1, \cite{Simm2016} found a relation for the fractional variability independent of wavelength and black hole mass, which we adopt as:
\begin{align}
F^2_\mathrm{var}=\mathrm{NEV}=\min(0.1, 10^{-1.43 - 0.74 \times l_{45}}).\label{eq:NEV}
\end{align}
Here, $l_{45}$ is the logarithm of the bolometric AGN BBB luminosity $L_\mathrm{bol}$ in units of $10^{45}\,\mathrm{erg/s}$. 
To summarise, the AGN variability uncertainty term depends on the AGN luminosity, with low-luminosity AGN having the largest variations. The practical implication is that photometry in similar wavelengths collected over several years from multiple surveys can be consistently included. Indeed, if the photometry shows statistically significant variability, this will make the model fit prefer an AGN-dominant model, because only the AGN components can induce additional variance.

\subsubsection{Bayesian sampling algorithm \label{sec:nestedsampling}}

The continuous and discrete parameters listed in \cref{tab:parameters}, plus the systematic model uncertainty parameter, define a challenging 21-dimensional parameter space. The model parameter space is likely to involve non-trivial degeneracies that should be fully explored. This is evident from the flexibility of the model illustrated in \cref{fig:AGNcomponents} pinned down by perhaps only a small number of photometric observations.

The nested sampling \citep{Skilling2004,Ashton2022} Monte Carlo algorithm can perform well in this setting. Nested sampling estimates the parameter posterior probability distribution and the Bayesian evidence (useful for model comparison), by maintaining a population of live points sampled from the prior. At each nested sampling iteration, the lowest likelihood live point is discarded, and the sampling continues, however, with the constraint that any new prior samples must exceed the likelihood of the discarded point. This has two effects: at each step, the likelihood threshold is raised such that the prior volume shrinks by a factor of K/(K+1), where K is the number of live points. Secondly, the population of live points initially globally samples the parameter space, but gradually focuses more towards the likelihood peak. The posterior probability distribution is estimated by posterior samples. These are the discarded points, weighted by the volume discarded at the corresponding iteration i, $V_i=1/K\times(K/(K+1))^i$, and the likelihood of the point, $w_i=\mathcal{L}_i\times V_i$. The evidence is estimated as $Z=\int{\mathcal{L}\times \pi(\theta)\,d\theta}\approx\sum_i w_i$. In practice, after many iteration the likelihood peak is reached and $\mathcal{L}$ plateaus while $V_i$ declines exponentially; thus the iterations can be stopped. The remainder of the live points is also added to the posterior and evidence integrals. For rapid progress towards the posterior peak, we choose K=50 live points.

Iterative progress is made in nested sampling by adding a new live point sampled from the prior, with the requirement that its likelihood is higher than that of the discarded live point. In the first few nested sampling iterations, the likelihood-restricted prior is similar to the entire prior. Therefore, it is efficient to sample directly from the prior using rejection sampling. We adopt the ellipsoidal rejection sampling technique of \cite{Mukherjee2006} for the first 10000 model evaluations, and then switch to slice sampling. Slice sampling \citep[e.g.][]{Jasa2012,Handley2015a} efficiently samples from a likelihood-restricted prior in high-dimensional parameter spaces. We start from a randomly selected live point and perform 20 slice sampling steps before the point is accepted as a new live point. A slice sampling step proposes a direction vector, and along this direction, proposal points are drawn until the proposal exceeds the likelihood threshold. This can be made efficient \citep[see][]{Kiatsupaibul2011} by a stepping-out procedure identifying slice end points. For a literature review of techniques, see \cite{Buchner2021a}. For proposing a direction vector, \cite{Buchner2023} compared various proposals with a diverse set of test problems. The most rapidly converging proposal alternates randomly between differential vectors of random pairs of live points and randomly chosen principal axes of the live point distribution. We adopt this slice sampling technique until the posterior weight of the remaining live points is negligible (<1\%).

The nested sampling procedure described above is implemented in the open-source Python library \texttt{UltraNest}\footnote{\url{https://johannesbuchner.github.io/UltraNest/}}. We connect GRAHSP to \texttt{UltraNest} to obtain posterior samples, Bayesian evidence estimates, diagnostics and visualisations of the posterior. \texttt{UltraNest} allows GRAHSP to resume from previously started runs, which enables modifying post-processing analysis outputs (see the next section) without refitting from scratch.

\subsubsection{Uncertainty quantification \label{sec:posterioranalysis}}

For each analysed source, an output folder is created. \texttt{UltraNest} places the posterior samples and other nested sampling outputs\footnote{see \url{https://johannesbuchner.github.io/UltraNest/performance.html\#output-files}} there.
From the posterior samples the corresponding SED model is calculated. For computational efficiency, by default GRAHSP process only 50 posterior samples (\texttt{-{-}num-posterior-samples} command line option).
Posterior predictions of the individual model components are made and visualised. For each photometric band selected in the configuration file, predicted model fluxes for the entire model, and the AGN and galaxy components alone, are computed. This allows comparisons to other datasets not used in the fit or predictions for future observations. 
The fit parameters and marginal posterior distributions of derived quantities, such as the star formation rate within the last 100\,Myr, the $12\,\si{\micro\meter}$ AGN luminosity or the NEV, are also created, and corner plots can be optionally created. For all visualisations, the information is stored in text files, so that the user can replot the data.
Finally, an output file listing all the analysed sources with the inferred parameter values and derived quantities, is created. The posterior probability distribution is summarised with the median (suffix \texttt{\_med} to the analysed parameter name), standard deviation (\texttt{\_std}), arithmetic mean (\texttt{\_mean}), geometric mean (\texttt{\_lmean}), and 2-sigma equivalent lower and upper quantiles (\texttt{\_lo}, \texttt{\_hi}). The geometric mean can be beneficial for log-quantities such as luminosities and masses.

\section{Data\label{sec:Data}}

We aim to thoroughly test our code and model with diverse data sets. These include pure galaxies (without detected AGN activity) from deep pencil-beam surveys, and large samples of AGN selected in X-ray, optical and infrared wavelengths. \Cref{tab:sampleoverview} in \cref{sec:sampleoverview} presents an overview of the sample sizes, photometric bands and depths.

\subsection{Dataset from COSMOS\label{sec:data:COSMOS}}

The 2\,deg$^2$ COSMOS survey field \citep{Scoville2007} is one of the prime extragalactic survey fields. Deep multiwavelength observations from UV/optical (down to 26 mag) and infrared enable the study of faint and distant galaxies.
In X-rays, it has been fully observed with XMM first \citep{Hasinger2008} and Chandra later \citep{Civano2015} down to a flux limit of $\qty{2.2d16}{\erg\per\centi\meter^2\per second}$ in the 0.5-2 keV band, making it easy to detect also faint AGN \citep{Brusa2010, Marchesi2016}.
Furthermore, extensive and deep spectroscopy campaigns (COSMOS Deep, zCOSMOS, MOSFIRE, DEIMOS, LEGA-C, just to name a few) were conducted over the years, making COSMOS the ideal benchmark for testing SED fitting techniques.
For the spectroscopy, we used the compilation that was available within the COSMOS collaboration in 2017 and considered only extragalactic sources with secure redshift (reliability 100\%), following the ranking introduced by \citet{Lilly2009zCOSMOS}.
We take UV to mid-IR photometry from the COSMOS2015 catalog \citep{Laigle2016}, namely near-UV from Galex, optical from Subaru broad (BVrizy), intermediate and narrow-band filters, near-infrared from the wide-field camera (WFCAM, J band), Canada-France Hawaii Telescope (CFHT, H and Ks band), and mid-infrared bands 1-4 of Spitzer's infrared camera (IRAC).
We further cleaned the sample by discarding all sources with a flag indicating any flux extraction problems such as saturation, and those with neighbors within 2 arc seconds in HST i-band imaging to avoid deblending problems. Fluxes from 3 arcsec apertures are preferred, which is sufficiently large to encompass most galaxies.  AB magnitudes are converted into fluxes. Magnitude errors are conservatively converted into flux errors by comparing the difference between the flux of the nominal magnitude and the flux obtained for a 1$\sigma$ lower magnitude.
Flux measurements are corrected for Milky Way extinction with the values listed in Table 3 of \cite{Laigle2016} for each band, multiplied by the catalogued galactic E(B-V) values.

After this basic cleaning, we have created two sub-samples: 
$\bullet$~{COSMOS-pure-galaxies}: in this sub-sample, all AGN candidates have been removed based on either (a) broad emission lines in the optical VLT \citep{Lilly2009zCOSMOS} and DEIMOS \citep{Hasinger18} spectra, (b) detection by XMM-Newton \citep{Brusa2010} or Chandra \citep{Civano2012, Marchesi2016}, (c) an entry in Simbad marking it as an AGN or AGN candidate, (d) the IRAC-based infrared AGN selection criterion of \cite{Donley2012}.
While some AGN may remain in this sample, they are likely limited to low luminosities. The final pure galaxy sample contains 3367 galaxies.\\
$\bullet$ X-ray detected AGN: The second sub-sample instead consists of X-ray selected AGN taken from the homogeneous catalog of \citet{Marchesi2016} based on the Chandra \citep{Civano2012} and Chandra-Legacy \citep{Civano2015} surveys of the COSMOS area. Out of the 4016 sources listed there, we retain 225 AGN.

The sample of X-ray-detected AGN is used to quantify the stellar mass estimation as a function of:
a) \textit{redshift accuracy}: Often spectroscopic redshifts are not available and photometric redshift are challenging to obtain for AGN \citep[e.g.,][]{Salvato2019}. For the spectroscopic redshift sample, we compute masses adopting the spectroscopic redshift. Additionally, to test the impact of uncertainty induced by photometric redshifts, we also estimated the masses adopting the photometric redshift value from \citet{Marchesi2016} computed as in \citet{Salvato2011}. Their accuracy and fraction of redshift outliers are at the few per cent level.
b) \textit{depth of the multiwavelength data}: Unlike deep pencil-beam surveys, wide-area surveys are characterized by a limited number of shallow photometric points. To emulate this situation, the photometry of these 225 AGN is modified in two steps: first, when the observed flux falls below the flux limits of current all-sky surveys, the measurement was discarded. The limiting depth in AB magnitudes was assumed to be: Galex/near-UV 20.8, B (23.15), r (22.70), i (22.20), z (20.70), yHSC (20.5+0.634\footnote{Vega to AB conversion}), J (20+0.938), Hw (18.8+1.379), Ksw (18.4+1.9), IRAC1 (16.6+2.699), IRAC2 (15.6+3.339), IRAC3 (11.3+5.174) and IRAC4 (8+6.620). Secondly, the photometric errors were increased by setting the flux errors to a fraction of the measured flux. That fraction is chosen from the mean fractional flux error in COSMOS of the respective band. This simple approach follows the assumption that both COSMOS and the emulated surveys are signal-to-noise limited, and the vast majority of sources are found at the faint end.

\begin{figure}
    \centering
    \includegraphics[width=\columnwidth]{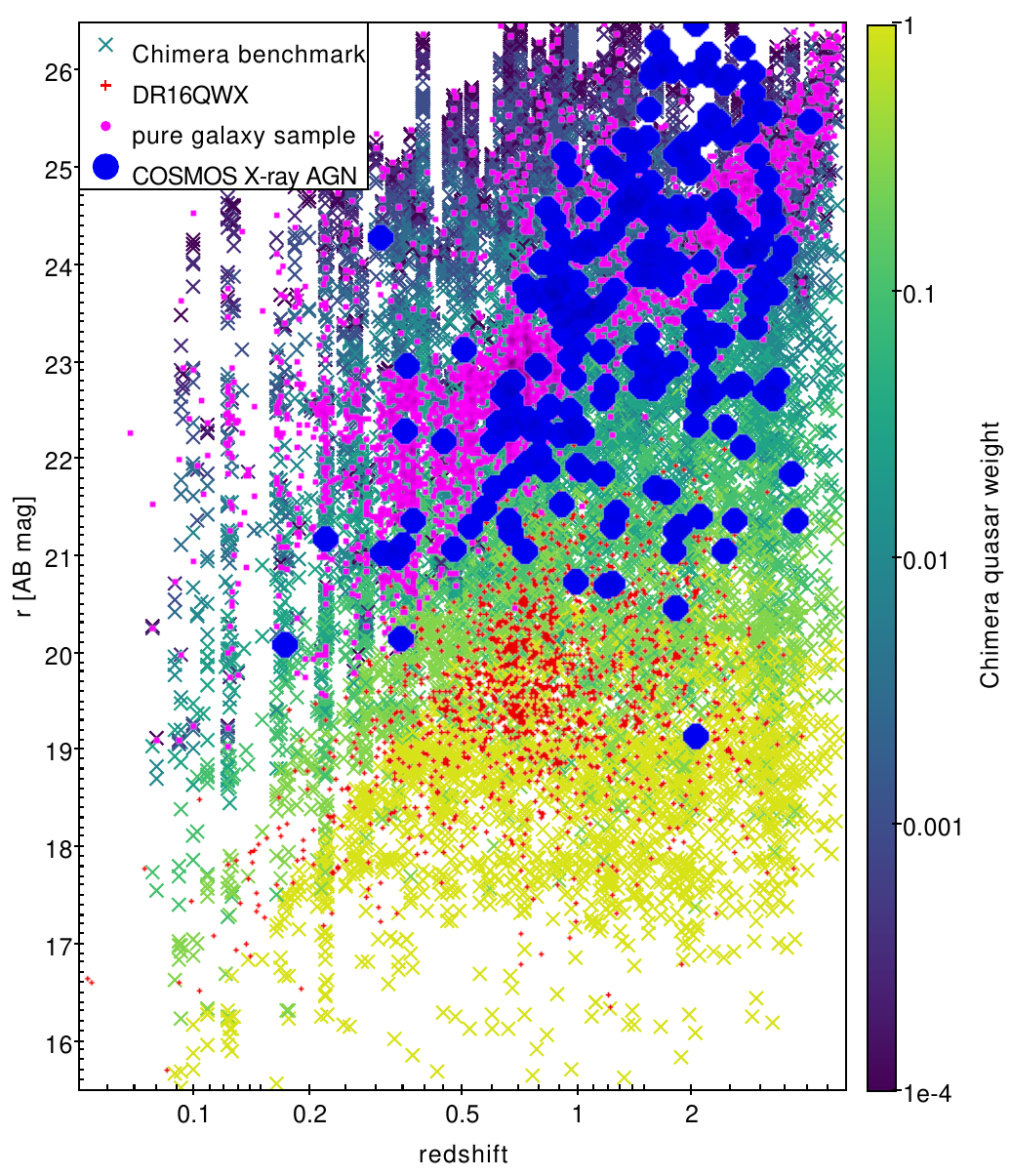}
    \caption{r magnitude as a function of redshift for the samples considered in this work. The COSMOS X-ray selected AGN (big blue circles) are, on average, fainter and at higher redshifts than all-sky WISE or X-ray selected quasars (red). The Chimera objects (crosses) combine COSMOS galaxies (pink) with different levels (colour bar) of quasar light (from optically selected quasars).}
    \label{fig:sampledistr}
\end{figure}
\begin{figure}
    \centering
    \includegraphics[width=\columnwidth]{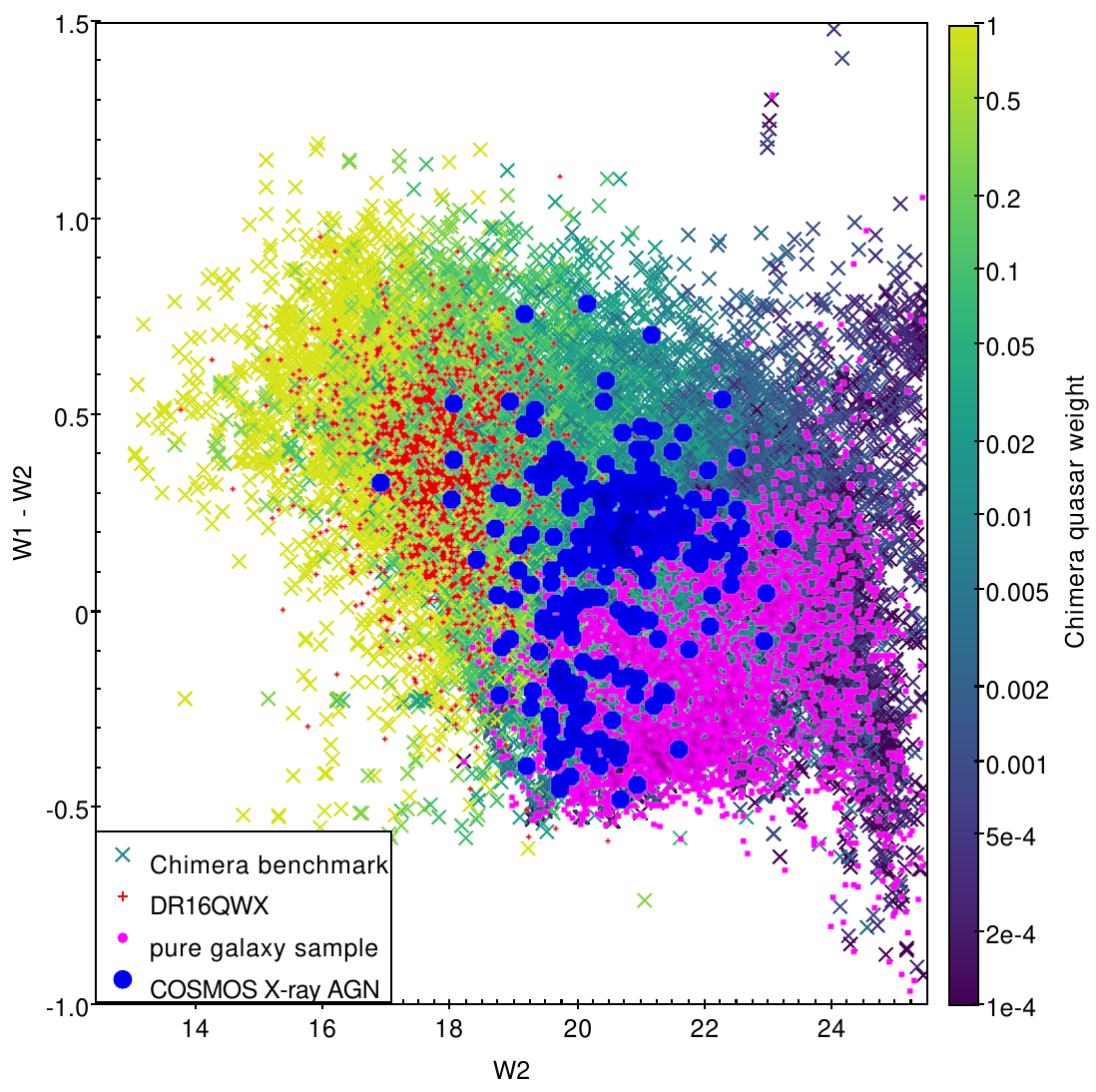}
    \caption{WISE diagnostic plot for the same samples as shown in \cref{fig:sampledistr}. Sources dominated by galaxy light tend to lie towards the bottom-right. With increasing AGN luminosities (colour bar), sources move to the upper-left. Magnitudes are in AB.}
    \label{fig:sampledistW}
\end{figure}

\subsection{All-sky selected AGN}

The AGN detected in X-ray pencil-beam surveys are not capturing the entire range of the AGN population, especially those with the most vigorously accreting black holes. For this reason, we test GRAHSP on AGN samples from wide- or all-sky surveys, selected in infrared, X-rays and optical. The following \cref{sec:data:DR16QWX} and \cref{sec:data:DR7Q} provide the details for the construction of the samples and the extraction of photometry.

\subsubsection{DR16QWX \& eFEDS: all-sky infrared and X-ray AGN\label{sec:data:DR16QWX}} 

In this section, we identify a complementary X-ray and mid-infrared sample of spectroscopic type 1 quasars.
For our DR16QWX sample, the starting point is the QSO data release 16 (DR16Q) from SDSS, presented in \citet{Lyke2020DR16Q}. The DR16Q lists quasars targeted because of infrared and X-ray selection, but also includes previously targeted quasars that have X-ray counterparts.
In terms of infrared selection, the DR16Q includes WISE \citep{Wright2010} AGN selected in the Stripe82X survey field \citep{LaMassa2019}.
In terms of X-ray selection, the DR16Q includes quasars that \cite{Salvato2019} identified as WISE counterparts to the X-ray all-sky ROSAT/2RXS \citep{Boller20162RXS} and XMMSLEW2 \citep{Saxton2008XMMSL1,XMMSSC2018XMMSlew2}. For combining XMMSLEW2 with DR16Q, we find the \cite{Salvato2019} counterpart by a 1'' position match of the respective WISE counterparts. For 2RXS, the \cite{Salvato2019} counterparts are listed directly in DR16Q.
In addition, the DR16Q also includes the optical counterparts to XMM sources from XMMPZCAT \citep{Ruiz2018} and the 3XMM-DR8, both based on the third version of the XMM-Newton Serendipitous Source Catalog \citep{Rosen2016}.
Therefore, from the DR16Q, we select quasars selected by WISE, ROSAT or XMM.
Only quasars \citep[classified as normal quasar or broad-absorption-line quasar in Table 2 of][]{Lyke2020DR16Q} were kept.
Furthermore, the b-band galactic extinction is required to be less than 0.1 magnitudes, which limits the sample to high galactic latitudes. Of the remaining sources, all show high redshift confidence (above 0.35). The vast majority of these (80\%) were not targeted by the optical selection of SDSS DR7 \citep{Schneider2010DR7}, making the selection complementary to bright, blue quasars. 
We restrict the sample to high quality redshifts (\texttt{ZWARNING}=0 and \texttt{SN\_MEDIAN\_ALL}>1.6, see e.g. \cite{Menzel2016}), and to not be in DR7 (\texttt{IS\_QSO\_DR7Q}<1). This leaves 2349 sources.
Finally, to focus on non-jetted AGN, sources within one arc second of source positions listed in the CRATES flat-spectrum radio source catalog \citep{Healey2007} or the BZCAT blazar catalog \citep{Massaro2015} are removed. Our final DR16QWX sample includes 2296 AGN.

For an eROSITA sample, we adopt the extragalactic sources from the eROSITA Final Equatorial-Depth Survey \citep[eFEDS][]{Brunner2021,Salvato2022}. We focus on sources with spectroscopic redshifts above 0.002, and remove jetted AGN as described above. 

Next, the UV to infrared photometry is collated with a one arcsecond search radius around the optical position. We carefully construct photometry from near-UV to mid-IR with matched apertures. Our implementation is released with this paper as RainbowLasso\footnote{\url{https://github.com/JohannesBuchner/RainbowLasso}}.
Near-UV magnitudes from Galex \citep{Bianchi2017Galex} are converted into fluxes (see \cref{sec:data:COSMOS}). Measurements below 20.8 AB mag, the nominal Galex survey depth, are discarded. Co-extracted optical (g, r, i, z) and WISE (W1-W4) photometry is obtained from the Legacy Survey data release 10 \citep[LS10][]{Dey2019}. 
Sources were discarded if the photometry extraction fit was forced, hit limits, or the parameters had to be held fixed\footnote{see \url{https://www.legacysurvey.org/dr10/bitmasks/\#fitbits}}. Heavily blended sources are also discarded. If a multi-source fit assigned less than 10 per cent of flux in the {\it g}, {\it r} or {\it W2} bands to other sources (\texttt{fracflux\_*} columns), the source is considered isolated. Final sample sizes are listed in the first row of \cref{tab:sampleoverview}.
Because of the large point-spread function in the {\it W3} and {\it W4} bands, we discard {\it W3} and {\it W4} photometry if other sources dominate in either band or the contribution in {\it W1} or {\it W2} exceeds 10\%.
We consider point sources those where a point-spread function model provided the best fit (\texttt{TYPE}=``PSF''), and the remainder extended sources.
For extended sources, we use fluxes extracted from 5 arcsecond apertures (\texttt{apflux} column 7 for the optical bands and column 5 for WISE bands), 
while for point sources, we use the model flux. 
Fluxes are corrected for galactic attenuation by dividing by the respective \texttt{MW\_TRANSMISSION} columns.
Near-infrared photometry is added from the Visible and Infrared Survey Telescope for Astronomy (VISTA) Hemisphere survey (VHS) data release 5 \citep{McMahon2013VHS,McMahon2013VHScat}, the United Kingdom Infrared Telescope (UKIRT) and InfraRed Deep Sky Survey (UKIDSS) \citep{Lawrence2007UKIDSS,Warren2007UKIDSSDR2}.
For point sources, we use aperture-corrected PSF model fluxes extracted from 2.8 arcsecond diameter apertures (\texttt{apermag4}), while for extended sources, we use not-aperture corrected fluxes from 5.6'' diameters (\texttt{aper*corr6}).
We also retain the X-ray flux from the source X-ray catalogs, where available, and convert it to a $2-10\,\si{\kilo\electronvolt}$ luminosity assuming a power law index of 1.8, as most sources are only mildly absorbed.

\subsubsection{SDSS-DR7Q: pure quasar sample\label{sec:data:DR7Q}}

We construct a pure quasar sample that complements the COSMOS pure-galaxy sample in \cref{sec:data:COSMOS}. By ``pure'' we mean here that the emission is dominated by AGN processes at essentially all wavelengths, and galaxy contribution is negligible. Ultraviolet and optically selected quasars were targeted by SDSS up to data release 7 catalogue \citep[DR7][]{Schneider2010DR7}. \citet{Shen2012SDSSDR7Q,Shen2012SDSSDR7Qcat} selected sources where the i-band absolute magnitude exceeds -22 and the SDSS spectra show at least one broad emission line. This resulted in 105,783 extremely luminous, unobscured quasars. We assume that these are completely dominated by the AGN process and can be treated as point sources.
We use the Milky Way-corrected AB magnitudes in the catalogue of \cite{Shen2012SDSSDR7Q}. Besides the optical bands (ugriz), the catalogue also provides near-infrared Vega magnitudes (JHK$_\mathrm{s}$) from the Two Micron All Sky Survey (2MASS) \citep{Skrutskie20062MASS} and infrared Vega magnitudes from WISE. The magnitudes and errors (see \cref{sec:data:COSMOS}) were converted into fluxes.

\section{The Chimera benchmark data set\label{sec:data:Chimera}}

To date, no benchmark exists for characterising the bias and variance of AGN host galaxy properties. These include fundamental physical quantities such as stellar masses, star formation rates, and stellar ages. The fundamental reason is that, unlike in the problem of estimating redshifts from photometry, true labels are generally not available even with more costly observations (spectroscopy).
Comparisons of outputs from various codes and different models indicate that the results vary with model assumptions \citep{Mobasher2015,Santini2012}.
Works on AGN hosts usually exclude from the analysis the unobscured AGN \citep[e.g.,][]{Aird2015}, because they present large uncertainties \citep[e.g.,][]{Bongiorno2012}.
\citet{Ciesla2015} demonstrated with simulated unobscured AGN how stellar mass and SFR are respectively over and under-estimated,  already with a 40 per cent light contribution by the AGN. Reliable retrieval becomes even more challenging the fewer photometric bands are available. However, because the model assumed for simulating data was the same as the model for inference, such tests provide limited insight into the true bias and scatter of estimated host parameters.

To account for these issues, we aim to design a benchmark dataset in which fiducial truth physical galaxy properties are \textit{known}.
The Chimera\footnote{In biology, a chimera is an individual consisting of tissues of diverse genetic constitution. In Greek mythology, a chimera is a fire-breathing she-monster having a lion's head, a goat's body, and a serpent's tail.} benchmark for AGN-galaxy decomposition is derived from real-world data without assuming a true model for AGN.
We rely on the inference of physical properties in pure galaxies (non-AGN), as validated in \cite{Mobasher2015}. We verified that the stellar masses determined by GRAHSP are consistent with those presented in the COSMOS2015 catalog (see the next section) and other codes (\cref{sec:othermethods}), so that any of these methods can be used as a reference value.
This establishes fiducial truth physical galaxy properties, in the sense that we know what would have been measured if the galactic nucleus were inactive.
In the second step, pure quasars from the DR7Q dataset (\cref{sec:data:DR7Q}) are randomly paired with pure galaxies (\cref{sec:data:COSMOS}), under the constraint that the \textit{spectroscopic} redshifts are within 0.01 of each other. The fluxes in each photometry band are then summed. For this step, the optical and near-infrared bands ({\it urizJHK}) can be summed directly, as the filter curves are in close agreement. For IRAC and WISE bands 1 and 2, tiny conversions from \citet{Stern2012} are applied. The higher IRAC bands substantially differ from WISE 3 and 4, and we do not consider them. The fluxes are summed, with a weighting factor applied to the quasar light, which controls how luminous a quasar is added.
Uncertainties are propagated assuming Gaussian noise. Fluxes are only provided for bands where both the galaxy and the quasar had a measurement.

The Chimera benchmark dataset is enriched with reference galaxy and AGN properties. These are the galaxy properties (stellar mass, SFR, stellar age, attenuation, etc.) from COSMOS2015 and AGN properties (bolometric and $5100\si{\angstrom}$ quasar luminosities, black hole mass) from \cite{Shen2012SDSSDR7Q}. The Chimera benchmark is released\footnote{\url{https://doi.org/10.5281/zenodo.8431646}} with this paper, so that it can be used as a reference for comparing physical parameter estimates for AGN in codes currently in use or under development.
\begin{figure}[ht]
    \centering
    \includegraphics[width=\columnwidth]{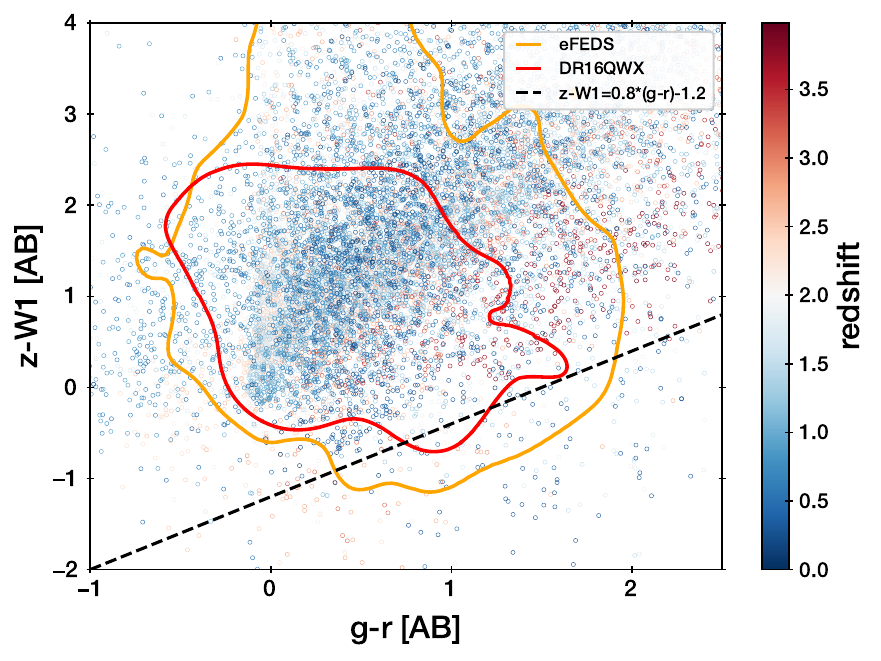}
    \caption{Colour diagnostic plot. The GRAHSP templates (coloured dots as a function of redshift) cover the entire plane.
    The observations are illustrated by coloured contours containing 99\% of each sample. At the centre-left, quasars dominate (red outline, DR16QWX), while the eFEDS sample includes a wider distribution.
    The dashed diagonal line demarks the separation of \cite{Salvato2022} between galaxies (above) and stars (below the line).
    }
    \label{fig:modelcolors}
\end{figure}

\section{Preliminary Tests on GRAHSP\label{sec:Test}}

\subsection{Model plausibility}

As a first check, model colours are verified against colours of observed samples of AGN and galaxies.
\Cref{fig:modelcolors} shows the {\it g-r-z-W1} colour-colour diagnostic plot presented in \cite{Salvato2022}. 
We draw random samples from the GRAHSP model prior, construct the template and present the model colours as dots in \cref{fig:modelcolors}. These cover the entire parameter space and beyond.
The model dots envelop the distribution of quasars from DR16QWX and eFEDS AGN, which are illustrated as contours in \cref{fig:modelcolors}. The locus of stars \citep[see][]{Salvato2022} is below the dashed line, and as expected, few galaxy model templates are in this region. Appendix~\ref{sec:modelphotometrychecks} extensively tests the model colours against four observational samples (DR16QWX, eFEDS, COSMOS AGN and COSMOS pure galaxies).
There it is shown that emission lines and the FeII template are an important contributor, without which the observed colours cannot be reproduced. This has also been noted in \citet[][]{Temple2021}, which studied the mean quasar colour over a wide redshift range. Here, we demonstrate that the GRAHSP model is capable of reproducing not only the mean colour but also the wide diversity of colours in the galaxy and AGN population.

\subsection{Retrieving stellar masses for COSMOS-pure-galaxies sample }

\begin{figure}
    \centering
    \includegraphics[width=\columnwidth]{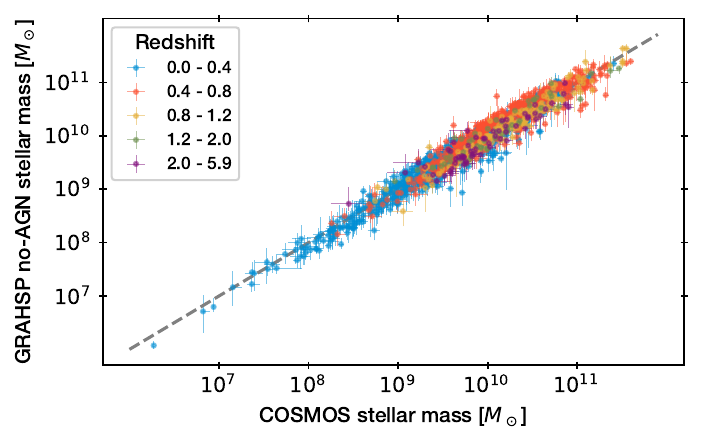}
    \includegraphics[width=\columnwidth]{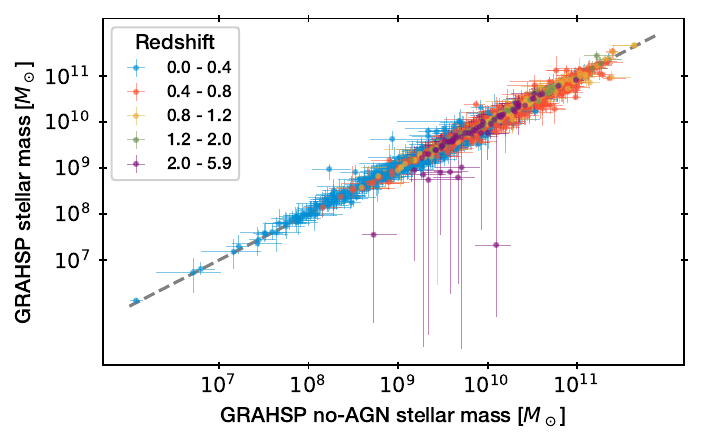}
    \includegraphics[width=\columnwidth]{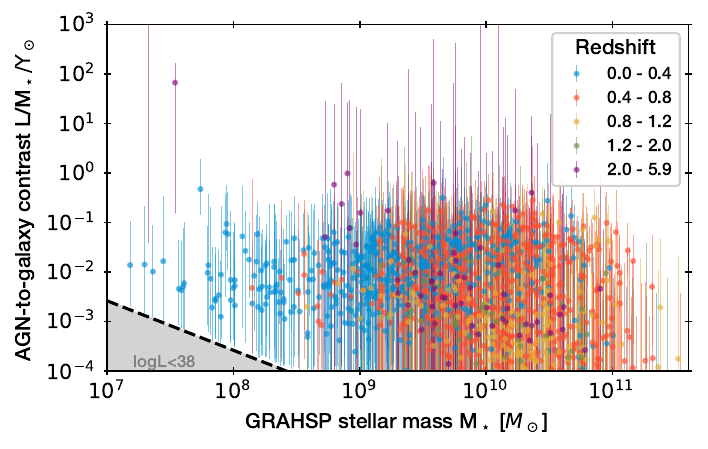}
    \caption{Stellar masses from COSMOS2015 are compared with those from GRAHSP, with the AGN component turned OFF (top panel) or ON (middle panel), for various redshift bins. In both cases, the agreement is extremely good, with very little scatter around the 1:1 relation. For the case with the AGN component turned ON, the GRAHSP AGN luminosity is plotted as a function of stellar mass. As expected, in all sources the AGN component is well below 10\% in all cases but two.
    }
    \label{fig:galaxyMgal}
\end{figure}

We first test GRAHSP on the COSMOS-pure-galaxy sample, using only galaxy models. The top panel of \cref{fig:galaxyMgal} compares the resulting stellar mass estimates to those published by \cite{Laigle2016}. To ensure consistent assumptions, we show only sources where their adopted redshift agrees with the spectroscopic redshift within 1 per cent. The stellar mass estimates agree well at all redshifts and at all stellar masses.

Secondly, we test the impact of introducing our AGN model components, even if it is not required. On the same COSMOS-pure-galaxy sample, the middle panel of \cref{fig:galaxyMgal} compares two runs: once without AGN from the previous experiment and once with the full GRAHSP setup described in section~\ref{sec:Method}. The estimates show close agreement, indicating that the added model complexity does not bias the measurement. In most cases, the upper error bar of the AGN luminosity is well below ${10}^{42}\,\mathrm{erg/s}$, allowing for low-luminosity AGN, at most. In the SED, the AGN is always sub-dominant compared to the host. This can be quantified with the AGN-to-galaxy light contrast ratio, $\LAGN/M_\star/\Psi$, where $\Psi$ is the solar light-to-mass ratio (see \cref{bolometric}). The bottom panel of \cref{fig:galaxyMgal} shows that the contrast lies well below 10\% in the vast majority of cases. From the entire sample, only few have a blue SED that can be fit with a star-burst or a luminous AGN, giving rise to large (but correct) uncertainties in the contrast ratio. These can also be seen in the middle panel.

\subsection{Template fit test}
\begin{figure}
    \centering
    \includegraphics[width=\columnwidth]{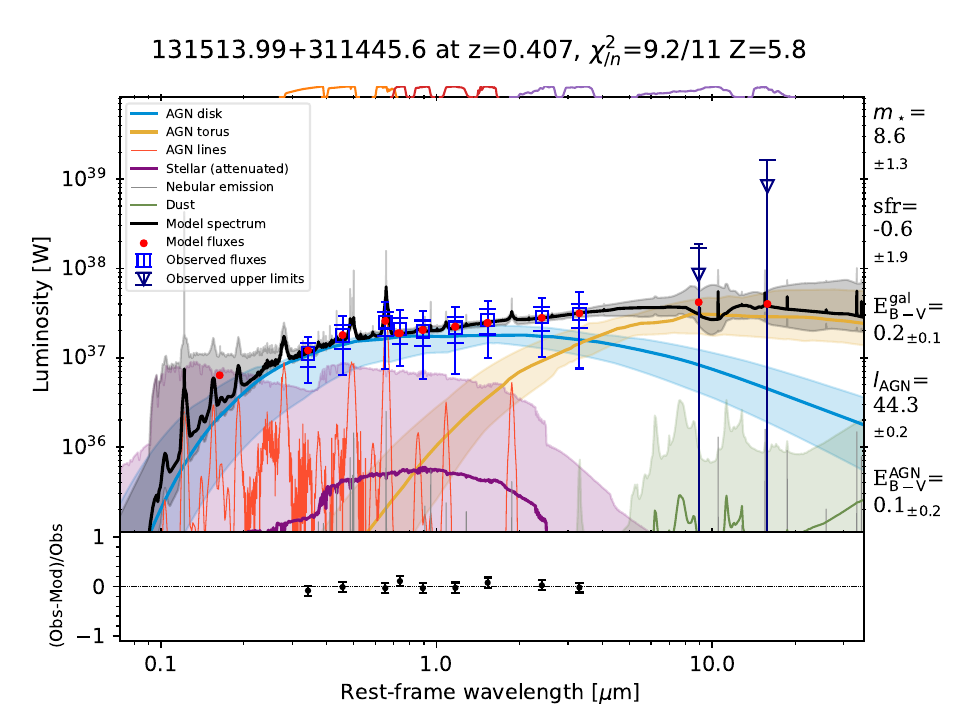}
    \caption{Example of an AGN-dominated SED fit from the DR16QWX sample. The total model (black) and the individual components (colours, see legend) are presented with the posterior mean (solid line) and 2 sigma equivalent uncertainties (shaded areas). Observed fluxes are shown as blue data points with blue squares and wide blue capped $3\sigma$ error bars. The overlaid blue error bars with thin caps show the enlarged total error budget after the fit. The filter curves of observed photometric bands are shown at the top, grouped to the same colours by instrument.
    Predicted model fluxes are shown as red points. These can deviate from the black curve because of the averaging over the filter curve.
    The bottom panel presents the relative residuals of model and observations. 
    The title lists the source ID, redshift, $\chi^2$ (\cref{eq:chi2}) per number of data points and Bayesian model evidence. On the right, key values are shown, including the stellar mass (in $M_\odot$), star formation rate (in $M_\odot$/yr), the galactic and AGN extinction, and the $5100\unit{\angstrom}$ AGN luminosity (in erg/s). Lower case variables are logarithmic.
    Below $2\unit{\micro\meter}$, the model is dominated by the AGN disk, and above by the torus. The third data point (z-band) is raised compared to the neighboring data points, which can be understood by the $H\alpha$ emission line at that redshift. The galaxy component has an extremely wide luminosity range (purple), i.e., the stellar mass is unconstrained (see the first number on the right) in this luminous AGN (see the fourth number on the right).
    }
    \label{fig:exampleagnsed}
\end{figure}
\begin{figure}
    \centering
    \includegraphics[width=\columnwidth]{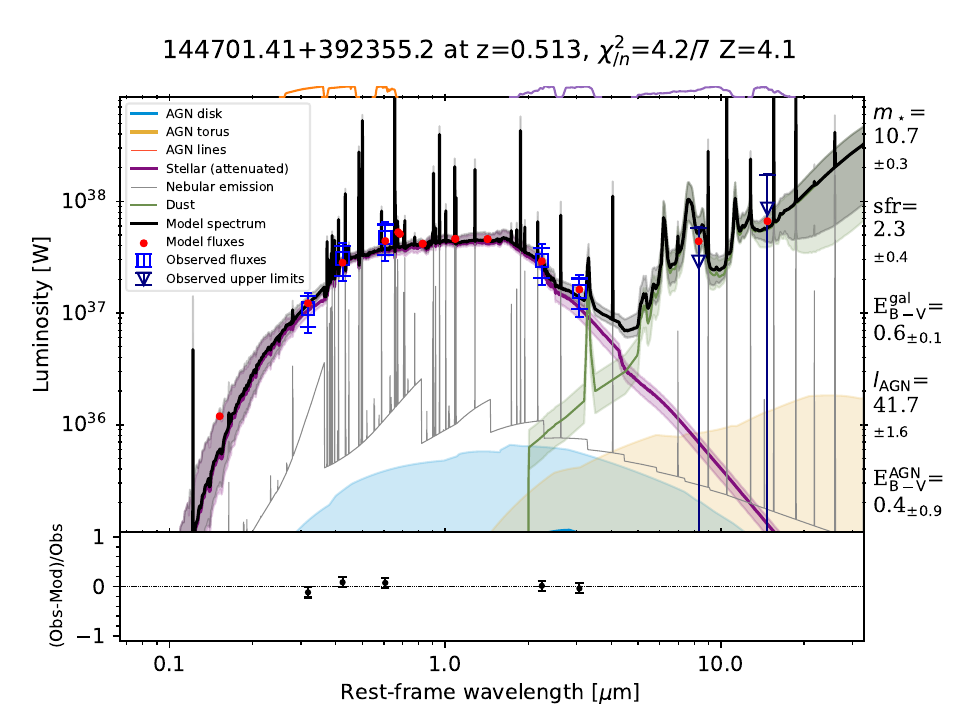}
    \caption{Example of a host-galaxy-dominated SED, with elements as in \cref{fig:exampleagnsed}. Here, the SED is dominated by galaxy components (purple and green curves), including nebular emission lines (grey). The stellar mass and star formation rate are well constrained (numbers on the right), while the AGN luminosity has more than an order of magnitude uncertainty around a low value of $10^{41}\unit{\erg\per\second}$. Here, model uncertainties (thin cap error bars) were only slightly increased without requiring an AGN component.
    }
    \label{fig:examplegalsed}
\end{figure}

\begin{figure*}
    \centering
    \includegraphics[width=\textwidth]{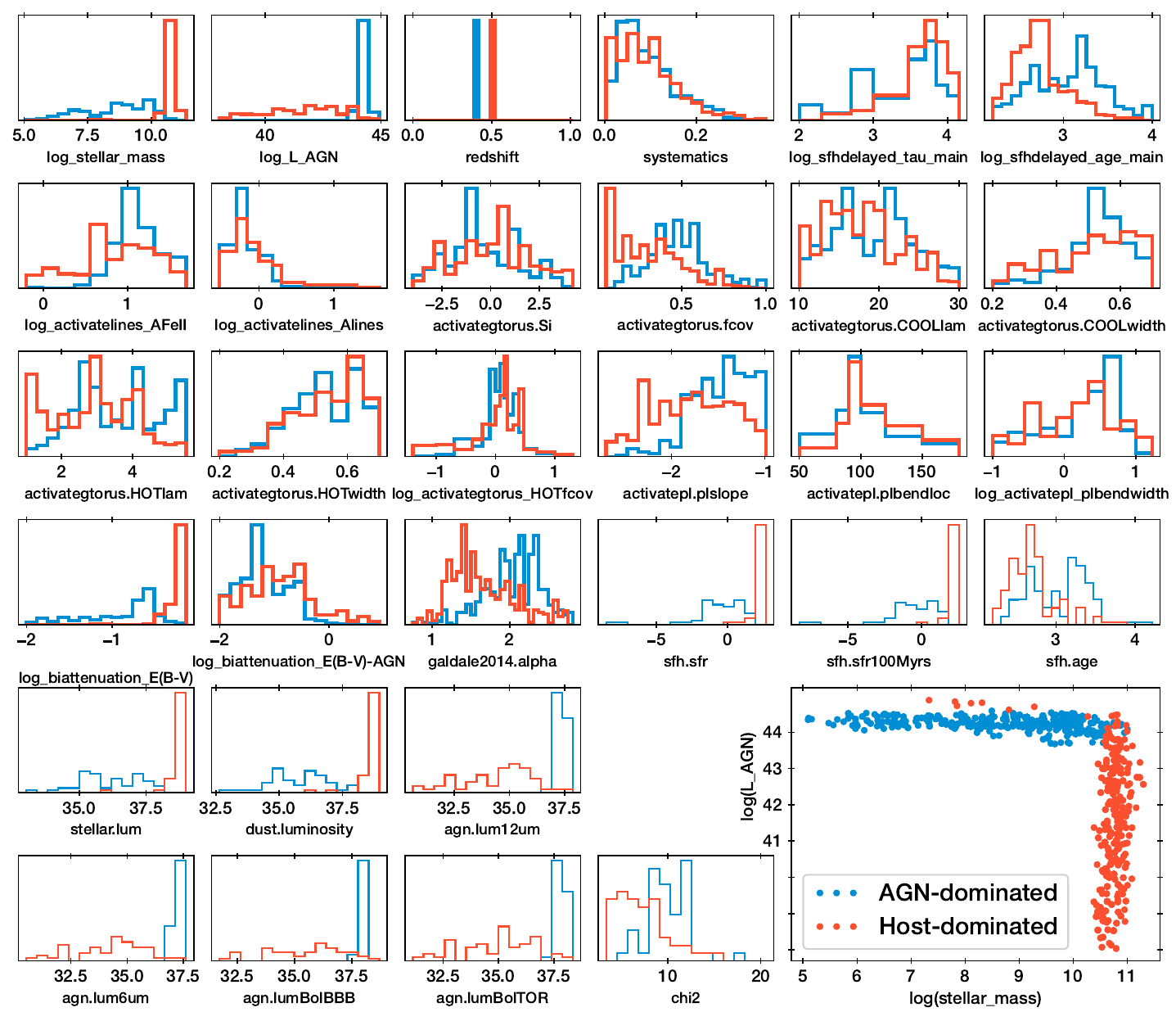}
    \caption{Parameter posterior probability distributions for two different objects: One AGN-dominated source (blue) of \cref{fig:exampleagnsed} and one host-galaxy-dominated source (red) of \cref{fig:examplegalsed}. Panels with thick histograms are model parameters. Panels with thin histograms are derived parameters. The large bottom-right panel shows the joint posterior distribution of AGN luminosity and stellar mass as posterior samples.}
    \label{fig:posteriors}
\end{figure*}

Next, the ability to describe SEDs from a diverse sample of AGN by the models in GRAHSP, is tested. We use here the DR16QWX sample of infrared and X-ray selected AGN (section \ref{sec:data:DR16QWX}) and show three extreme cases. Firstly, \cref{fig:exampleagnsed} shows an AGN-dominated case. The model follows the data points closely. The H$\alpha$ emission line at 0.653$\mu$m substantially contributes to the measured flux in the relevant filter (blue square), which is also reflected in the model flux (red point). This demonstrates the importance of emission features in the templates. 
\Cref{fig:exampleagnsed} shows the total model and its components. Because the fit is fully Bayesian with continuous amplitude parameters, the posterior (shaded areas) is smooth. The corresponding parameter constraints are shown in \Cref{fig:posteriors} as blue histograms. In particular, the stellar mass (top left panel) posterior is a flat distribution reaching up to approximately $10^{10} M_\odot$, indicating that for this AGN-dominated source only an upper limit can be estimated.
A host-galaxy-dominated source is presented in \cref{fig:examplegalsed} as a second example. The galaxy template is constrained well. Comparing the thin cap error bars with the wide cap blue error bars, we see that the data uncertainties were slightly enlarged by the fit (see \cref{sec:posterioranalysis}).
The parameter constraints are shown in red in \cref{fig:posteriors}. The panels marked `log\_stellar\_mass` and `sfh.sfr` show that stellar mass and star formation rate could be constrained, respectively. The AGN luminosity (second panel in the top row) is constrained to less than $\qty{e44}{\erg\per\second}$.
Degeneracies between all model parameters can also be explored. The bottom right panel of \cref{fig:posteriors} only shows the conditional distribution of stellar mass and AGN luminosity, with the obtained posterior samples.

\begin{figure}
    \centering
    \includegraphics[width=\columnwidth]{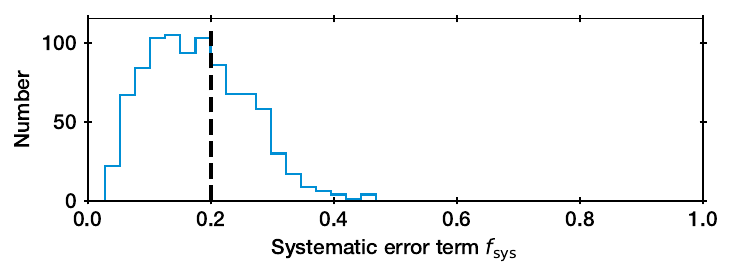}
    \caption{Distribution of posterior medians of the systematic model error parameter $f_\mathrm{sys}$ (see eq.\ref{eq:errorbudget}).}
    \label{fig:syserrdist}
\end{figure}

We investigate this a bit closer by looking at the inferred systematic uncertainty parameter values. \Cref{fig:syserrdist} presents the distribution of posterior medians on the systematic model error parameter f$_{sys}$ for the entire DR16QWX sample. The median of the prior is 0.2 (vertical black dashed line). This indicates that, for most sources, no additional model uncertainty had to be introduced during fitting. For reference, the cases from \cref{fig:exampleagnsed} and \cref{fig:examplegalsed} have a median f$_{sys}$ of approximately 0.08.

\subsection{Extension to X-ray data}
\label{subsec:xray}
\begin{figure}
    \centering
    \includegraphics[width=\columnwidth]{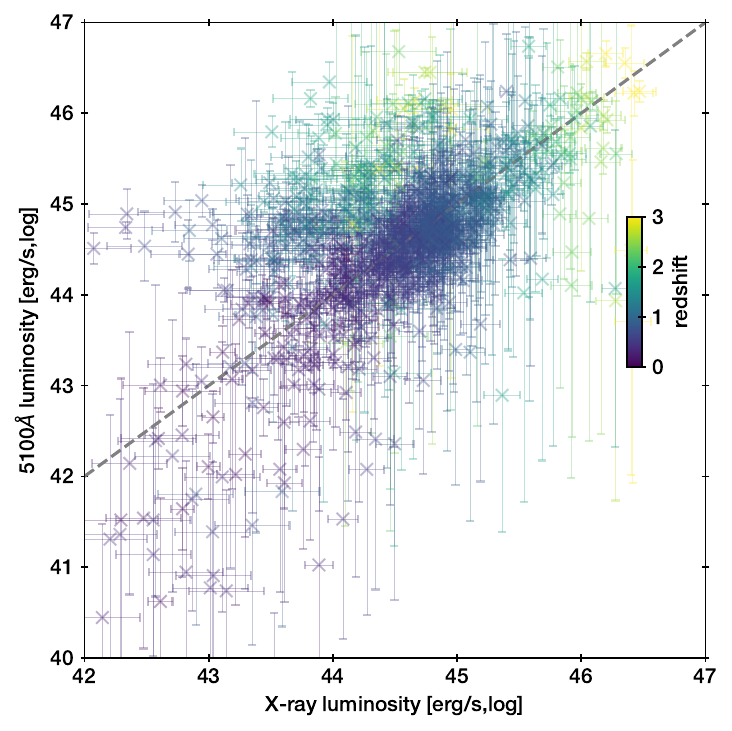}
    \includegraphics[width=0.9\columnwidth]{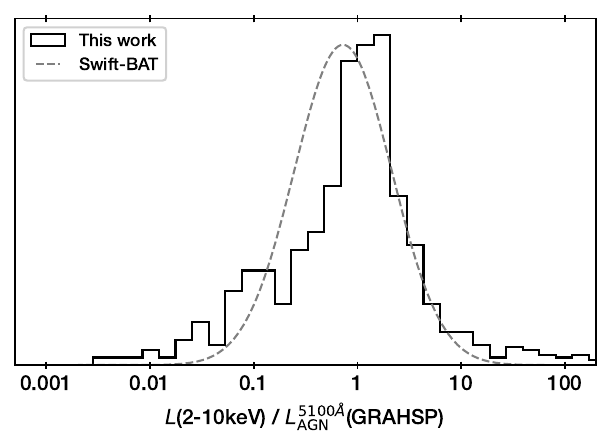}
    \caption{
    \textit{Top panel}: The AGN luminosity at $5100\si{\angstrom}$ inferred from GRAHSP is compared to the $2-10\si{\kilo\electronvolt}$ luminosity from X-ray observations in the DR16QWX sample. Points are colour-coded by redshift, with yellow indicating the highest redshifts.
    \textit{Bottom panel}: The ratio is shown as a black histogram.
    For comparison, the grey dashed curve represents the mean ratio and scatter observed by \cite{Koss2018}.
    }
    \label{fig:LLdist}
\end{figure}

\cite{newcigale2022ApJ...927..192Y} investigated the use of X-ray constraints to improve the model fit. As illustrated with the galaxy-dominated case (\cref{fig:examplegalsed}), in particular low-luminosity and obscured AGN sources can benefit from additional constraints on the AGN component. GRAHSP supports placing priors on the AGN luminosity. Here we investigate whether X-ray observations are appropriate to constrain the $5100\si{\angstrom}$ luminosity parameter of our model. 

The upper panel of \cref{fig:LLdist} shows the correlation between SED-inferred $5100\unit{\angstrom}$ luminosities and X-ray-based luminosities. The X-ray luminosities were not used in the fit and are taken by conversion from X-ray survey catalog fluxes (see section~\ref{sec:data:DR16QWX}). They show a 1:1 correlation with some scatter, particularly due to sources with large uncertainties. We investigate the intrinsic scatter by considering the luminosity ratio (blue histogram in the bottom panel of \cref{fig:LLdist}). The distribution is compared to results from the local Universe in the literature, where \cite{Koss2018} reported a $0.43 \pm 0.47$ log-ratio distribution between the $5100\si{\angstrom}$ and $14-195\si{\kilo\electronvolt}$ luminosities, which the authors relate to the $2-10\si{\kilo\electronvolt}$ luminosity with a shift of 0.42~dex.
Fortunately, the two shifts almost cancel out, making the $5100\si{\angstrom}$ luminosity approximately equal the $2-10\si{\kilo\electronvolt}$ luminosity, with a scatter of 0.5~dex (gray distribution in the bottom panel of \cref{fig:LLdist}). This agrees with our distribution of mean ratios. This demonstrates that measured X-ray fluxes, blurred with a 0.5~dex Gaussian, can be used as AGN luminosity priors in GRAHSP fits. Practically, this is achieved by providing the necessary columns in the input file (see \cref{tab:parameters}).

\subsection{Literature model setups\label{sec:othermethods}}

In the subsequent sections we test GRAHSP. We compare the results from GRAHSP to those derived from three alternative methodologies that have been described in the literature. All of these are based on CIGALE, and for our analysis we adopt the latest version of that code (2022.1). We adopt broadly the same setups that are described in the literature except that in all cases we use a Planck cosmology \cite{Planck18} and the \cite{maraston2005MNRAS.362..799M}  SPS model, noting however that adopting the \cite{Bruzual2003} models produces similar results.

The first model, ``D14'', adds the AGN model of \cite{dale2014ApJ...784...83D} to the stellar population. The slope is allowed to vary (values: 0.0625, 0.5, 1.0, 1.5, 2.0, 2.5, 3.0, 3.5, 4.0), and so is the AGN fraction (values: 0.01, 0.1, 0.15, 0.2, 0.25, 0.3, 0.35, 0.4, 0.45, 0.5, 0.55, 0.6, 0.65, 0.7, 0.75, 0.8, 0.85, 0.9, 0.99).
Since the AGN model has only two parameters, the galaxy parameters can be evaluated on a fine grid while retaining an acceptable computational cost, with $\tau$ taking values of 1, 50, 500, 1000, 2000, 2500, 3000, 3500, 4000, 4500, 5000, 5500, 6000, 7000, 8000 Myr and the age ranges from 100 to 13\,Gyr in 15 logarithmic steps. A modified starburst attenuation law is applied.

The second model, ``C15'', follows the setup of \citep{Ciesla2015}. The AGN is modelled with the \citet{fritz2006MNRAS.366..767F} template library. The full list of parameters is given in \cref{sec:cigalesetups}. Due to the high number of AGN parameters, the galaxy parameters need to be more coarsely gridded in CIGALE, with $\tau$ taking values of 500, 1000, 3000, 5000, 10000 Myr and age taking values of 500, 1000, 3000, 5000, 7000, 10000. A Calzetti-Leiterer attenuation law is applied \citep{Calzetti2000,leitherer2002ApJ...574..114L}.

The third model, ``Y19'', follows the setup of \cite{Yang2019}. The AGN model is SKIRTOR \citep{Stalevski2016}, which also has many free parameters listed in \cref{tab:othermodelsparams} in the appendix. 
The galaxy parameters are coarsely gridded, with $\tau$ taking values of 100, 500, 1000, 5000 Myr and age taking values of 500, 1000, 3000, 5000, 7000.

The comparison of GRAHSP and the three CIGALE model setups is shown in \cref{sec:Results}. 
For a fair comparison, we compute the Chimera fiducial truth from COSMOS with the galaxy model setup exactly matching the respective model setup. However, adopting the COSMOS2015 results as fiducial truth does not change the results, as there is good agreement on pure galaxy inference (see, e.g., the top panel of \cref{fig:galaxyMgal}).

GRAHSP has many more free parameters than the above setups do. For example, SKIRTOR is the model with the largest number of free parameters, but for computationally feasible fits with CIGALE, many need to be fixed. \Cref{sec:brown-skirtor-test} investigates whether having all parameters free can resolve any differences between GRAHSP and CIGALE. In particular, similar to \cref{subsec:calibration}, we fit SKIRTOR against the \cite{Brown2019} Atlas of AGN spectra in \Cref{sec:brown-skirtor-test}, and find substantial residuals.

\subsection{Validation of the Chimera benchmark\label{sec:chimera:lowLvalidation}}

\begin{figure}
    \centering
    \includegraphics[width=\columnwidth]{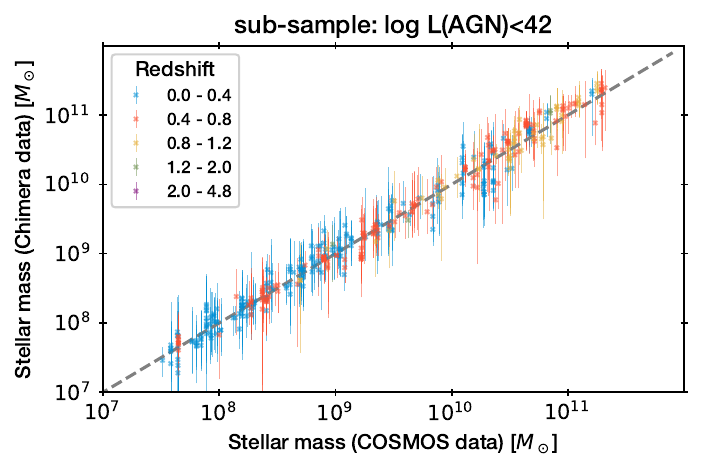}
    \caption{Comparison of stellar masses measured from COSMOS-pure-galaxies sample to that measured by GRAHSP for the Chimeras built using the same galaxies and with low AGN contribution. The measurements are consistent; for some cases, the uncertainties are very large.}
    \label{fig:chimera:lowLresults}
\end{figure}

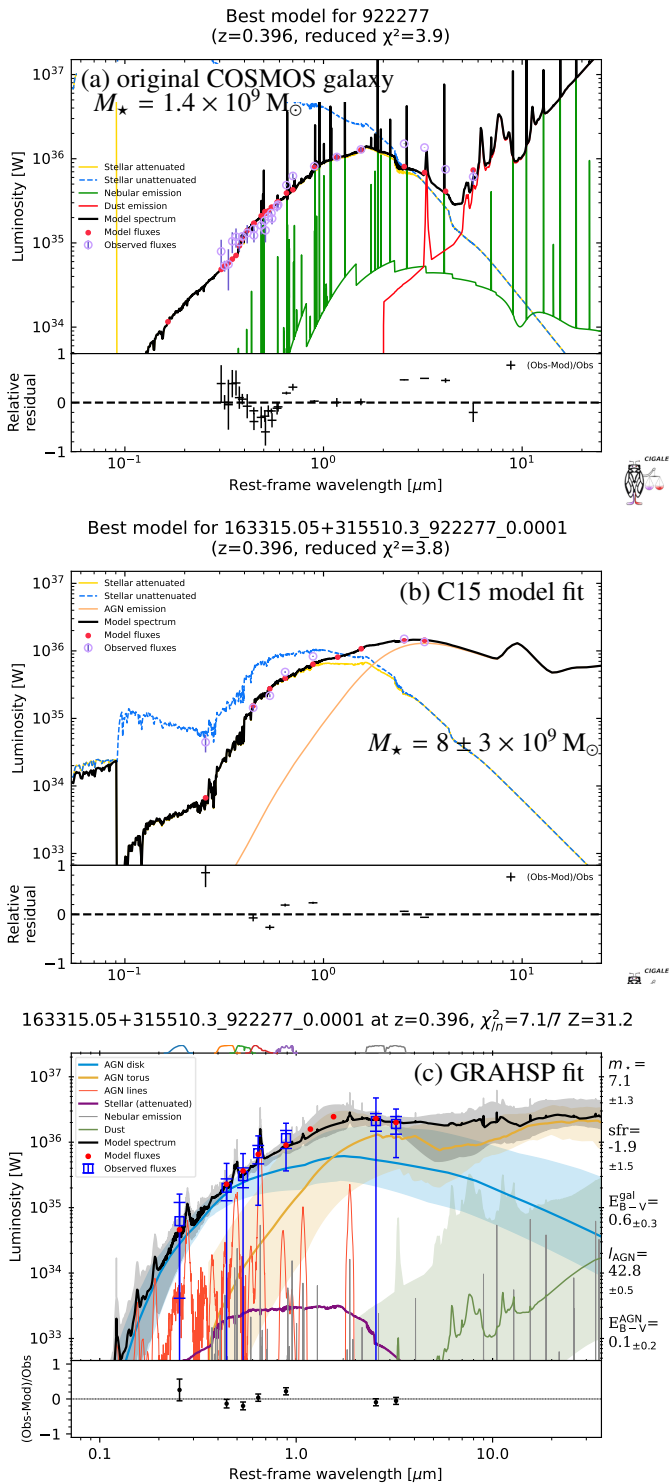
\begin{figure}
    \centering
    \begin{tikzpicture}
        \draw (0, 0) node[inner sep=0] {\includegraphics[width=\columnwidth]{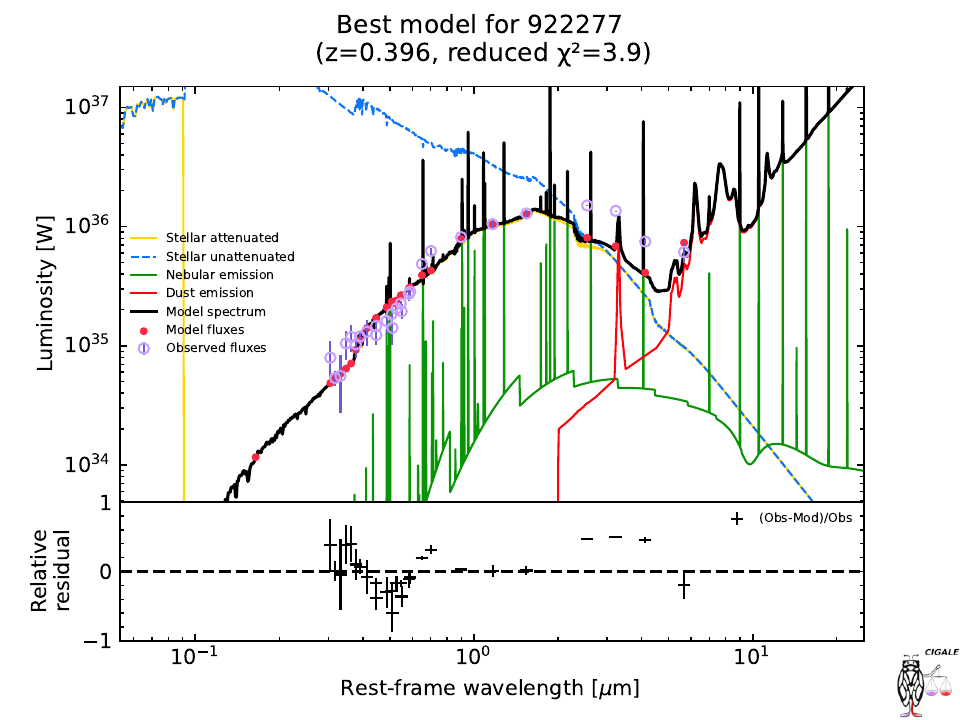}};
        \fill[fill=white, semitransparent] (0.5cm, 2cm) rectangle (-3.3cm, +2.5cm);
        \draw (1cm, 2cm) node[anchor=south east] {(a) original COSMOS galaxy};
        \draw (-0.2cm, 1.7cm) node[anchor=south east] {$M_\star=\qty{1.4e9}{M_\odot}$};
    \end{tikzpicture}
    \vspace{-0.5cm}
    \begin{tikzpicture}
        \draw (0, 0) node[inner sep=0] {\includegraphics[width=\columnwidth]{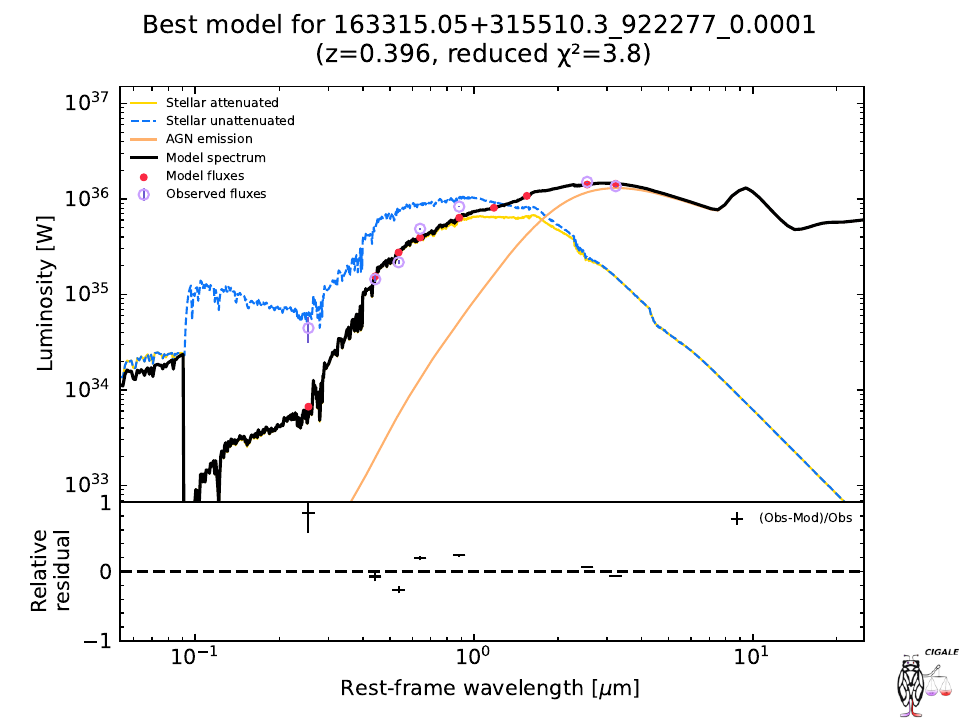}};
        \draw (3.5cm, 2cm) node[anchor=south east] {(b) C15 model fit};
        \draw (3.7cm, 0cm) node[anchor=south east] {$M_\star={8\pm3}\times\qty{e9}{M_\odot}$ };
    \end{tikzpicture}
    \begin{tikzpicture}
        \draw (0, 0) node[inner sep=0] {\includegraphics[width=\columnwidth]{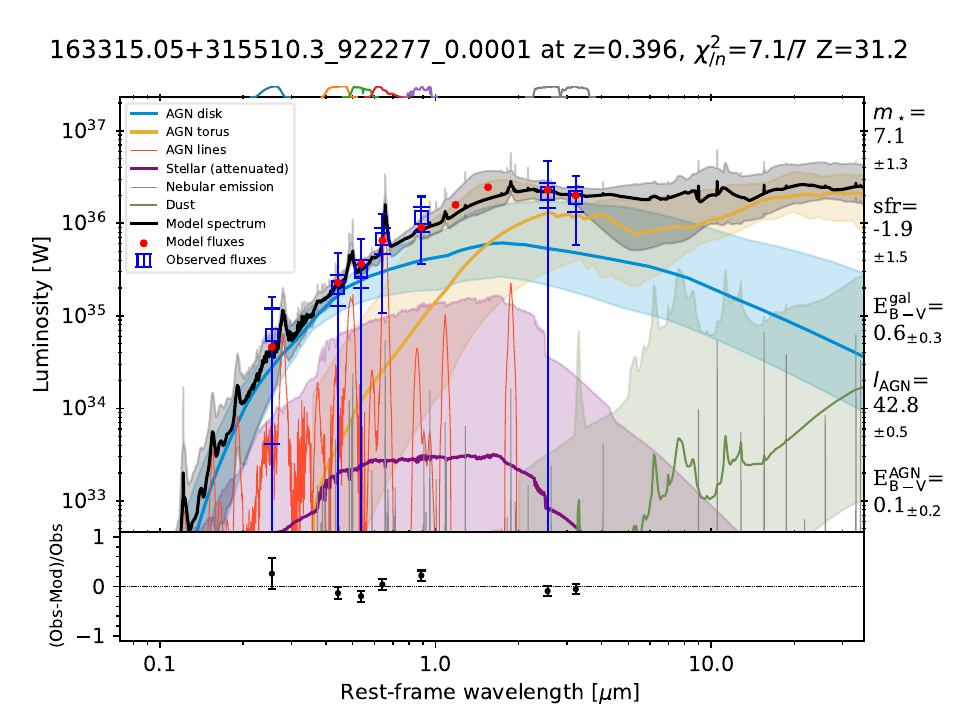}};
        \draw (3.5cm, 1.9cm) node[anchor=south east] {(c) GRAHSP fit};
    \end{tikzpicture}
    \caption{Example of an outlier at the lowest Chimera quasar weight (0.0001), which should be galaxy-dominated with very low AGN luminosity.
    The \textit{top panel} shows the original COSMOS galaxy. The observed fluxes (purple) are reproduced with a high reduced $\chi^2$ (see title) by the galaxy model (black curve), with substantial residuals in the near-infrared and optical wavelengths.
    The \textit{middle panel} presents the constructed Chimera fluxes (purple), fitted with the C15 model's galaxy (yellow curve) and AGN (orange curve) mixture model, using CIGALE. The AGN component models the data in the infrared.
    The stellar mass has become eight times larger than in the top panel.
    The \textit{bottom panel} presents the analysis of the same Chimera data with GRAHSP. Here the fit is dominated by a low-luminosity AGN, and the stellar mass is poorly constrained. The x-axis range is wider in this panel.
    }
    \label{fig:chimera:lowLoutliers}
\end{figure}

Before applying GRAHSP to the Chimera benchmark and evaluating the results, we perform a sanity check on the created low-luminosity AGN objects. At luminosities of $L(\mathrm{AGN})<{10}^{42}\,\mathrm{erg/s}$ (i.e., typically corresponding to quasar flux weights of 0.1 per cent), the Chimera objects have essentially the same fluxes as those from the original COSMOS-pure-galaxy sample, but with two differences. Firstly, the number of filters is much more limited compared to the COSMOS filter set used to establish the fiducial truth stellar mass. Secondly, the filter curves are taken from SDSS, which do not exactly correspond to the COSMOS filters where the galaxy fluxes are from (Subaru and CFHT). To establish that this is not a concern, and that the Chimeras have usable fluxes, \cref{fig:chimera:lowLresults} compares the stellar mass computed by GRAHSP on these Chimeras to the original stellar mass from \citet{Laigle2016}.
The stellar masses from the two analyses are consistent, with a mean of 0.0 dex and a scatter of 0.2 dex. Similar results are found for the comparison of CIGALE with the D14, C15 and Y19 models. There are a few (0.2\%) outliers, however. 

One outlier is shown in \cref{fig:chimera:lowLoutliers}. The top panel shows the analysis of the original COSMOS data with CIGALE. The subsequent panel shows the Chimera object with CIGALE (middle) and with GRAHSP (bottom panel). 
The COSMOS data are fitted with a strongly star-forming galaxy template, which is required because of the longest-wavelength infrared data points. Heavy attenuation then suppresses the UV (compare the blue curve to the black curve). The model is not a good fit, with a reduced $\chi^2$ of 3.9. The infrared colours are not indicative of an AGN, following \cite{Assef2013}. Instead, this indicates that the galaxy model may be incomplete. The stellar mass in units of $10^{10}\,M_\odot$, when analysing the COSMOS photometry, is 5 in the COSMOS2015 catalogue, but 0.14 in CIGALE and 0.18 in GRAHSP when run without an AGN component.
When an AGN component is used to fit the Chimera object, the infrared excess is modelled with an AGN component (\cref{fig:chimera:lowLoutliers}, CIGALE: middle panel; GRAHSP: bottom panel). This differs slightly in shape between the D14, C15 and Y19 models, but the results are similar. The middle panel of \cref{fig:chimera:lowLoutliers} demonstrates this point: galaxy light is assigned to the AGN component. The resulting stellar masses are now an order of magnitude higher, between 1 and $\qty{2.2e10}{M_\odot}$, as much less attenuation is needed.

We include such cases where even the galaxy is not perfectly modelled. These are realistic, and would be difficult to recognise if the uncertainties were larger or fewer bands were used. They are relevant as the inference is unstable when measuring a stellar mass of galaxies for comparison to AGN samples. Nevertheless, such cases are rare. An order of magnitude or more difference to the COSMOS reference masses are present in four per cent of the Chimeras in the case of CIGALE, and six per cent with GRAHSP, as shown in (\cref{fig:chimera:lowLoutliers}). Their rarity implies that the COSMOS data analysis is valid as a fiducial truth for Chimera objects. In the subsequent sections, larger AGN luminosities are also considered, and the impact on the stellar mass estimate reliability is determined.
\section{Results\label{sec:Results}}
\begin{figure*}
    \centering
    \includegraphics[width=\textwidth]{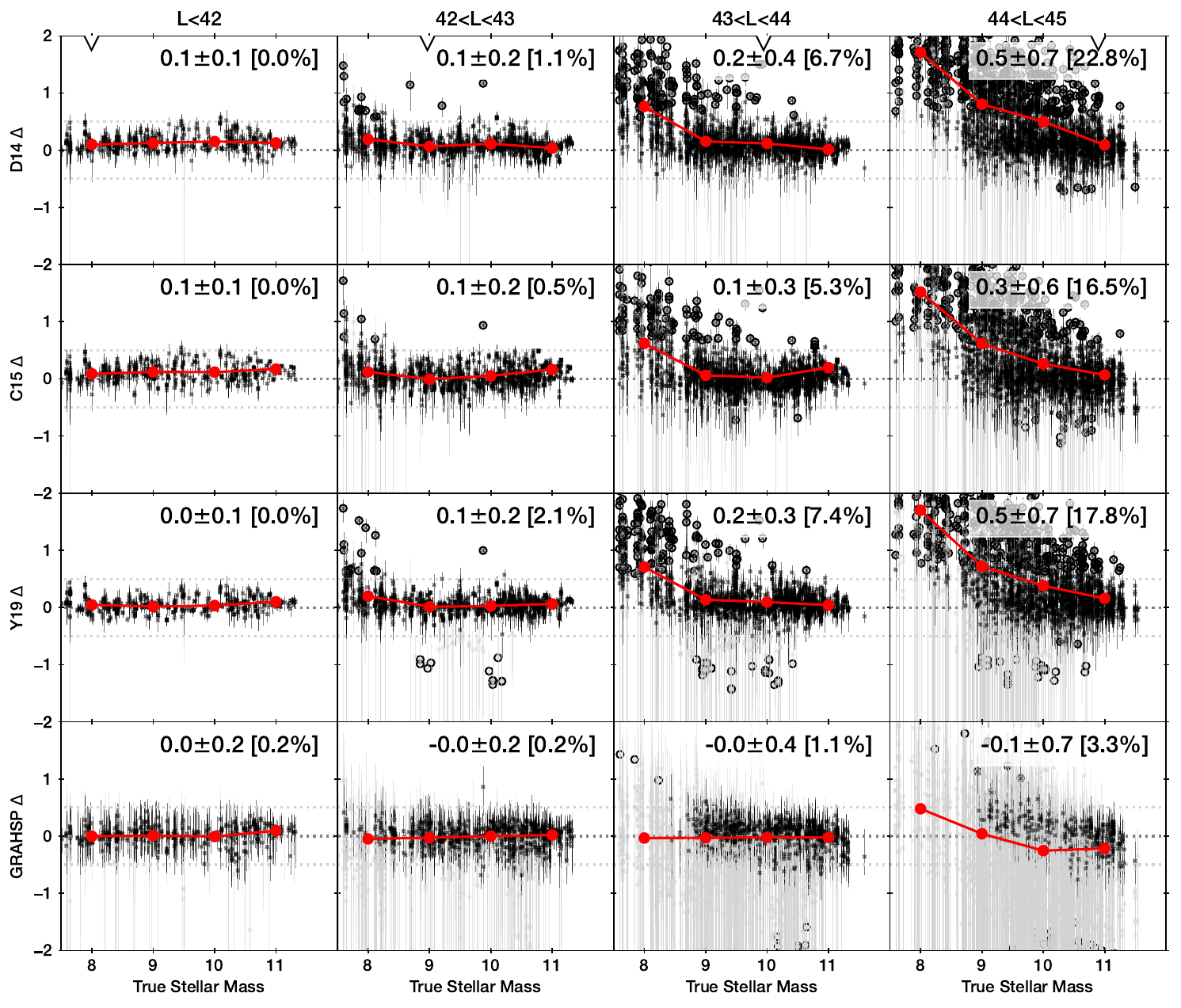}
    \caption{Retrieval of stellar masses on the Chimera benchmark data set. The y-axis always shows the ratio of the estimated stellar mass to the true stellar mass (x-axis), in log units of $M_\odot$.  The first three rows are AGN models in CIGALE, the last row is GRAHSP. 
    The sample is split into four bins (panel columns) of true AGN bolometric luminosity (in erg/s, log).
    Results with large error bars ($>1$ dex) are in grey.
    If the measured stellar mass deviates from the expectation (dotted grey horizontal lines) by more than 0.5 dex, it is considered an outlier and indicated with a circle.
    In each bin, the mean and standard deviation are quoted, together with the fraction of outliers in brackets.
    CIGALE models show a 0.2-0.5 dex bias at L$>$43. For GRAHSP, the bias is below 0.1 dex. The GRAHSP fraction of outliers is also lower.
    The downward-pointing triangles at the very top of the figure indicate where the AGN and galaxy luminosities are approximately equal ($\lambda=1$ in \cref{eq:lambda}). The CIGALE models start to deviate to the left of these markers, corresponding to $\lambda>1$.
    }
    \label{fig:chimeraresultsM}
\end{figure*}

\begin{figure*}
    \centering
    \begin{tikzpicture}
        \draw (0, 0) node[inner sep=0] {\includegraphics[width=\columnwidth,clip,trim={0 0.2cm 1.5cm 1.4cm}]{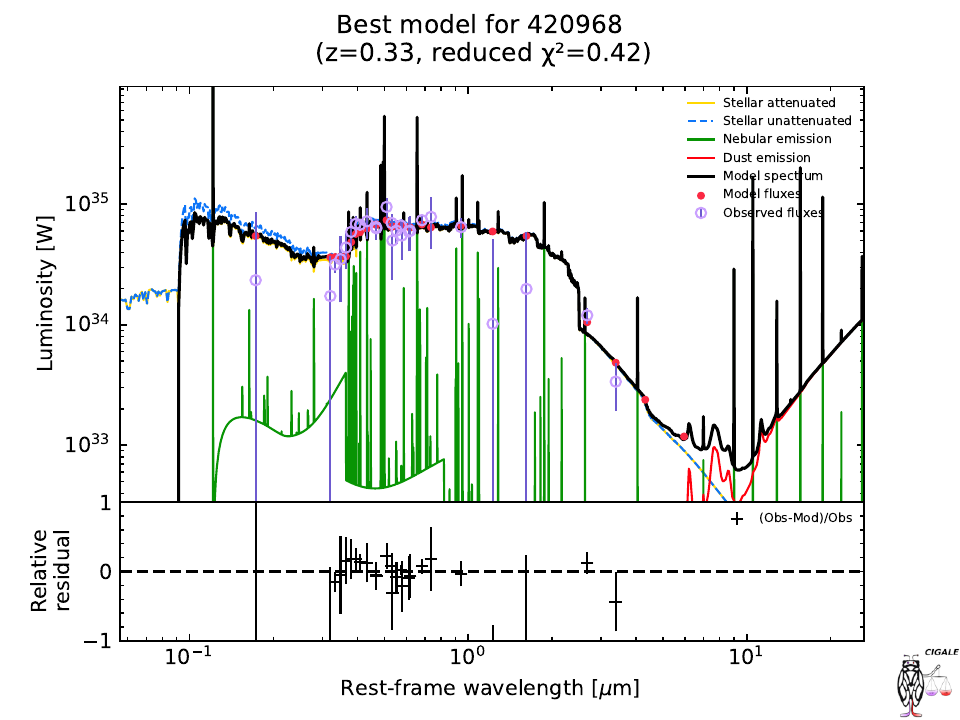}};
        \fill[fill=white, semitransparent] (1.2cm, 2.5cm) rectangle (-3.2cm, +3.0cm);
        \draw (1.2cm, 2.5cm) node[anchor=south east] {(a) original COSMOS galaxy};
        \draw (-1cm, 2.1cm) node[anchor=south east] {$M_\star=10^{8}\,M_\odot$};
    \end{tikzpicture}
    \begin{tikzpicture}
        \draw (0, 0) node[inner sep=0] {\includegraphics[width=\columnwidth,clip,trim={0 0.2cm 1.5cm 1.4cm}]{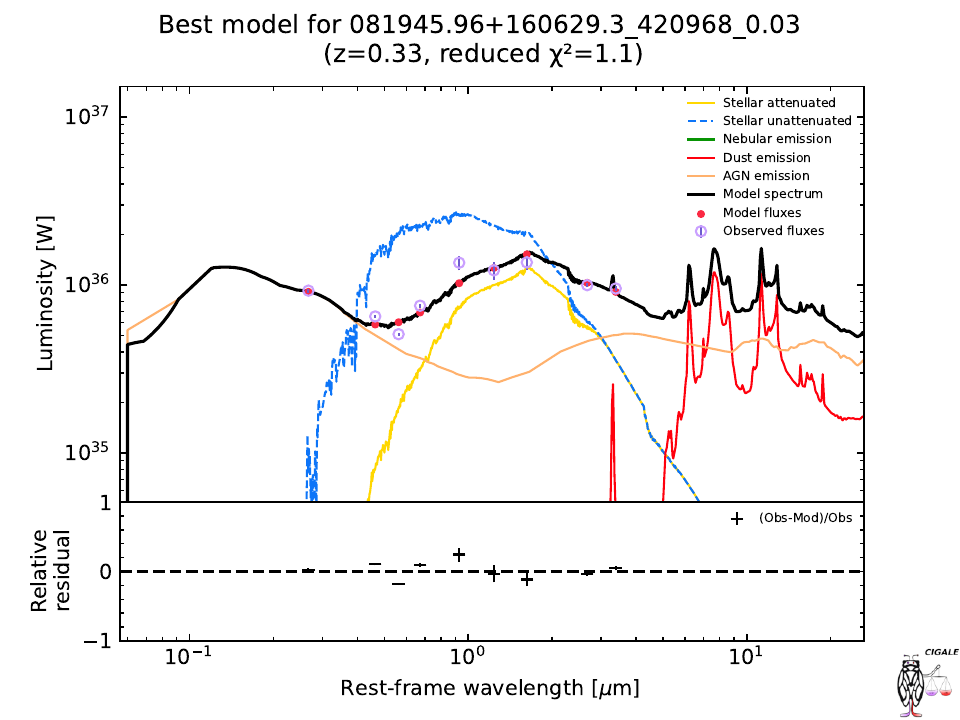}};
        \draw (1.2cm, 2.5cm) node[anchor=south east] {(b) D14 model fit, $\chi^2_\mathrm{red}=1.11$};
    \end{tikzpicture}
    \begin{tikzpicture}
        \draw (0, 0) node[inner sep=0] {\includegraphics[width=\columnwidth,clip,trim={0 0.2cm 1.5cm 1.4cm}]{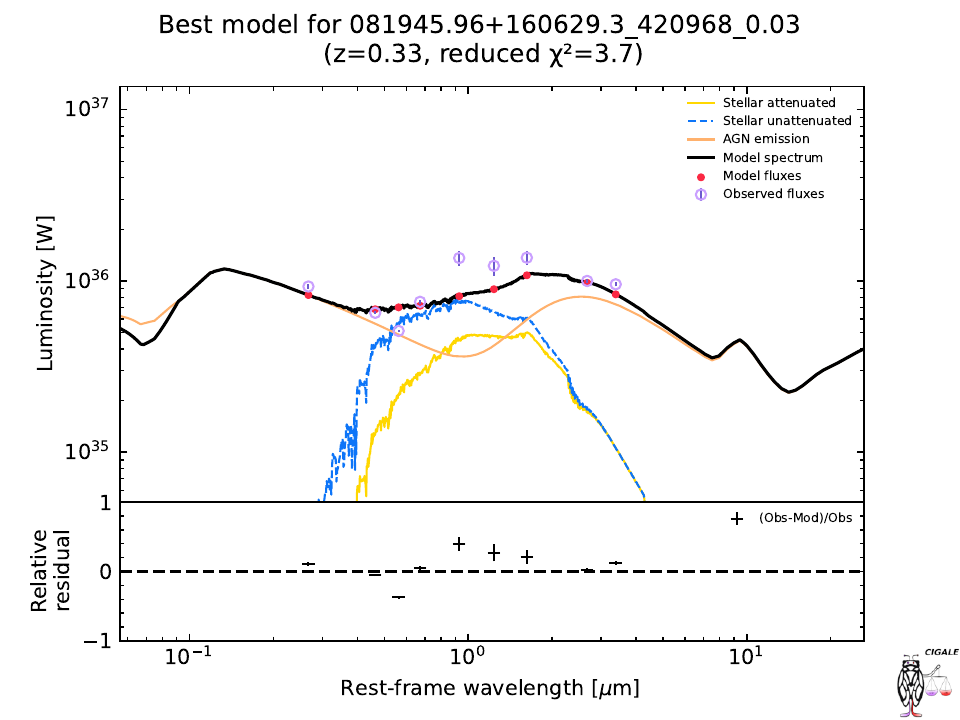}};
        \draw (1.2cm, 2.5cm) node[anchor=south east] {(c) C15 model fit, $\chi^2_\mathrm{red}=1.68$};
    \end{tikzpicture}
    \begin{tikzpicture}
        \draw (0, 0) node[inner sep=0] {\includegraphics[width=\columnwidth,clip,trim={0 0.2cm 1.5cm 1.4cm}]{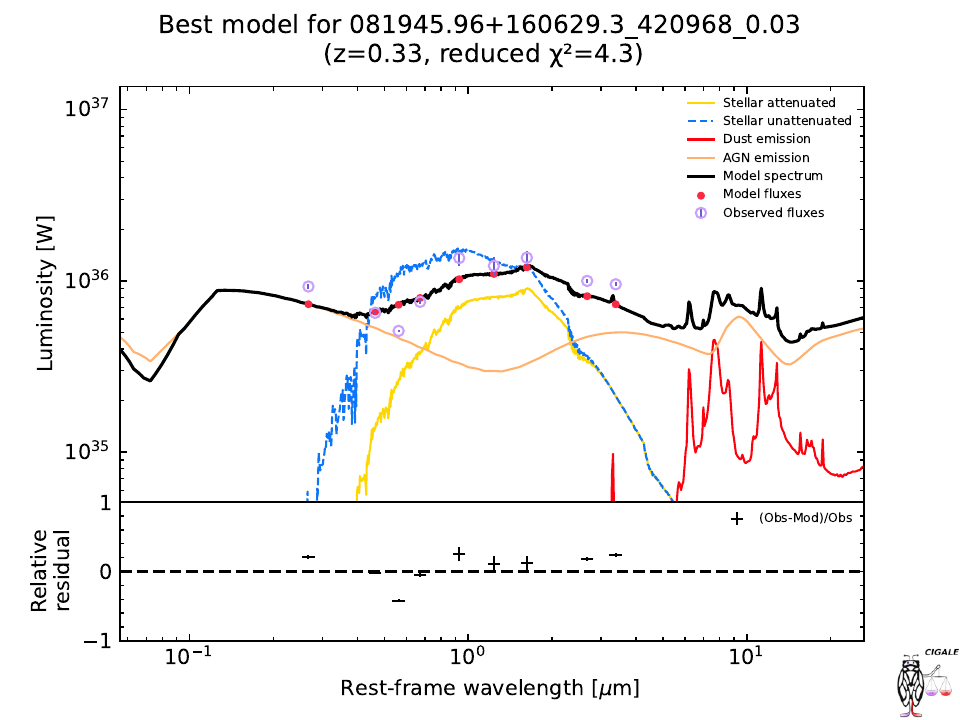}};
        \draw (2.2cm, 2.5cm) node[anchor=south east] {(d) Y19 model Cigale fit, $\chi^2_\mathrm{red}=4.26$};
    \end{tikzpicture}
    \begin{tikzpicture}
        \draw (0, 0) node[inner sep=0] {\includegraphics[width=1.4\columnwidth,clip,trim={0 0.2cm 0cm 0.2cm}]{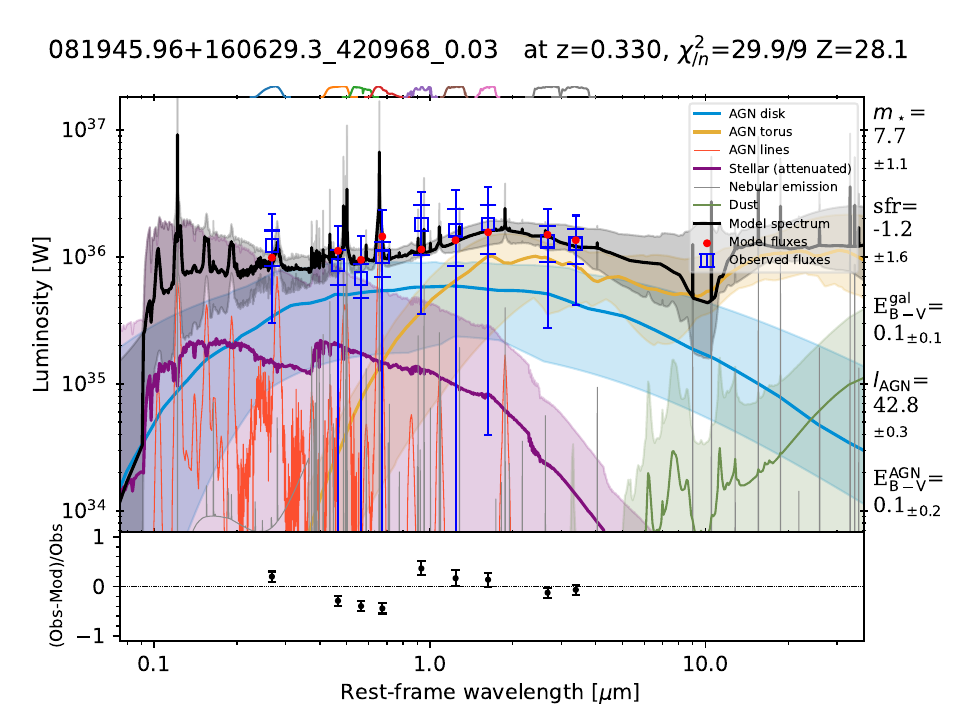}};
        \draw (-2.2cm, 2.8cm) node[anchor=south east] {(e) GRAHSP fit};
    \end{tikzpicture}
    \caption{Example of outlier in CIGALE-based analyses from the $10^{43-44}\,\mathrm{erg/s}$ AGN luminosity bin. 
    \textit{Panel (a)} shows the original COSMOS data (CosmosId: 420968 at z=0.33), where a $M_\star=10^{8}\,M_\odot$ galaxy-only model (black) reproduces the observed fluxes (purple) with small residuals ($\chi^2_\mathrm{red}=0.4$). Panels (b), (c) and (d) present CIGALE fit to the Chimera fluxes (purple) with D14, C15 and Y19 settings.
     In all three models, the galaxy template is incorrectly placed at ${10}^{36}\,\mathrm{W}$ instead of $<{10}^{35}\,\mathrm{W}$ as in panel (a). In the C15 and Y19 models, the best-fit model is dominated by the AGN component (orange). The \textit{bottom panel (e)} shows the fit by GRAHSP for the same object. The galaxy component has very large uncertainties (purple shade at the bottom), but includes the true stellar contribution (compare the purple curve here to the black curve in the top left panel). The error budget has slightly increased during the analysis from the blue error bars with wider caps to those with thin caps. The power-law AGN model (blue) dominates the fit at all wavelengths.
    }
    \label{fig:chimera:midLoutliers}
\end{figure*}

The Chimera benchmark described in \cref{sec:data:Chimera} provides galaxy properties true values that can be used for testing the capabilities of SED fitting methods. In this section we test GRAHSP. We also compare it to CIGALE with the three model setups (D14, C15, Y19; described in \cref{sec:comparison-literature}).

\subsection{Stellar mass retrieval}

First, we investigate the retrieval of stellar masses. We split the Chimera sample into bins of true AGN bolometric luminosity. In \cref{fig:chimeraresultsM}, we compare the true values against results from runs of GRAHSP and three CIGALE setups. At low luminosities (first column), all estimates have relatively small uncertainties, and centre around the true value (near zero deviation $\Delta=\log M_\mathrm{estimated}/M_\mathrm{true}$). The content of the bottom left panel was already presented in \cref{fig:chimera:lowLresults} as a correlation. In the following, the mean of the deviations is referred to as the bias, and the standard deviation as the scatter of the estimator. We weight the mean and standard deviation by the error bar size. Finally, we mark outliers with circles. These are defined as cases where the error bar excludes the 1 dex wide band centred around the true value, illustrated with grey dotted horizontal lines. \Cref{sec:chimera:lowLvalidation} already discussed one example outlier in the lowest luminosity bin from the D14, C15 and Y19 models.

The bias, scatter and outlier fraction are noted in the upper-right corner of each panel.
At higher luminosities (last column of \cref{fig:chimeraresultsM}), the 
settings most commonly used in literature show a positive bias of +0.5 dex. This is most pronounced at lower stellar masses, where the mass is over-estimated by an order of magnitude or more. The outlier fraction is 20 per cent. In contrast, GRAHSP (bottom row), has outlier fractions below 5 per cent and near-zero bias ($\leq0.2\,\mathrm{dex}$). The scatter is comparable across all methods.

An example fit result for such an outlier is illustrated in \cref{fig:chimera:midLoutliers}. The top-left panel shows the true COSMOS galaxy, modelled with the many photometric bands in COSMOS. In the subsequent panels, the constructed Chimera object is shown, with all luminosities about one order of magnitude higher ($\qty{1e36}{\watt}$ instead of $\qty{1e35}{\watt}$), due to the AGN.
Even though the galaxy should therefore be 10x fainter than the AGN,
the Chimera object is fitted by the D14, C15 and Y19 models (see panels) with approximately equal luminosity between AGN (orange) and galaxy (yellow) contribution. The bottom panel shows the result from GRAHSP. The more flexible AGN model of GRAHSP, together with its fine parameter space sampling and consideration of systematic uncertainties, can describe the data with an AGN-dominated combination (bottom panel of \cref{fig:chimera:midLoutliers}), where the host galaxy stellar mass is consistent with that of the COSMOS fiducial truth.

We caution that such failures could not be recognised from the fit alone. Firstly, \cref{fig:chimera:midLoutliers} shows $\chi^2$ values comparable to unity for some of the models. Secondly, we implemented the cleaning criteria proposed by \cite{Mountrichas2021b}, which require the maximum-likelihood stellar mass to agree with the posterior mean stellar mass within a factor of five. This occurs in 100 per cent of the sample. Therefore, the criterion does not improve the outlier fractions, which range from 5 to 20 per cent. Thirdly, agreement between different model assumptions also does not resolve incorrect solutions, as illustrated by the example of \cref{fig:chimera:midLoutliers} and \cref{fig:chimera:lowLoutliers}, which give across three different CIGALE AGN models a consistent but incorrect high stellar mass.

\begin{figure*}
    \centering
    \includegraphics[width=\textwidth]{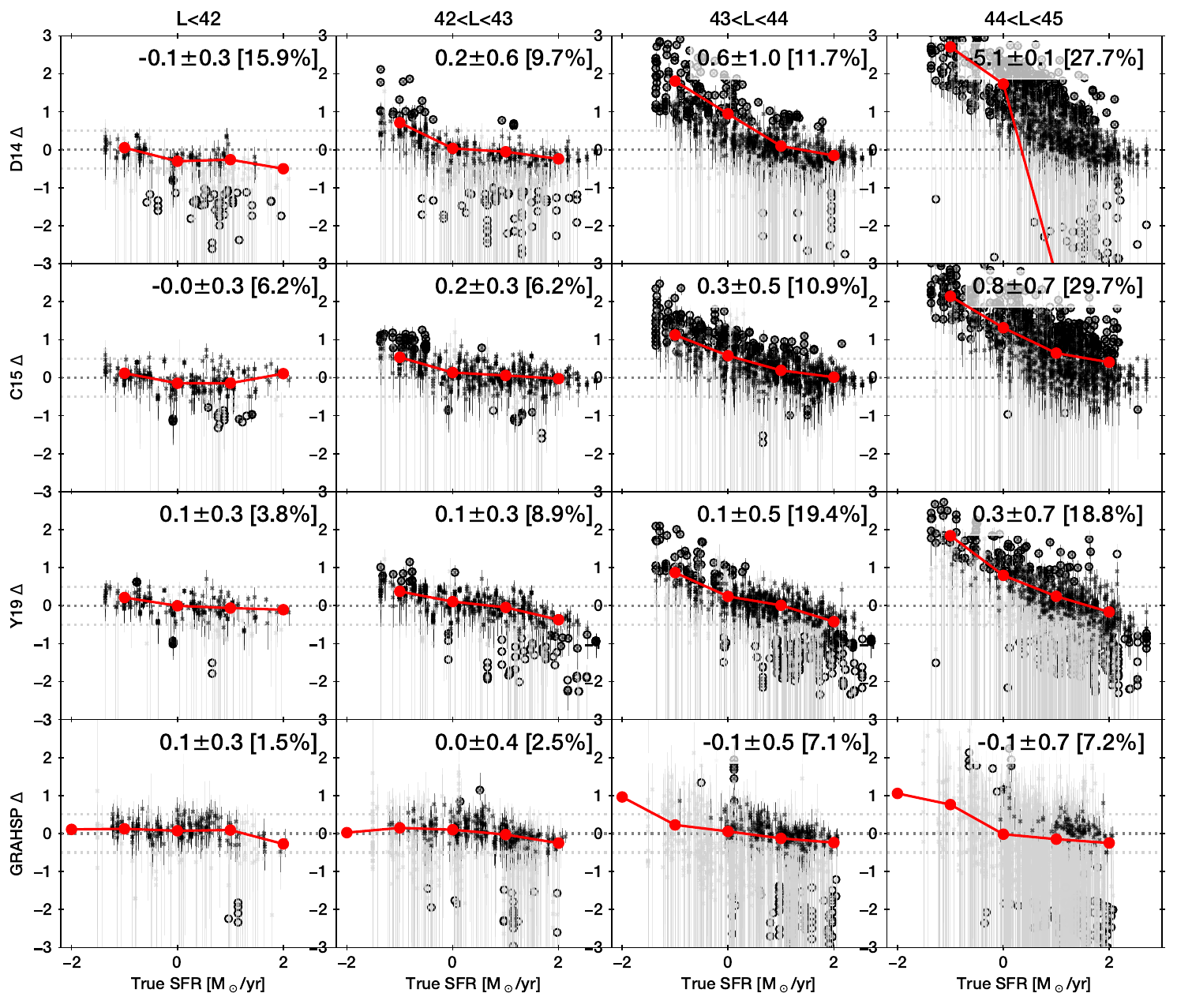}
    \caption{Star formation rate retrieval on the Chimera benchmark data set. As in \cref{fig:chimeraresultsM}.
    Each panel shows the deviation of the estimated SFR from the true SFR (in log units of $M_\odot/\mathrm{yr}$, x-axis).
    The first three rows' right panels show the performance of current models at moderate AGN luminosities. Over-estimates by 1-2 dex are common at low star formation rates, with outlier fractions between 10-30 per cent. The bias of GRAHSP (last row) is lower and the outlier fraction consistently below 10 per cent.
    }
    \label{fig:chimeraresultsS}
\end{figure*}

\subsection{Star formation rate}

The next host galaxy property considered is the star formation rate (SFR). SFR is known to be difficult to estimate in individual galaxies \citep[e.g.,][]{Ciesla2015}. Nevertheless, works such as \cite{Aird2019} rely on sample-averaged stellar masses, so quantifying the bias and scatter of estimators is important. Here, the SFR in the last Myr is expressed in $M_\odot/\mathrm{yr}$, and again compared to the true SFR from the Chimera benchmark sample in \cref{fig:chimeraresultsS}.
We only consider those objects where in the reference COSMOS galaxy, the fiducial truth SFR could be constrained to better than an order of magnitude by all models.

At the lowest true SFRs, the inferred SFR is overpredicted by one to three orders of magnitude. At the highest SFR ($>10\,M_\odot$, the retrieval is unbiased at least to the right order of magnitude.
At the highest AGN luminosities, GRAHSP shows the lowest bias.
The D14 model (top panels) shows an erratic behaviour at the highest SFR for AGN luminosities of $\log L>43$. The other two CIGALE models, have a larger positive bias than GRAHSP.
In the CIGALE models, the outlier fraction lies between 20 and 30 per cent, while the outlier fractions of GRAHSP remain below 10 per cent even at the highest luminosity bin.

\subsection{Luminosity-dependence of estimators}

\begin{figure}
    \centering
    \includegraphics[width=\columnwidth]{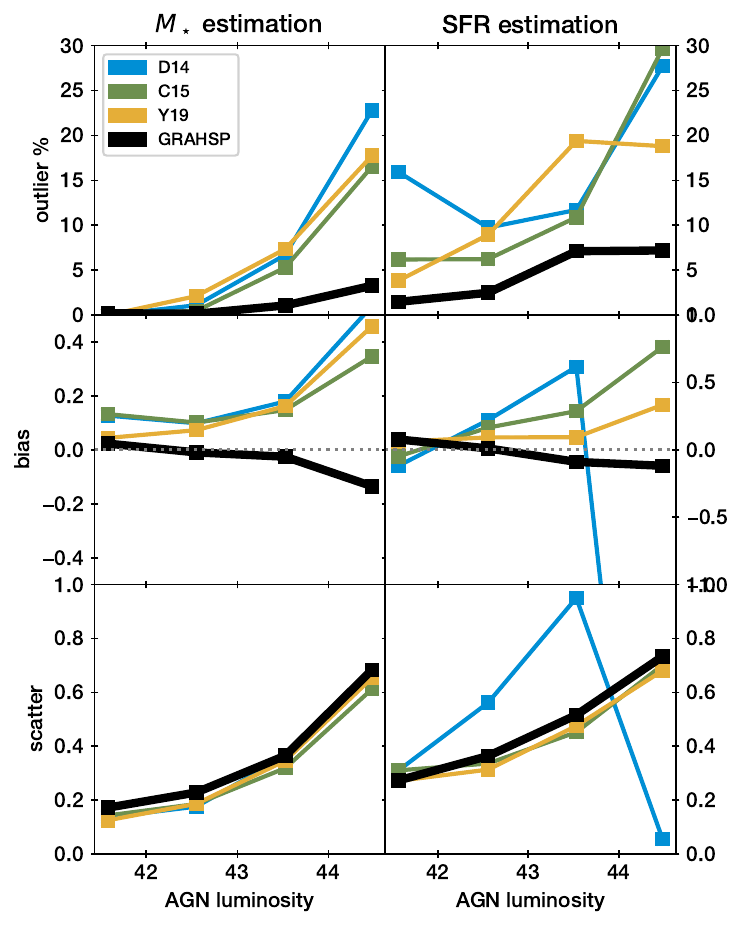}
    \caption{Comparison of codes and models on the Chimera benchmark data sets. The panel rows show the outlier fraction (see \cref{fig:chimeraresultsM}), the bias and scatter in dex.
    The first column shows stellar mass estimation (from \cref{fig:chimeraresultsM}), the second column star-formation rate estimation (from \cref{fig:chimeraresultsS}).
    Each curve as a function of AGN bolometric luminosity (log, in erg/s) represents one method.
    The outlier fraction of GRAHSP is lower than those of other methods.
    The bias of GRAHSP is near-zero. Other methods systematically overestimate the masses in luminous AGN (middle left panel), and star-formation rates (middle right panel).
    }
    \label{fig:chimeratrends}
\end{figure}

\begin{figure}
    \centering
    \includegraphics[width=0.9\columnwidth]{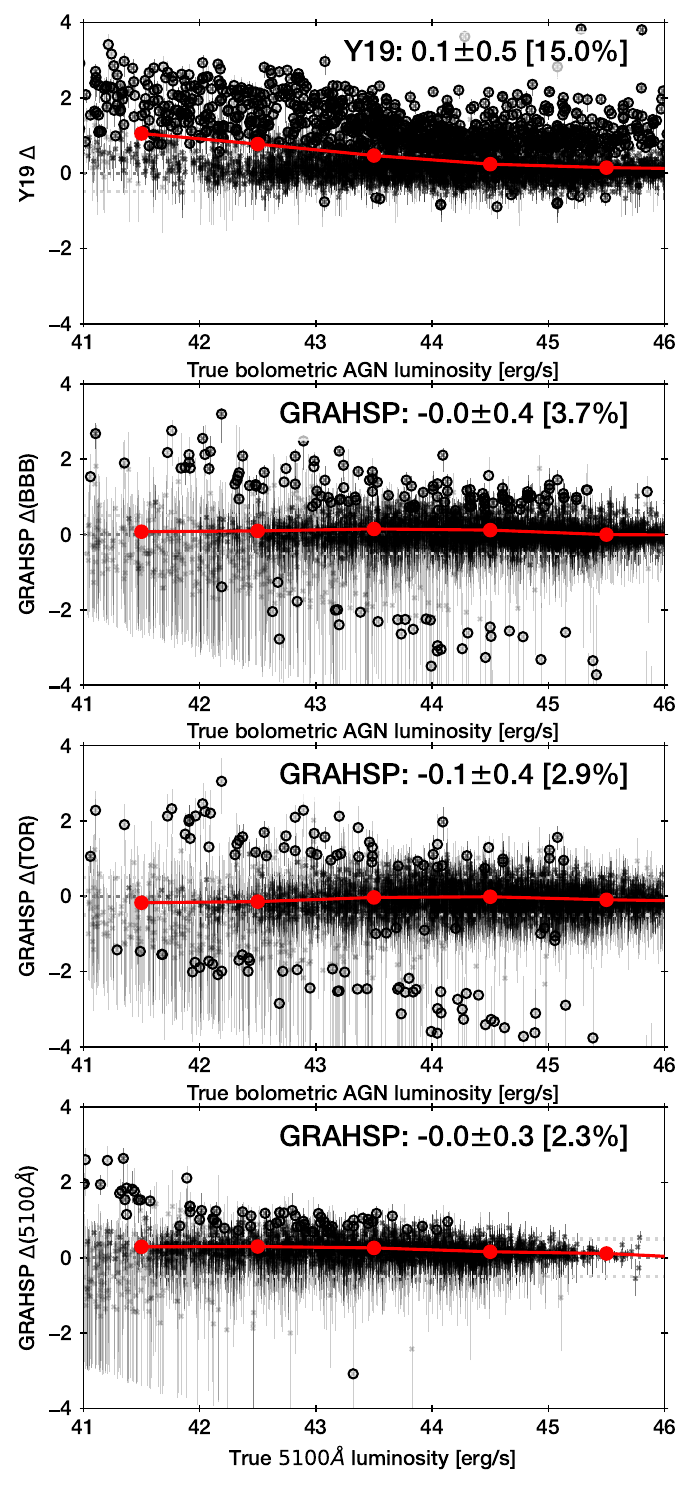}
    \caption{AGN luminosity retrieval on the Chimera benchmark data set. The top panel shows the deviation from the bolometric luminosity (in log units of erg/s).
    Current methods (top panel) show an upward bias at $L_\mathrm{bol}<\qty{e43}{\erg\per\second}$, and an outlier fraction of 15 per cent. The bolometric luminosity (see \cref{bolometric}) from the BBB and the torus are presented in the second and third panels, respectively. The bottom panel compares the retrieval of $5100\AA$ luminosity. In all cases, the average bias of GRAHSP is close to zero, with an outlier fraction lower than 5 per cent.
    }
    \label{fig:chimeraresultsL}
\end{figure}

The trends with AGN luminosity are summarised in \cref{fig:chimeratrends}. The left panels show the outlier fraction, bias and scatter of the stellar mass estimator, from top to bottom panel. The outlier fraction is much lower in GRAHSP than the other methods, and does not strongly increase with luminosity. Indeed, the outlier fraction (top left panel in \cref{fig:chimeratrends}) approximately corresponds to the probability expected outside the error bars. This is not the case for the Y19, C15 and D14 CIGALE models, for which the outlier rates rise up to 20 per cent at high luminosities. While the bias (centre left panel in \cref{fig:chimeratrends}) of the D14, C15 and Y19 methods starts near zero in low-luminosity AGN, it rises steeply with luminosities, up to 0.4 dex. This means that any sample-averaged stellar masses are systematically raised by this amount.

The right column of \cref{fig:chimeratrends} presents trends of the SFR estimator. The range (right axis) of the bias panel had to be increased compared with the stellar masses (left column), indicating how much more uncertain this measurement is. While the scatter is comparable between methods, the bias and outlier fraction are lowest in GRAHSP. The latter indicates that CIGALE-based reported SFR error bars are too small, whereas GRAHSP produces realistic error bars.

\subsection{AGN luminosity}

The last comparison from the Chimera benchmark compares the retrieval of luminosities. The top panel of \cref{fig:chimeraresultsL} shows the retrieval of bolometric luminosity by the Y19 model across all Chimeras. We excluded cases where the fiducial truth luminosity listed in \cite{Shen2012SDSSDR7Qcat} is larger than 1 dex. The bias is small (0.1 dex) with small scatter (0.5 dex), with 19 per cent lying outside the 1 dex-wide band of the true value. Some of this scatter is likely due to differences in modelling assumptions of the \cite{Shen2012SDSSDR7Q} bolometric computation and AGN model. The second panel shows the results for the BBB luminosity of GRAHSP. The bias is slightly negative and luminosity-independent. The outlier fraction is only 3 per cent. The third panel shows the bolometric luminosity estimated from the torus, with comparable results.

The bottom panel of \cref{fig:chimeraresultsL} evaluates the $5100\unit{\angstrom}$ luminosity retrieval of GRAHSP. The estimator has a very low bias (-0.1 dex) with a scatter of 0.4 dex. A low fraction, three per cent, are outliers. 

\section{Discussion\label{sec:Discussion}}

This work makes two major contributions to the field of host galaxy studies of AGN. Firstly, the Chimeras, a data-driven benchmark dataset with fiducial truth of the AGN and galaxy decomposition, is made publicly available to the community. Secondly, GRAHSP, an SED analysis method that is unbiased in the estimation of stellar mass and AGN luminosity, is presented. We now discuss these contributions in turn.

\subsection{The Chimera benchmark data set}

The Chimera benchmark data set covers an enormous range of galaxies that could hypothetically exist in the Universe. The AGN luminosities range from ${10}^{41-48}\,\mathrm{erg/s}$, the redshifts from 0.08 to 3.5, the stellar masses from $10^7$ to $10^{11.2}\,M_\odot$, the star formation rates from 0.01 to 100 $M_\odot/\mathrm{yr}$, and the black hole masses from $10^7$ to $10^{10}\,M_\odot$. The galaxies and AGN are randomly mixed, which likely results in situations that do not, or only extremely rarely, occur in nature. However, only if our analysis methods provide reasonable inferences in these situations can we be certain that these situations do not occur.

The Chimera benchmark is a novel data-driven, model-agnostic approach. Generating mock observations from models \citep[e.g.,][]{Ciesla2015} tests the internal validity of these models, and the constraining power of data under the assumption that the model is correct.
However, in reality, the diversity of AGN and galaxy phenomenology is not covered by any current model (see e.g., \cref{fig:chimera:lowLoutliers}), since it requires understanding stellar population evolutionary models and re-emission by the inter-stellar medium, and even less certain for AGN, the emission of the accretion flow \citep[see e.g.][]{Antonucci2018} and the infrared re-processing by a multi-scale, multi-temperature nuclear obscurer \citep[see e.g.][]{Netzer2015,Hoenig2016}.

Alternative approaches to vet the inferred stellar mass and SFR have been proposed. These include considering the fit quality (reduced $\chi^2$) and the asymmetry of the posterior, and consistency between several codes with different assumptions. These are incomplete indicators, as demonstrated in \cref{fig:chimeraresultsM}, as multiple codes and models may give comparable, but incorrect decomposition, with acceptable fit quality.

Generating a benchmark data set of AGN-galaxy hybrids from observational data alone sidesteps these issues. Fits with simplified, computationally feasible models can then be evaluated without assuming a generating model.

Currently, the Chimera benchmark focuses on the mixture of type 1 quasars with typical galaxies. This is the most challenging situation, and thus the centre of this work. A future modification could be to introduce a variety of empirical attenuation laws onto the type 1 quasar SED to emulate type 2 AGN.

At low AGN contamination the galaxy stellar mass is retrieved correctly for the vast majority of Chimera objects by all methods. This validates two assumptions of the Chimera benchmark. Firstly, it demonstrates that the stellar mass inferred with fewer bands (the Chimera bands) agrees with the assumed fiducial truth measured from many bands (COSMOS, including intermediate bands). Secondly, that it is acceptable to sum fluxes from similar bands (ugrz filters of SDSS vs CFHT/Subaru) without correcting for the slightly different filter curves.

The Chimera benchmark could also be used to train a machine-learning model to infer host and AGN properties. However, two serious concerns need to be considered: Firstly, the method would be limited to the bands of Chimera, and thus would not be able to handle slightly different filter bands. Secondly, to reliably characterise the performance of such a method would require an independent test data set with known fiducial truth. A random hold-out subset of Chimera would not reflect the capability to unseen samples. An example independent test data set with fiducial truth is shown below in \cref{sec:comparison-literature}.

\subsection{Unbiased estimates of galaxy stellar mass and SFR with GRAHSP}

In this work we consider an estimator \textit{unbiased} despite contamination of AGN light, if the bias is negligible compared to other sources of uncertainty. For example, \cite{Mobasher2015} demonstrated that the stellar mass estimation of non-AGN galaxies is uncertain to 0.14~dex due to variations in assumptions by different methods, but increases to 0.3 if the nebular emission or population model is varied. The scatter needs to be considered when analysing individual sources, and may make the bias negligible. In sample studies however, especially those comparing AGN to non-AGN samples, the bias is crucial.

We demonstrate that GRAHSP surpasses the capabilities of existing SED analysis methods. GRAHSP is unbiased in estimating stellar masses up to high AGN luminosities. The scatter and catastrophic outlier rates are drastically smaller at all AGN luminosities compared to commonly used CIGALE models. We speculate that other SED fitting packages not tested here suffer from the same biases if they employ limited (single-parameter) BBB and torus models, or the Fritz and Dale AGN templates, as compared with ours. With the release of the Chimera benchmark, any code developer or user can assess the validity of their code for host galaxy characterization.

We confirm previous results indicating that estimating star formation rates is challenging in active galaxies. The large scatter present in all methods indicates that for individual galaxies the SFR can be overestimated by 1-2 orders of magnitude. GRAHSP provides unbiased estimates on average. In individual bins of SFR and AGN luminosity, GRAHSP is also unbiased over a wider range of SFRs and luminosities. However, in the high AGN luminosity bins at SFR below $1M_\odot/\mathrm{yr}$, all methods overestimate the SFR.

The method characterizations could also be inverted to correct methods for their biases. For this, one would need to estimate the bias and scatter in bins of inferred properties instead of the fiducial truth values, since the latter are not available when applying to real data. This, in turn, requires assuming a sample distribution.

GRAHSP opens the door to a new field of study connecting host galaxies with their black holes. For the vast majority of AGN, the properties of super-massive black holes, such as black hole mass or spin can only be measured through spectropscopy of broad lines, i.e., in type 1. Now, this information can be reliably connected to the host galaxy properties, which are demonstrated to be measured by GRAHSP in an unbiased fashion. This will enable answering the long-standing question of whether type 1 AGN live in different host galaxies than those of type 2.

\subsection{Comparison with literature results for eROSITA AGN\label{sec:comparison-literature}}

The 140 square degree eROSITA full-depth equatorial field (eFEDS) was observed by eROSITA in Dec 2019 \citep{Brunner2021}. This revealed approximately 20,000 luminous AGN \citep{Salvato2022}. We computed the stellar masses with GRAHSP for AGN with known spectroscopic redshifts, based on the catalogues of \cite{Salvato2022}, combined with multiwavelength photometry collated in the same way as described in \cref{sec:data:DR16QWX}. For consistency, we ran GRAHSP with \cite{Bruzual2003} stellar population models assuming a Chabrier initial mass function.

\begin{figure}
    \centering
    \includegraphics[width=\columnwidth]{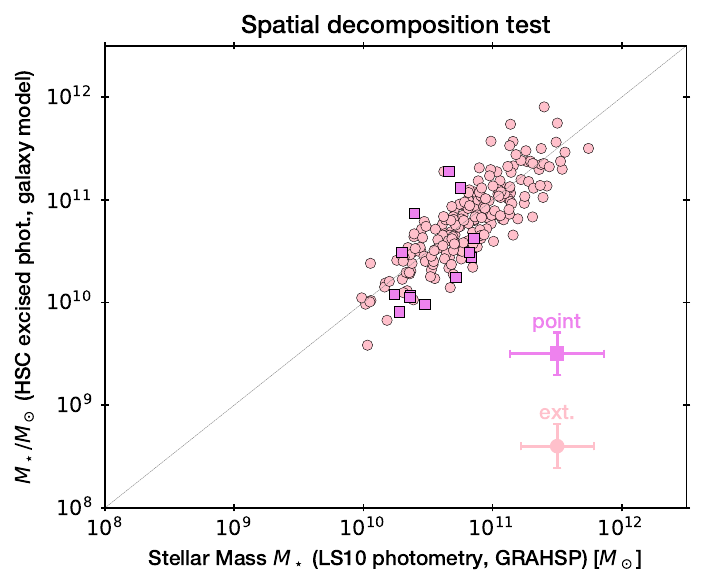}
    \caption{Comparison of GRAHSP stellar mass estimates (x-axis) to those based on deeper and spatially decomposed HSC photometry modelled with a galaxy (y-axis). The stellar masses are consistent. Squares and circles indicate point-like and extended sources, respectively, in LegacySurvey.}
    \label{fig:hsccomparison}
\end{figure}

A subset of eFEDS was observed by the Hyper Suprime-Cam (HSC) Subaru Strategic Program \citep{Aihara2019}. \cite{LiJunyao2021HSC} developed a spatial decomposition technique that separates nuclear, PSF-like light (including the AGN) from extended emission (the majority of the host galaxy). The isolated host galaxy photometric fluxes were then assembled into an SED and fitted with a galaxy model by \cite{LiJunyao2024eFEDS}. We compare their stellar masses, for overlapping sources with the same spectroscopic redshifts, 
in \cref{fig:hsccomparison}. These show consistent stellar masses.
Only cases where the GRAHSP uncertainties are smaller than 1 dex are shown. The remainder is also consistent with the 1:1 line within the uncertainties.

\begin{figure}
    \centering
    \includegraphics[width=\columnwidth]{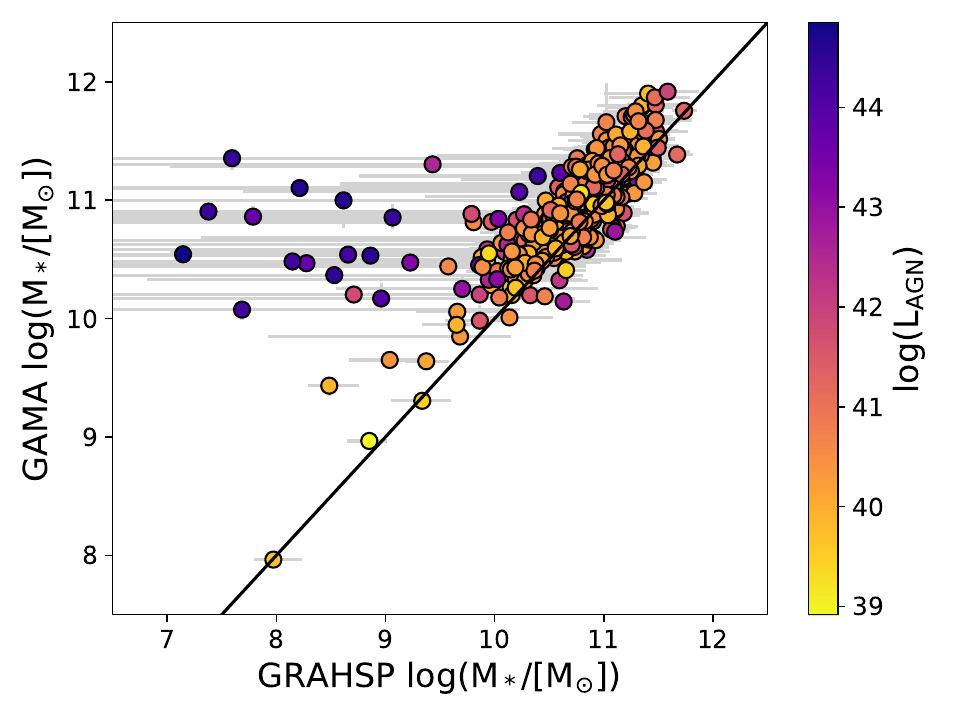}
    \caption{Comparison of GRAHSP stellar mass estimates (x-axis) to those based on a galaxy-only model with deeper GAMA photometry (y-axis). The color bar indicates the $5100\unit{\angstrom}$ AGN luminosity from GRAHSP. For host-dominated sources (yellow), the stellar mass estimates are in agreement. For more AGN-dominated sources (purple), the stellar mass of GAMA (without an AGN model) lies significantly above that inferred by GRAHSP.} 
    \label{fig:gamacomparison}
\end{figure}

Secondly, a subset of eFEDS was observed by the Galaxy And Mass Assembly (GAMA) survey field G09 \citep{Driver2022}. The optical (ugriz) photometry data reach 19th magnitude and are complemented by multiwavelength data \citep[GALEX UV, KiDS NIR, WISE MIR, Herschel FIR][]{Bellstedt2020, Bellstedt2021}, to compute stellar masses with ProSpect \citep{Robotham2020}. The assumed model does not include an AGN component.
We match their sources by position to the catalogue of \cite{Salvato2022}, discard sources where different redshifts were assumed, and focus again on the ones where the stellar mass could be constrained with GRAHSP.
\Cref{fig:gamacomparison} shows that the GAMA stellar masses are consistent with those inferred by GRAHSP, if the AGN is faint compared to the galaxy (yellow). In more luminous AGN (purple), the GAMA stellar masses are much higher. Similar to the example of \cref{fig:chimera:midLoutliers}, this is likely due to the GAMA model fit attributing AGN light to the stellar population.

\subsection{The importance of unbiased inference}

Understanding biases in the inference of key galaxy parameters is crucial to making progress in galaxy evolution. We briefly discuss some potential impacts of the systematic over-estimation of stellar masses and SFR in the presence of AGN. Firstly, such biases can induce a spurious correlation between AGN luminosity and stellar mass and between AGN luminosity and SFR. The former may be interpreted as the galaxy's gravitational potential being conducive to feeding the AGN, the latter as a co-feeding of stars and black holes from the same gas reservoir.
Secondly, if the stellar mass in AGN is over-estimated at high AGN luminosities, then $\mathrm{SAR}=L_\mathrm{bol}(AGN)/M_\star$ will be under-estimated. A resulting dearth of sources with high SAR may appear as a bend or break of the SAR distribution. This may be interpreted as a feedback effect preventing high accretion rates. A related effect is that dwarf galaxies hosting luminous AGN may be systematically absent from SED fitting catalogues. For the high redshift Universe, \cite{Silva2023} recently argued that JWST-discovered extremely massive galaxies may be caused by SED fitting issues related to modelling of the AGN component.
Finally, there are second-order effects.
For example, a selection that picks up a relatively narrow range of Eddington ratios, combined with a bias of increased stellar mass with AGN luminosity, can induce a spurious black hole mass - stellar mass correlation.

With these complications in mind, it is essential to tackle the problem of unbiased inference. Unbiased inference enables a clear interpretation of the complex, multi-dimensional parameter space of AGN-galaxy co-evolution, which includes $L_\mathrm{bol}(AGN)$, $M_\star$, SFR, redshift, obscuration, and other parameters. Realistic uncertainties are also essential. Because the aforementioned parameters are inferred from the same SED fit, they can have complex degeneracies. For example, the photometry fluxes set the sum of AGN luminosity and luminosity by stars (SFR or stellar mass parameter), see e.g. \cref{fig:posteriors}. To ensure the secure discovery of parameter correlations, the covariance of these parameters needs to be considered. As pointed out in \cref{sec:galmodel}, uninformative SFR inference can spuriously introduce a main sequence in AGN-hosts. Such effects can be avoided by carefully considering uncertainties and their degeneracies, which can indicate whether $\tau$, SFR, $M_\star$ could indeed be constrained.

While the identified difficulties in the inference of host galaxy parameters can lead to artificial correlations, this does not imply that the conclusions drawn in the literature have to be reversed. Rather, the absence of such biases needs to be demonstrated. Addressing inference biases, together with cautious interpretation and careful data preparation, is a continued effort to refine our understanding of AGN-galaxy co-evolution.

\section{Summary \label{sec:Conclusions}}

The long-standing problem of understanding the host galaxy's stellar population of AGN is tackled in this work. Firstly, in \cref{sec:data:Chimera} we carefully designed a benchmark of galaxy-AGN hybrid objects that allows us to investigate galaxy parameter retrieval introduced by the contamination by AGN light. We demonstrate that previous methods are systematically biased towards high stellar masses in \cref{fig:chimeraresultsM} and high star formation rates in \cref{fig:chimeraresultsS}, especially when the AGN light is on par with or dominates the stellar emission. This is important for interpreting the AGN host galaxy literature to date.

Secondly, we develop a new code, GRAHSP, which is unbiased and has lower variance in estimating basic galaxy and AGN properties. The SED-fitting combines (1) a modern galaxy stellar population and nebular emission model with self-consistent dust re-emission dependent on its attenuation, (2) a flexible, empirically validated AGN model consisting of a BBB with emission lines  and an infrared component, (3) sources of data variance including AGN variability and systematic model uncertainty, (4) a fast Bayesian inference algorithm capable of exploring the extensive parameter space degeneracies.

The AGN model of GRAHSP, \texttt{activate}, is validated on a battery of five stringent observational tests: 
1) The 0.1 to $50\unit{\micro\meter}$ spectra of a diverse set of 41 local AGN, 2) stacked average optical and infrared AGN spectra,
3) spectral polarimetry up to 1$\mu$m for the transition between BBB and torus,
4) the range of photometric colours from redshift z=0 to 6 and 5) typical emission line equivalent width ratios.
Our model passes these tests and is able to reproduce the AGN population with deviations in flux below 20 per cent. Crucial for robust host galaxy analysis, our model not only reproduces a mean SED or mean colours at each redshift, but also the full diversity observed.

A variety of tests in real-world applications demonstrates that GRAHSP genuinely retrieves the AGN host stellar population. The improved modelling will enable next-generation AGN host galaxy and co-evolution studies.

\section*{Acknowledgements}
JB thanks Sophia G. H. Waddell for improving the acronym. AG acknowledges support from the Hellenic Foundation for Research and Innovation (HFRI) project "4MOVE-U" grant agreement 2688, which is part of the programme "2nd Call for HFRI Research Projects to support Faculty Members and Researchers". BL acknowledges funding from the EU H2020-MSCA-ITN-2019 Project 860744 “BiD4BESt: Big Data applications for black hole Evolution STudies” 

\subsection*{Services}
This research uses services or data provided by the Astro Data Lab at NSF's National Optical-Infrared Astronomy Research Laboratory. NOIRLab is operated by the Association of Universities for Research in Astronomy (AURA), Inc., under a cooperative agreement with the National Science Foundation. 

This research has made use of the SIMBAD database \cite{Wenger2000} and the VizieR catalogue access tool \cite{Ochsenbein2000vizier}, operated at CDS, Strasbourg, France.

\subsection*{Software}
astropy \citep{AstropyCollaboration2013,AstropyCollaboration2018}, topcat \citep{topcat}, stilts (\url{https://www.star.bris.ac.uk}), Sciserver \citep[implemented at MPE, following]{SciserverJHU2020},  ultranest \citep{Buchner2021}, matplotlib \citep{matplotlib}, scipy \citep{scipy}.

\subsection*{Photometric survey data}

This work is based on data from eROSITA, the soft X-ray instrument aboard SRG, a joint Russian-German science mission supported by the Russian Space Agency (Roskosmos), in the interests of the Russian Academy of Sciences represented by its Space Research Institute (IKI), and the Deutsches Zentrum für Luft- und Raumfahrt (DLR). The SRG spacecraft was built by Lavochkin Association (NPOL) and its subcontractors and is operated by NPOL with support from the Max Planck Institute for Extraterrestrial Physics (MPE). The development and construction of the eROSITA X-ray instrument was led by MPE, with contributions from the Dr. Karl Remeis Observatory Bamberg \& ECAP (FAU Erlangen-Nuernberg), the University of Hamburg Observatory, the Leibniz Institute for Astrophysics Potsdam (AIP), and the Institute for Astronomy and Astrophysics of the University of Tübingen, with the support of DLR and the Max Planck Society. The Argelander Institute for Astronomy of the University of Bonn and the Ludwig Maximilians Universität Munich also participated in the science preparation for eROSITA.

The UKIDSS source tables were provided by the WIde Field Astronomy Unit (WFAU) of the University of Edinburgh and originate from UKIDSS DR11 of the WFCAM science archive \citep{Hambly2008WFCAM}.
The UKIDSS project is defined in \citep{Lawrence2007UKIDSS}. UKIDSS uses the UKIRT Wide Field Camera (WFCAM; \citep[WFCAM;][]{Casali2007UKIRT} and a photometric system described in \cite{Hewett2006UKIRT}.
The VISTA Hemisphere Survey is based on observations collected at the European Organisation for Astronomical Research in the Southern Hemisphere at the La Silla or Paranal Observatories under programme ID(s) 179.A-2010(A), 179.A-2010(B), 179.A-2010(C), 179.A-2010(D), 179.A-2010(E), 179.A-2010(F), 179.A-2010(G), 179.A-2010(H), 179.A-2010(I), 179.A-2010(J), 179.A-2010(K), 179.A-2010(L), 179.A-2010(M), 179.A-2010(N), 179.A-2010(O).
This publication makes use of data products from the Two Micron All Sky Survey, which is a joint project of the University of Massachusetts and the Infrared Processing and Analysis Center/California Institute of Technology, funded by the National Aeronautics and Space Administration and the National Science Foundation.

The Legacy Surveys consist of three individual and complementary projects: the Dark Energy Camera Legacy Survey (DECaLS; Proposal ID \#2014B-0404; PIs: David Schlegel and Arjun Dey), the Beijing-Arizona Sky Survey (BASS; NOAO Prop. ID \#2015A-0801; PIs: Zhou Xu and Xiaohui Fan), and the Mayall z-band Legacy Survey (MzLS; Prop. ID \#2016A-0453; PI: Arjun Dey). DECaLS, BASS and MzLS together include data obtained, respectively, at the Blanco telescope, Cerro Tololo Inter-American Observatory, NSF’s NOIRLab; the Bok telescope, Steward Observatory, University of Arizona; and the Mayall telescope, Kitt Peak National Observatory, NOIRLab. Pipeline processing and analyses of the data were supported by NOIRLab and the Lawrence Berkeley National Laboratory (LBNL). The Legacy Surveys project is honored to be permitted to conduct astronomical research on Iolkam Du’ag (Kitt Peak), a mountain with particular significance to the Tohono O’odham Nation.

NOIRLab is operated by the Association of Universities for Research in Astronomy (AURA) under a cooperative agreement with the National Science Foundation. LBNL is managed by the Regents of the University of California under contract to the U.S. Department of Energy.

This project used data obtained with the Dark Energy Camera (DECam), which was constructed by the collaboration of the Dark Energy Survey (DES). Funding for the DES Projects has been provided by the U.S. Department of Energy, the U.S. National Science Foundation, the Ministry of Science and Education of Spain, the Science and Technology Facilities Council of the United Kingdom, the Higher Education Funding Council for England, the National Center for Supercomputing Applications at the University of Illinois at Urbana-Champaign, the Kavli Institute of Cosmological Physics at the University of Chicago, Center for Cosmology and Astro-Particle Physics at the Ohio State University, the Mitchell Institute for Fundamental Physics and Astronomy at Texas A\&M University, Financiadora de Estudos e Projetos, Fundacao Carlos Chagas Filho de Amparo, Financiadora de Estudos e Projetos, Fundacao Carlos Chagas Filho de Amparo a Pesquisa do Estado do Rio de Janeiro, Conselho Nacional de Desenvolvimento Cientifico e Tecnologico and the Ministerio da Ciencia, Tecnologia e Inovacao, the Deutsche Forschungsgemeinschaft and the Collaborating Institutions in the Dark Energy Survey. The Collaborating Institutions are Argonne National Laboratory, the University of California at Santa Cruz, the University of Cambridge, Centro de Investigaciones Energeticas, Medioambientales y Tecnologicas-Madrid, the University of Chicago, University College London, the DES-Brazil Consortium, the University of Edinburgh, the Eidgenossische Technische Hochschule (ETH) Zurich, Fermi National Accelerator Laboratory, the University of Illinois at Urbana-Champaign, the Institut de Ciencies de l’Espai (IEEC/CSIC), the Institut de Fisica d’Altes Energies, Lawrence Berkeley National Laboratory, the Ludwig Maximilians Universitat Munchen and the associated Excellence Cluster Universe, the University of Michigan, NSF’s NOIRLab, the University of Nottingham, the Ohio State University, the University of Pennsylvania, the University of Portsmouth, SLAC National Accelerator Laboratory, Stanford University, the University of Sussex, and Texas A\&M University.

BASS is a key project of the Telescope Access Program (TAP), which has been funded by the National Astronomical Observatories of China, the Chinese Academy of Sciences (the Strategic Priority Research Program “The Emergence of Cosmological Structures” Grant \#XDB09000000), and the Special Fund for Astronomy from the Ministry of Finance. The BASS is also supported by the External Cooperation Program of the Chinese Academy of Sciences (Grant \#114A11KYSB20160057), and the Chinese National Natural Science Foundation (Grant \#12120101003, \#11433005).

The Legacy Survey team makes use of data products from the Near-Earth Object Wide-field Infrared Survey Explorer (NEOWISE), which is a project of the Jet Propulsion Laboratory/California Institute of Technology. NEOWISE is funded by the National Aeronautics and Space Administration.

The Legacy Surveys imaging of the DESI footprint is supported by the Director, Office of Science, Office of High Energy Physics of the U.S. Department of Energy under Contract No. DE-AC02-05CH1123, by the National Energy Research Scientific Computing Center, a DOE Office of Science User Facility under the same contract, and by the U.S. National Science Foundation, Division of Astronomical Sciences under Contract No. AST-0950945 to NOAO.

The Hyper Suprime-Cam (HSC) collaboration includes Japan, Taiwan's astronomical communities, and Princeton University. The HSC instrumentation and software were developed by the National Astronomical Observatory of Japan(NAOJ), the Kavli Institute for the Physics and Mathematics of the Universe (Kavli IPMU), the University of Tokyo, the High Energy Accelerator Research Organization (KEK), the Academia Sinica Institute for Astronomy and Astrophysics in Taiwan (ASIAA), and Princeton University. Funding was contributed by the FIRST program from the Japanese Cabinet Office, the Ministry of Education, Culture, Sports, Science and Technology (MEXT), the Japan Society for the Promotion of Science (JSPS), Japan Science and Technology Agency (JST), the Toray Science Foundation, NAOJ, Kavli IPMU, KEK,ASIAA, and Princeton University.

\subsection*{Spectroscopic survey data}

Funding for the Sloan Digital Sky Survey (SDSS) has been provided by the Alfred P. Sloan Foundation, the Participating Institutions, the National Aeronautics and Space Administration, the National Science Foundation, the US Department of Energy, the Japanese Monbukagakusho, and the Max Planck Society. The SDSS Web site is http://www.sdss.org/. 

The SDSS is managed by the Astrophysical Research Consortium (ARC) for the Participating Institutions. The Participating Institutions are The University of Chicago, Fermilab, the Institute for Advanced Study, the Japan Participation Group, The Johns Hopkins University, Los Alamos National Laboratory, the Max-Planck-Institute for Astronomy (MPIA), the Max-Planck-Institute for Astrophysics (MPA), New Mexico State University, University of Pittsburgh, Princeton University, the United States Naval Observatory, and the University of Washington.

\bibliographystyle{aa}
\bibliography{ref}

\appendix

\section{Literature CIGALE setups}
\label{sec:cigalesetups}

For the comparison with CIGALE, Table~\ref{tab:othermodelsparams} specifies the AGN model configuration of the setups considered. These are described in more detail in Section~\ref{sec:othermethods}.

\begin{table}
\centering
\caption{CIGALE model setups from the literature. The AGN model parameters and their allowed values are listed. }
\label{tab:othermodelsparams}
\begin{tabular}{l p{5cm}}
     \hline
     \hline
     \multicolumn{2}{l}{Fritz model} \\
     \texttt{r\_ratio} & 30, 100 \\
     \texttt{tau} & 0.3, 3, 6, 10 \\
     \texttt{beta} & -0.5 \\
     \texttt{gamma} & 0 \\
     \texttt{opening\_angle} & 100 \\
     \texttt{psy} & 0.001, 50.1, 89.99 \\
     \texttt{disk\_type} & 1 \\
     \texttt{delta} & -0.36 \\
     \texttt{fracAGN} & 0.0, 0.05, 0.1, 0.15, 0.2, 0.25, 0.3, 0.4, 0.5, 0.6, 0.7, 0.8 \\
     \texttt{law} & 0 \\
     \texttt{EBV} & 0.03 \\
     \texttt{temperature} & 100 \\
     \texttt{emissivity} & 1.6 \\
     \hline
     \hline
     \multicolumn{2}{l}{SKIRTOR model} \\
     \texttt{t} & 7 \\
     \texttt{pl} & 1 \\
     \texttt{q} & 1 \\
     \texttt{oa} & 40 \\
     \texttt{R} & 20 \\
     \texttt{Mcl} & 0.97 \\
     \texttt{i} & 30, 70 \\
     \texttt{disk\_type} & 1 \\
     \texttt{delta} & -0.36 \\
     \texttt{fracAGN} &  0.01, 0.1, 0.2, 0.3, 0.4, 0.5, 0.6, 0.7, 0.8, 0.9, 0.99 \\
\end{tabular}
\end{table}



\section{The effect of photometric redshifts \label{sec:photozchecks}}

In this section, we investigate the effect of photometric redshifts on stellar mass estimates. The SED fitting to characterise the host galaxy, as presented in this paper, relies on knowledge of the redshift. Dedicated methods have been developed to estimate photometric redshifts for AGN \cite[see the review of][]{Salvato2022}. The quality of photometric redshift estimation is typically quantified with a spectroscopic test sample through a catastrophic outlier fraction ($|z_\mathrm{phot}-z_\mathrm{spec}|/(1+z_\mathrm{spec})>0.15$) and a scatter around the spectroscopic redshift:
$\sigma_\mathrm{NMAD}=1.48\times\mathrm{median}(|z_\mathrm{phot}-z_\mathrm{spec}|/(1+z_\mathrm{spec})$.
Depending on the precision and number of photometric filter bands, the scatter can range from 0.01-0.02 \citep[low to mid-luminosity Chandra-COSMOS AGN][]{Salvato2011}, all the way to more than 0.1 \citep[eFEDS luminous AGN survey, outside KiDS and HSC survey areas][]{Salvato2021}.
In this section, we investigate how such a scatter in the redshift estimate impacts the stellar mass estimate.

\begin{figure}
    \centering
    \includegraphics{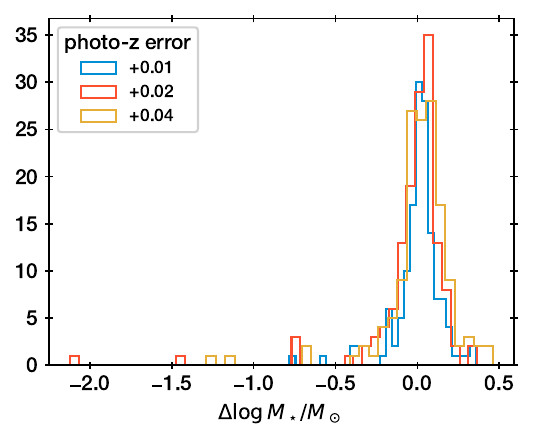}
    \caption{Change in the stellar mass estimate when the redshift is inaccurate. The x-axis shows the change in stellar mass from a reference run at the spectroscopic redshift, and another run assuming $z+\Delta\times(1+z)$ as the redshift, with $\Delta$ values of 0.01 (blue), 0.02 (red) and 0.04 (yellow). The red and yellow histograms are similar, with a tail to the left. The blue histogram is narrower and lacks the long tails present in the other two histograms.}
    \label{fig:photozimpact}
\end{figure}

For this purpose, we use the COSMOS AGN sample which features many photometric bands. We modify the assumed redshifts by $z'=z_\mathrm{spec}+0.01\times\times(1+z_\mathrm{spec})$, and repeat this exercise with shifts of 0.02 and 0.04. We re-run GRAHSP with these redshifts and compare the estimated stellar masses. We only consider the cases where the 2$\sigma$ uncertainties in the reference fit were smaller than an order of magnitude.
\Cref{fig:photozimpact} presents the impact on stellar masses. Here, the difference in posterior mean stellar mass is shown, without considering the uncertainty further. There are 5 (4) cases that have changed in stellar mass by more than a factor of 3, when the photometric redshift is changed by 0.04 (0.02) times (1+z). Surprisingly, the stellar mass predominantly becomes smaller, although the same source fluxes are explained by a source placed at larger distances. At a photometric redshift change of 0.01 times (1+z), there are only 2 such outliers. Also the distribution is narrower, with a standard deviation of 0.14 dex rather than 0.26 (0.21) dex for the 0.02 (0.04) photometry error case.
The results underline that precision photometric redshifts are important for SED fitting studies. Such precision photometric redshifts will enable the application of robust SED fitting across complete samples without requiring costly follow-up spectroscopy. Alternatively, the effect of photo-z deviations can be modeled by considering the distribution presented in \cref{fig:photozimpact}.

\section{Overview of samples}
\label{sec:sampleoverview}
\Cref{tab:sampleoverview} presents an overview of the samples fitted in this work. For each sample and filter, the table lists the number of sources with flux measurements and the median magnitudes.

\begin{table*}[]
    \centering
    \caption{For each subsample analysed in the main text, the median value in AB magnitude in every photometric band is listed. Between parentheses, the total number of sources with that particular photometric point available is also reported.}
    \label{tab:sampleoverview}
\begin{tabular}{l|ccccc}
Filter name & Chimera benchmark & COSMOS pure galaxies & COSMOS X-ray AGN & DR16QWX & eFEDS\&HSC common sample \\
\hline
\hline
\textit{(total)} & \textit{(18168)} & \textit{(3367)} & \textit{(225)} & \textit{(929)} & \textit{(554)}\\ 
\hline
NUV &  & 24.5 (2186) & 25.3 (118) & 19.8 (260) & 19.9 (37)\\
u\_sdss & 24.3 (18157) &  &  &  & \\
r\_sdss & 22.8 (18168) &  &  &  & \\
i\_sdss & 22.6 (18168) &  &  &  & \\
z\_sdss & 22.1 (18168) &  &  &  & \\
SUBARU\_V &  & 23.5 (3362) & 23.9 (225) &  & \\
SUBARU\_r &  & 23.1 (3362) & 23.7 (223) &  & \\
SUBARU\_i &  & 22.5 (3367) & 23.1 (224) &  & \\
SUBARU\_z &  & 22.2 (3367) & 22.5 (224) &  & \\
decam\_g &  &  &  & 19.8 (924) & 20.7 (554)\\
decam\_r &  &  &  & 19.6 (925) & 20.1 (554)\\
decam\_z &  &  &  & 19.3 (924) & 19.5 (554)\\
UV\_Y &  &  &  & 18.5 (7) & 19.2 (75)\\
UV\_J &  &  &  & 19.4 (288) & 19.0 (154)\\
UV\_H &  &  &  & 19.2 (274) & 18.8 (71)\\
UV\_K &  &  &  & 19.1 (278) & 18.5 (145)\\
J\_2mass & 21.5 (12633) &  &  &  & \\
H\_2mass & 21.2 (11548) &  &  &  & \\
Ks\_2mass & 20.9 (12020) &  &  &  & \\
WFCAM\_Y &  &  &  & 19.2 (95) & 19.3 (502)\\
WFCAM\_J &  & 22.0 (3359) & 21.8 (217) & 18.9 (95) & 19.1 (519)\\
WFCAM\_H &  &  &  & 18.7 (95) & 18.8 (504)\\
WFCAM\_Ks &  &  &  & 18.4 (98) & 18.5 (531)\\
CFHT\_H &  & 21.8 (3361) & 21.4 (224) &  & \\
CFHT\_K &  & 21.6 (3362) & 21.1 (224) &  & \\
HSC\_y &  & 22.1 (3367) & 22.4 (225) &  & \\
IB427\_SCam &  & 24.2 (3343) & 24.7 (223) &  & \\
IB464\_SCam &  & 23.9 (3315) & 24.4 (222) &  & \\
IB484\_SCam &  & 23.9 (3357) & 24.4 (224) &  & \\
IB505\_SCam &  & 23.7 (3357) & 24.2 (224) &  & \\
IB527\_SCam &  & 23.7 (3358) & 24.2 (223) &  & \\
IB574\_SCam &  & 23.5 (3357) & 24.0 (224) &  & \\
IB624\_SCam &  & 23.2 (3362) & 23.6 (224) &  & \\
IB679\_SCam &  & 22.7 (3364) & 23.3 (224) &  & \\
IB709\_SCam &  & 22.7 (3367) & 23.3 (224) &  & \\
IB738\_SCam &  & 22.6 (3366) & 23.2 (224) &  & \\
IB767\_SCam &  & 22.5 (3367) & 23.2 (224) &  & \\
IB827\_SCam &  & 22.4 (3366) & 22.9 (224) &  & \\
NB711\_SCam &  & 22.8 (3364) & 23.3 (224) &  & \\
NB816\_SCam &  & 22.5 (3367) & 22.9 (225) &  & \\
IRAC1 & 20.7 (18168) & 21.8 (3361) & 20.6 (225) &  & \\
IRAC2 & 20.6 (18141) & 22.0 (3360) & 20.5 (225) &  & \\
IRAC3 &  & 22.0 (2625) & 20.4 (225) &  & \\
IRAC4 &  & 21.8 (1622) & 20.6 (225) &  & \\
WISE1 &  &  &  & 18.1 (929) & 18.4 (554)\\
WISE2 &  &  &  & 17.8 (929) & 18.6 (554)\\
WISE3 &  &  &  & 16.6 (614) & 16.9 (76)\\
WISE4 &  &  &  & 15.0 (614) & 15.1 (76)\\
MIPS1 &  & 18.7 (880) & 18.3 (159) &  & \\
PACS\_green &  & 21.6 (118) & 21.7 (20) &  & \\
PACS\_red &  & 21.0 (76) & 20.8 (9) &  & \\
PSW &  & 21.6 (281) & 21.5 (54) &  & \\
PMW &  & 21.4 (75) & 21.4 (35) &  & \\
PLW &  & 20.9 (21) & 21.0 (12) &  & \\
\hline
\hline
\end{tabular}
\end{table*}

\section{Model photometry plots \label{sec:modelphotometrychecks}}

\begin{figure}
    \centering
    \includegraphics[width=\columnwidth]{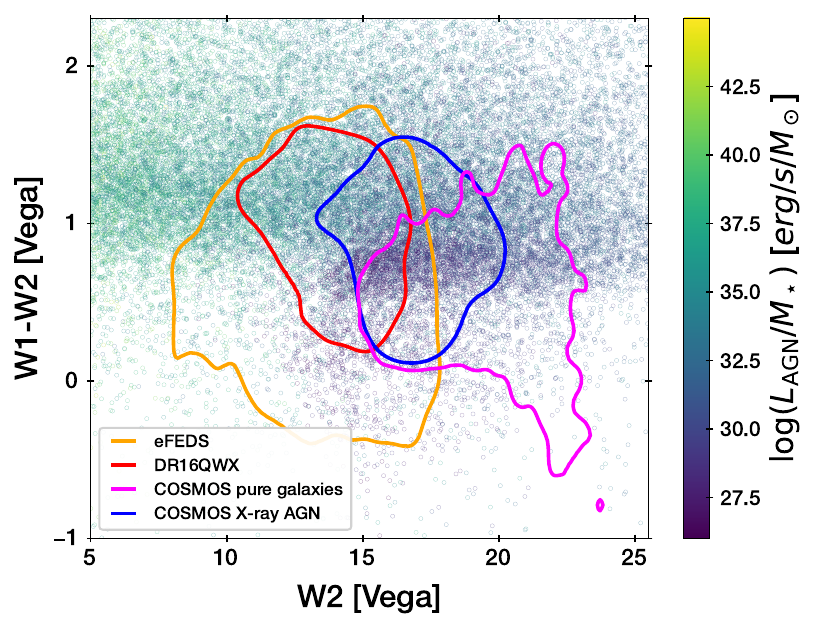}
    \caption{WISE W1-W2 colour diagnostic plot. Observed colours for different samples are shown in contours encompassing 99 per cent of the sample. The GRAHSP templates (coloured dots) cover the area of the samples.}
    \label{fig:modelWISEcolors}
\end{figure}

\begin{figure}
    \centering
    \includegraphics[width=\columnwidth]{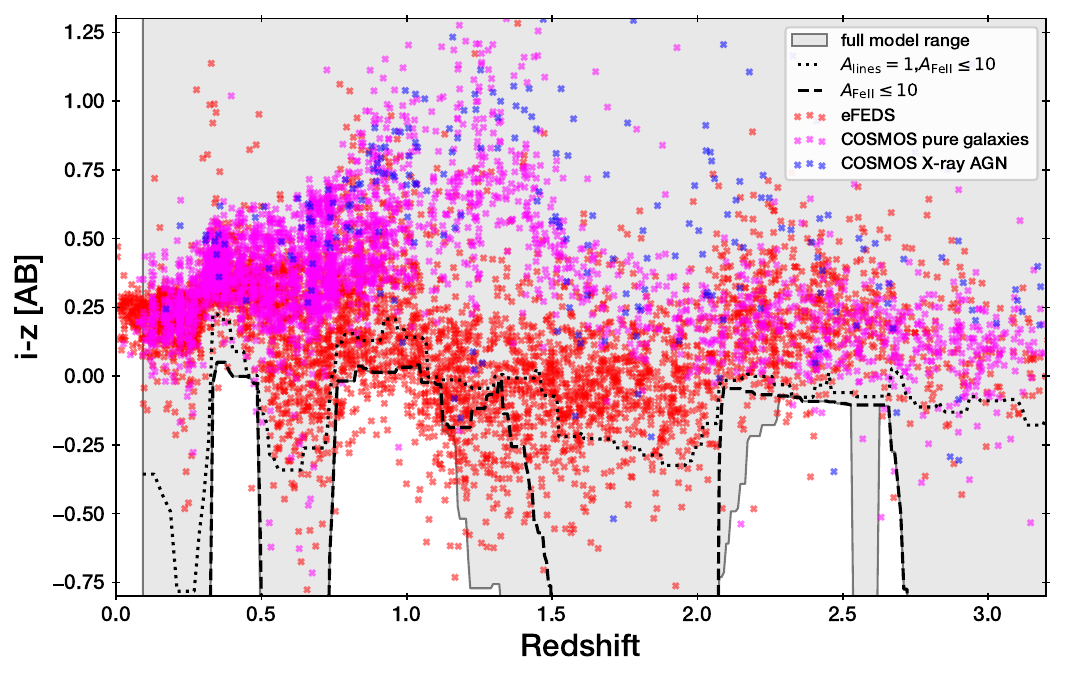}
    \caption{Testing the i-z colour range evolution of the model against observations (coloured dots). The model range is shown in grey, for three configurations: the dotted curve is a fiducial model with fixed FeII template and emission line strength. There is good overlap, except near z=0.6, z=1.2 and z=1.5 at negative i-z colours.
    Increasing the emission line strength range creates an overlap at z=0.6 and z=1.5 (dashed curve). Additionally increasing the FeII template strength range (grey curve, full model range) also produces a good overlap at z=1.2. However, at z=0.8, some observed eFEDS colours are lower than the model.
    }
    \label{fig:modelizcolorevol}
\end{figure}

\begin{figure}
    \centering
    \includegraphics[width=\columnwidth]{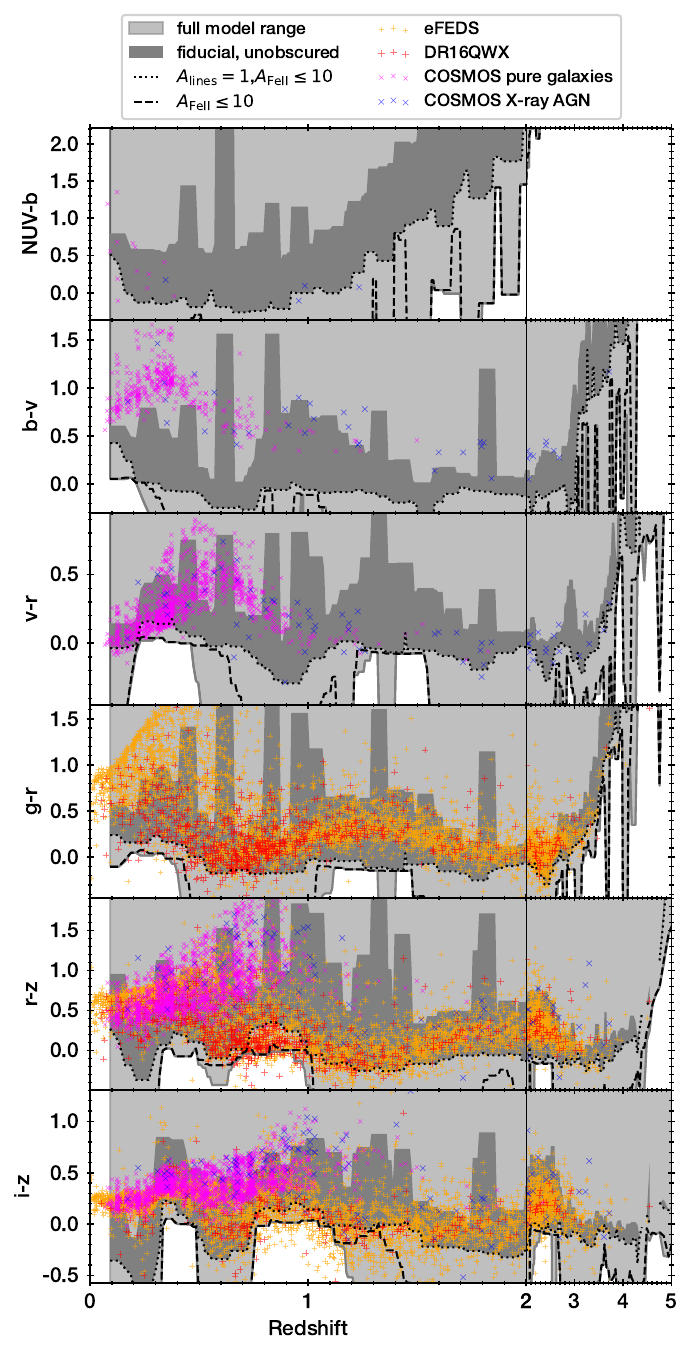}
    \caption{Colour evolution with redshift. Each row shows one colour. 
    Observations are shown in dots for different samples. 
    At each redshift, the model templates cover the observed range, illustrated by the grey area.
    }
    \label{fig:modeloptUVcolorevol}
\end{figure}
\begin{figure}
    \centering
    \includegraphics[width=\columnwidth]{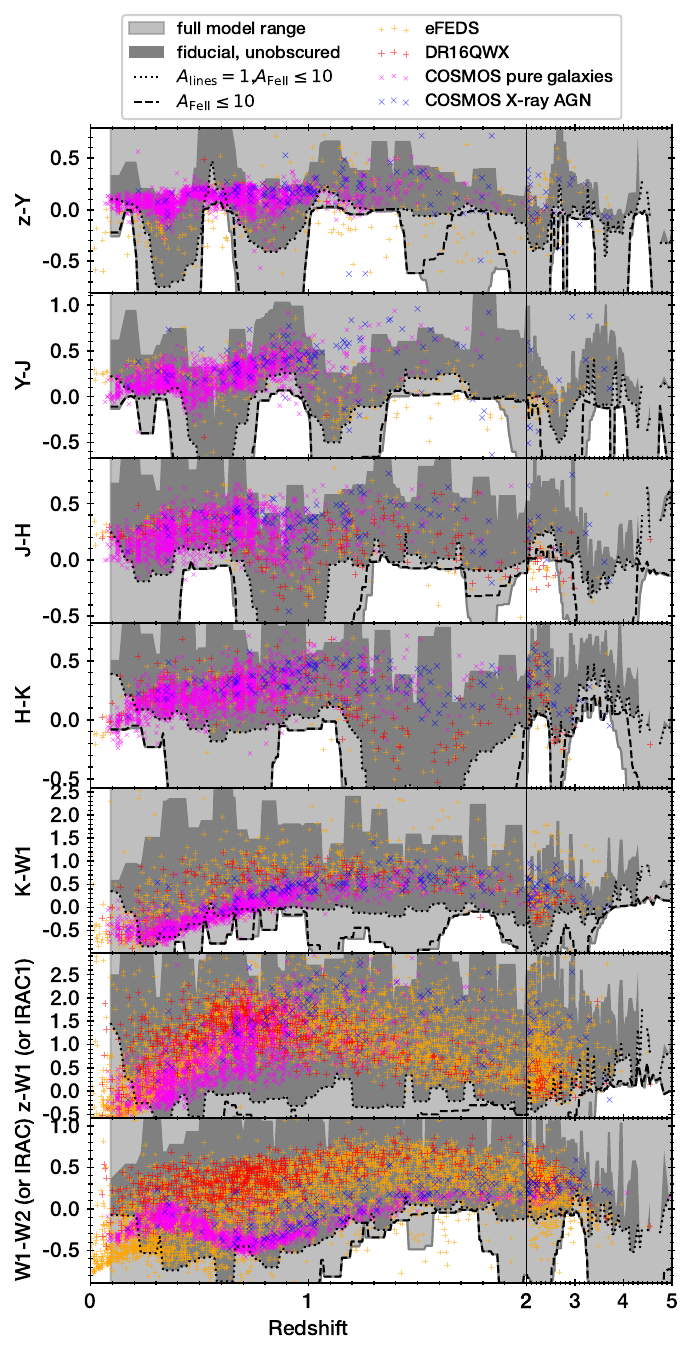}
    \caption{Colour evolution with redshift. Each row shows one colour.
    Observations are shown in dots for different samples. 
    At each redshift, the model templates cover the observed range, illustrated by the grey area.
    Filters are from DECAM/UltraVista/WISE for eFEDS, DR16QWX and the model, but from 
    Subaru/HSC/WFCAM/CFHT/IRAC for COSMOS, making the comparison slightly inconsistent.
    }
    \label{fig:modelIRcolorevol}
\end{figure}
This section compares colours derived from the GRAHSP model templates to those observed in the samples used in this work. 
\Cref{fig:modelWISEcolors} presents the WISE W1-W2 colour diagnostic plot. The entire space relevant for AGN and galaxies is covered well by the model. Most DR16QWX objects that lie above the W1-W2>0.7 infrared quasar selection \citep{Assef2013}. There is an absence of model and data points near 0 at the left side of the plot, where stars typically dominate the surveys (not shown).

Figure~\ref{fig:modelizcolorevol} shows a notable disagreement between data and model in the i-z colour with a fiducial model (grey area above the dotted curve) using the empirical emission line strengths listed in \Cref{tab:linelist} and the FeII template. At redshifts $z=0.6-1.5$, much lower i-z colours are observed, namely that the i-band is brighter than the z-band.
This can be explained by emission lines. The dashed curve shows the results of allowing the FeII template amplitude to be higher, which eases the tension at all redshifts. Some low colours remain, which can be partially improved by also allowing the line emission strength to increase. However, there is still room for improvement at z=0.8. This could be addressed by varying the adopted FeII template. The remaining disagreements are of the order of 10 per cent in magnitude and therefore also flux, a typical systematic uncertainty of models, as also shown in \Cref{subsec:calibration}.

\Cref{fig:modeloptUVcolorevol,fig:modelIRcolorevol} presents the colour evolution as a function of redshift 
for a high number of bands. Only objects with spectroscopic redshift are shown, and all magnitudes are in AB.
Filters are from DECAM/UltraVista/WISE for eFEDS, DR16QWX and the model, but from 
Subaru/HSC/WFCAM/CFHT/IRAC for COSMOS, making the comparison inconsistent because the filter curves differ slightly.
The fourth panel of \cref{fig:modeloptUVcolorevol} shows g-r colours, which reach negative values at redshift $z=0.6$.
Values below -0.2 can only be achieved when the emission line strength ($A_\mathrm{lines}$) is allowed to vary. This can be seen by comparing the dotted curve, which marks the lower end of the model range with $A_\mathrm{lines}=1$, with the dashed curve, where $A_\mathrm{lines}$ can vary.
At most redshifts, the model covers the data well. The unobscured, fiducial model range (dark grey), with both E(B-V) parameters below 0.03 and a limited line diversity ($A_\mathrm{lines}=1$, $A_\mathrm{FeII}\leq10$), follows the colour evolution of the bulk of the observations. There are some data points outside this range, but they are within the full model range (light grey).
There are some small remaining differences in the r-z colours near $z=0.8$ and in the J-H colours (third panel of \cref{fig:modelIRcolorevol}) near $z=2.5$. These correspond to differences of less than 20 per cent in flux.
In the W1-W2 panel of Figure~\ref{fig:modelIRcolorevol}, a small number of galaxies at low W1-W2 are outside the model colour range at $z=1.5-3$. These can also be seen in Figure~\ref{fig:modelWISEcolors} on the bottom right end. The model templates thin out, but appear to still cover the necessary colour-magnitude space.

With the caveat that colours can be contaminated by nearby objects, the photometry test shown here corroborates the findings of \Cref{subsec:calibration} that the model captures the diversity of the AGN-host galaxy population at most redshifts.

\section{Testing CIGALE/SKIRTOR on the Brown SED Atlas}
\label{sec:brown-skirtor-test}
In this section, we test the fidelity of CIGALE's SED model against the Brown SED Atlas.

We start with the same approach as in \cref{subsec:calibration}, but use the Y19 model setup. Unlike in \cref{sec:comparison-literature}, all SKIRTOR parameters (t, pl, q, oa, R, i, delta, fracAGN, E(B-V)-polar, temperature and emissivity) are left free to vary with all its permitted values. Further parameters include $M_\star$, $\tau$, age, $\alpha$ and E(B-V). This gives a total of 16 free parameters. To find the best-fit parameter values for each source, we minimise a $\chi^2$ statistic as in \cref{subsec:calibration} with \texttt{scipy}'s differential evolution algorithm with default settings.

\begin{figure*}
    \centering
    \includegraphics[width=0.9\textwidth]{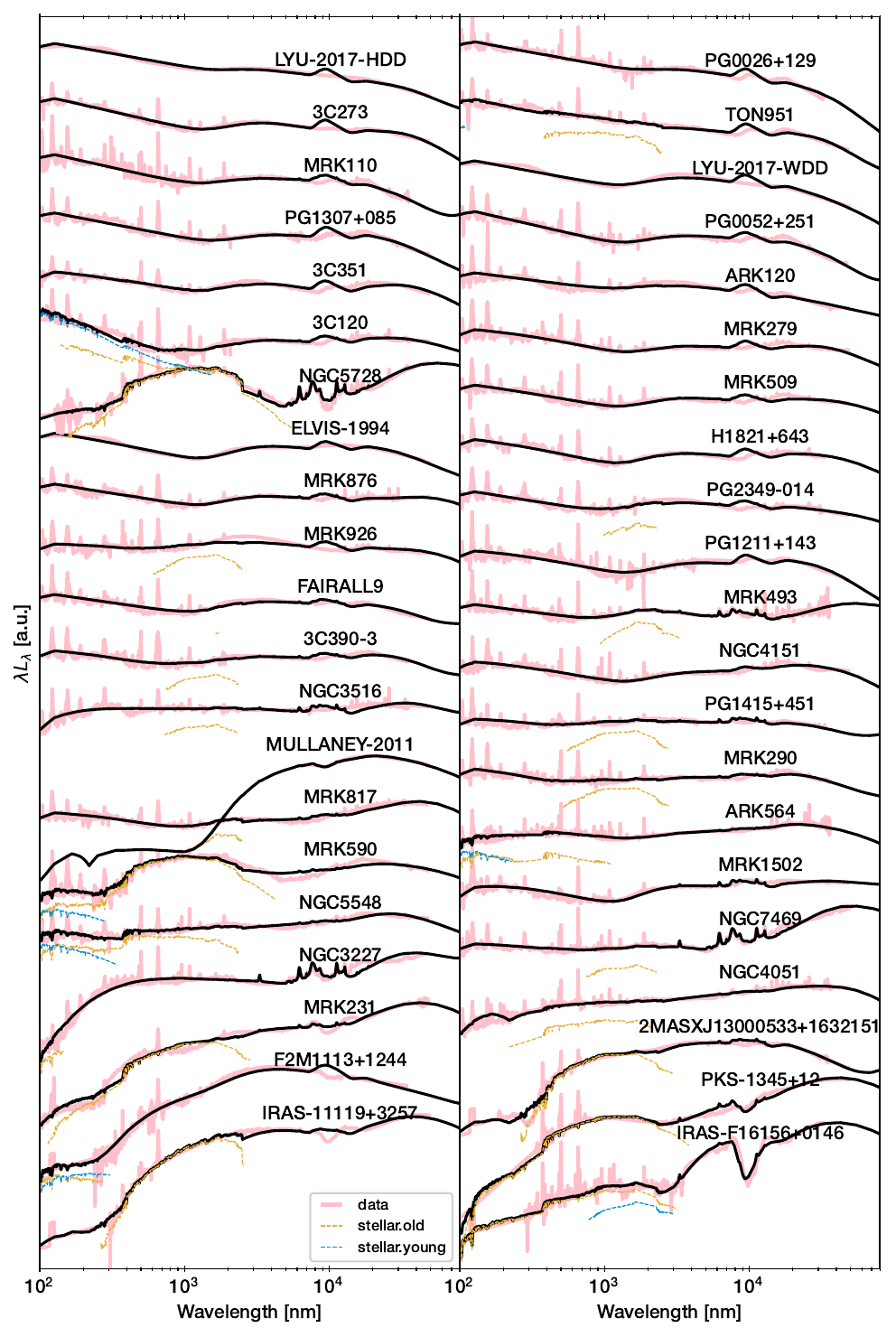}
    \caption{Like \cref{fig:browncomparison}, but comparing CIGALE's SKIRTOR model (black) with the Atlas of AGN spectra (red) from \cite{Brown2019}.
    }
    \label{fig:calibration-brown-SKIRTOR}
\end{figure*}

\Cref{fig:calibration-brown-SKIRTOR} presents the best fits. Because this model lacks AGN emission lines, in the UV-opt range, the lines are not reproduced. Further, at 10$\si{\micro\meter}$, the model typically over-estimating the emission. For some sources (such as Mrk926, 3C351, Mrk590, Ark120) the continuum also has deviations in the 2-10$\si{\micro\meter}$ range, which are not present in \cref{fig:browncomparison}.

\begin{figure}
    \centering
    \includegraphics[width=\columnwidth]{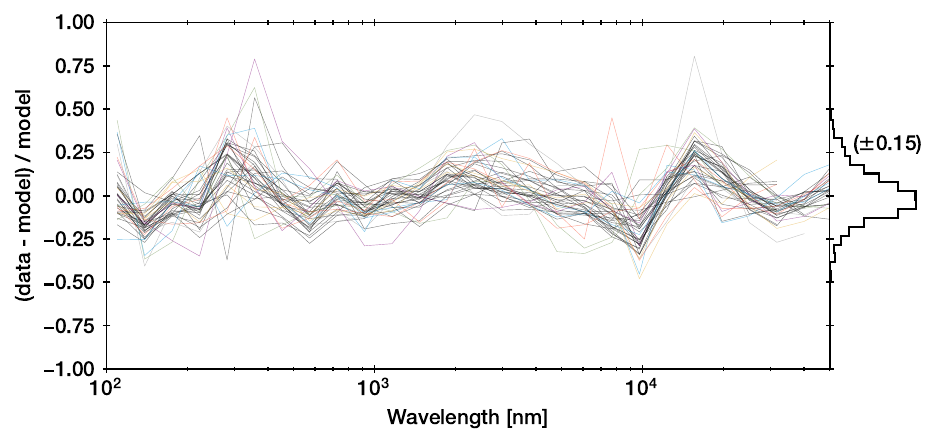}
    \caption{Like \cref{fig:brownresid}, but with CIGALE's SKIRTOR model.}
    \label{fig:calibration-brown-SKIRTOR-resid}
\end{figure}

The model residuals are presented in \cref{fig:calibration-brown-SKIRTOR-resid}. Here, discrepancies of 25 per cent are apparent from 10 (negative) to 15$\si{\micro\meter}$ (positive), indicating that the Si feature is imperfectly modelled. Near 2-4$\si{\micro\meter}$, a hump is apparent, indicating that the hot dust model is incomplete. Near 200-400$\si{\nano\meter}$, structure is also apparent in the residual, possibly related to unmodelled emission lines (FeII forest), giving rise to a 20 per cent discrepancy. 

In contrast to the residuals shown in \cref{fig:browncomparison}, the residuals in \cref{fig:calibration-brown-SKIRTOR-resid} are not random from object to object, but show a systematic pattern. This implies that photometric model colours will be systematically shifted with CIGALE when modelling galaxies hosting AGN. As discussed in \cref{sec:Method}, we suspect such residuals to be the cause for the stellar mass and SFR bias identified in the CIGALE results of this paper. The correlated residuals cannot be fully resolved by adding independent uncertainties to each measured flux. For the reduced bias of GRAHSP when compared to CIGALE, the number of parameters are not the only difference. Here, we have left 16 parameters free to vary but residuals remain. However, the computation and memory usage of CIGALE cannot handle such a high number of parameters when fitting photometry.

\listoftodos[]

\end{document}